\documentclass[aps,superscriptaddress,nofootinbib,showpacs,eqsecnum,preprint,tightenlines]{revtex4}
\usepackage{hyperref}
\usepackage{epsfig,rotating}
\usepackage{amsmath,amssymb}
\usepackage{dsfont}
\usepackage{bbm}
\usepackage{slashed}
\numberwithin{equation}{section}

\newcommand{\be}{\begin{equation}}
\newcommand{\ee}{\end{equation}}
\newcommand{\bea}{\begin{eqnarray}}
\newcommand{\eea}{\end{eqnarray}}

\newcommand{\vx}{\vec{x}}

\newcommand{\vp}{\vec{p}}

\newcommand{\vq}{\vec{q}}

\newcommand{\vk}{\vec{k}}

\begin{document}

\title{The case for mixed dark matter from sterile neutrinos.}

\author{Louis Lello}
\email{lal81@pitt.edu}

\affiliation{Department of Physics and Astronomy, University of Pittsburgh, Pittsburgh, PA 15260, USA}
\affiliation{Department of Physics, Brookhaven National Laboratory, Upton, N.Y., 11973, USA}

\author{Daniel Boyanovsky}
\email{boyan@pitt.edu}

\affiliation{Department of Physics and Astronomy, University of Pittsburgh, Pittsburgh, PA 15260, USA}

\date{\today}

\begin{abstract}
Sterile neutrinos are $SU(2)$ singlets that mix with active neutrinos via a mass matrix, its diagonalization leads to mass eigenstates that couple via standard model vertices.   We study the cosmological production of heavy neutrinos via \emph{standard model charged and neutral current vertices}  under a minimal set of assumptions: i)   the mass basis contains a \emph{hierarchy of heavy neutrinos}, ii)    these have very small mixing angles with the  active (flavor) neutrinos, iii)   standard model particles, including light (active-like) neutrinos are in thermal equilibrium.   If kinematically allowed, the same weak interaction processes that produce  active-like neutrinos also produce the heavier species.    We introduce the quantum kinetic equations that describe their production, freeze out and decay and discuss the various processes that lead to their production in a wide range of temperatures assessing their feasibility as dark matter candidates. The final distribution function at freeze-out is a \emph{mixture} of the result of the various production processes.
We identify processes in which finite temperature \emph{collective excitations} may lead to the production of the heavy species. As a specific example, we consider the production of heavy neutrinos in the mass range $M_h \lesssim 140 \,\mathrm{MeV}$ from pion decay shortly after the QCD crossover including finite temperature corrections to the pion  form factors and mass. We consider the different decay channels that  allow for the production of heavy neutrinos showing that their frozen distribution functions exhibit effects from ``kinematic entanglement'' and argue for their viability as mixed dark matter candidates. We discuss abundance, phase space density and stability constraints and argue that heavy neutrinos with lifetime $\tau> 1/H_0$ freeze out of local thermal equilibrium, and \emph{conjecture} that those with lifetimes $\tau \ll  1/H_0$ may undergo  cascade decay   into lighter DM candidates and/or inject non-LTE neutrinos into the cosmic neutrino background. A comparison is made between production through pion decays with the production of non-resonant production via active-sterile mixing.

\end{abstract}

\keywords{Cosmological Neutrinos, Dark Matter Theory, Neutrino Theory, Physics of the Early Universe}

\maketitle

\section{Introduction}

The present understanding of the large scale cosmological evolution is governed by unknown quantities, dark matter and dark energy \cite{book1}, where the commonly accepted wisdom points towards a dark matter candidate which is a weakly-interacting, cold thermal relic (WIMP) \cite{wimpreview}. Dark matter candidates alternative to the WIMP paradigm attract a fair amount of attention\cite{axionreview,sterilereview} due to several observational shortcomings of the standard cold dark matter cosmology ($\Lambda CDM$) at small scales. N-body simulations of cold dark matter produce dark matter profiles that generically feature cusps   yet observations suggest a smooth-core  profile \cite{corecusp,dalcantonhogan}(core-cusp problem). The same type of N-body simulations also predict a large number of dark matter dominated satellites surrounding a typical galaxy which is inconsistent with current observations \cite{toobig} (missing satellites problem). Both the missing satellites and core-cusp problem can be simultaneously resolved by allowing some fraction of the dark matter to be ``warm" (WDM) \cite{wdm1,wdm2,wdm3,wdm4,wdm5,wdm6} with a massive ``sterile" neutrino being one popular candidate \cite{warmdm,dodwid,dodwid2,abazajian3,sterilesexperiment} - other examples include Kaluza-Klein excitations from string compactification or axions \cite{anupam,axionreview}. The ``hotness" or ``coldness" of a dark matter candidate is discussed in terms of its free streaming length, $\lambda_{fs}$, which is the cut-off scale in the linear power spectrum of   density perturbations. Cold dark matter (CDM) features small ($\lesssim$ pc) $\lambda_{fs}$ that brings about cuspy profiles whereas WDM features $\lambda_{fs} \sim \mbox{few kpc}$ which would lead to cored profiles. Recent WDM simulations including velocity dispersion suggest the formation of cores but do not yet reliably constrain the mass of the WDM candidate in a model independent manner since the distribution function of these candidates is also an important quantity which determines the velocity dispersion and thereby the free streaming length\cite{padu}.

For most treatments of sterile neutrino dark matter, a nonthermal distribution function is needed in order to evade cosmological bounds \cite{planck}. The mechanism of sterile neutrino production in the early universe through oscillations was originally studied in ref. \cite{dolgovenqvist} (see also the reviews \cite{dolgovreview}) and in \cite{dodwid,shifuller,asaka,laine,abanew} sterile neutrinos are argued to be a viable warm dark matter candidate produced out of LTE in the absence (Dodelson-Widrow, DW) or presence (Shi-Fuller, SF) of a lepton asymmetry. Models in which a standard model Higgs scalar decays into a pair of sterile neutrinos at the electroweak  scale  (or higher) also yield an out-of-equilibrium distribution suitable for a sterile neutrino dark matter candidate \cite{boyan1,kusenko1,petraki,merle}. Observations of the Andromeda galaxy with Chandra led to tight constraints on the (DW) model of sterile neutrinos\cite{casey}, and more recently the observation of a  3.5 keV signal from the XMM Newton X-ray telescope has been argued to be due to a 7 keV sterile neutrino\cite{bulbul,boyarsky}, a position which has not gone unchallenged \cite{jeltema,malysh,anderson,sekiya}. The prospect of a keV sterile DM candidate continues to motivate theoretical and observational studies \cite{merle,abazajian0,abazajian,abazajian2,kaplinghat2,kaplinghat3,kusenko1,lellolightsterile,added2,abanew}. In all of these production mechanisms, the assumption of a vanishing initial population is implemented and, as we discuss, there are a wide array of processes which are ignored with this simplification - a problem which is starting to become appreciated \cite{merleini}.

On the particle physics front, neutrino masses, mixing and oscillations are the first evidence yet of physics beyond the standard model. A significant world-wide experimental program has brought measurements of most of the parameters associated with  neutrino mass \cite{pdg,neutrinoexperiments} with several significant remaining questions poised to be answered in the near future \cite{upcomingexperiments}.   Short baseline   neutrino oscillation experiments (LSND, MiniBooNE) \cite{lsnd,miniboone} present a picture of the neutrino sector which may require an additional sterile neutrino species of mass $\sim 1 eV$ \cite{sterilesexperiment,giunti2,mirizzi} but there remains tension with other experiments \cite{nosterilesexperiment} and a definitive resolution of these anomalies will require further experiments \cite{lasserre,decay,lello1,shrock,barger,giunti}.  An interpretation of short baseline experiment anomalies as a signal for sterile neutrinos leads to a   relatively light mass $\sim eV$ for the sterile neutrino; however,  it has been argued that sterile neutrinos with mass on the order of $\mathrm{MeV}$ or larger \cite{gninenko} can decay and could also explain the short baseline anomalies or, alternatively, heavy sterile neutrinos produced through rare decay channels could also explain the anomaly \cite{heavysterile}. Well motivated proposals make the case for a robust program to  search for  multiple heavy neutrinos \cite{heavyoscil,boyancascade} in a wide range of experiments including  hadron colliders \cite{added1,added3,han,pilafsis}. Furthermore, it has been argued that heavy sterile neutrinos in the mass range $100-500 \,\mathrm{MeV}$   can decay nonthermally and enable evasion of cosmological and accelerator bounds\cite{fullkuse}.

Several current experimental programs seek direct detection of sterile neutrinos : the KATRIN (Karlsruhe Tritium Neutrino Experiment) experiment\cite{drexlin,mertens} searches for sterile neutrinos with masses up to $\lesssim 18 \,\mathrm{keV}$ in tritium beta decay, the MARE experiment\cite{mare} (Microcalorimeter Arrays for a Rhenium Experiment) explores the mass range $\lesssim 2.5 \,\mathrm{keV}$ in the beta decay of Rhenium 187, the ECHo experiment (Electron Capture $^{163}\,Ho$ Experiment)\cite{echo} searches for sterile neutrinos in the mass range $\lesssim 2  \,\mathrm{keV}$ in beta decay of $^{163}\,Ho$. Various recent proposals make the case for searches of \emph{heavy}   neutrinos at the Large Hadron Collider\cite{added1,added3,han,pilafsis} and current and future underground neutrino detectors may be able to probe dark matter candidates with $\simeq \mathrm{few}\,\mathrm{MeV}$\cite{pospelov}. In extensions beyond the standard model, the possibility of a hierarchy of heavy neutrinos would be  natural: there is a (very wide) hierarchy of quark masses, charged lepton masses, and as is now clear a hierarchy of light massive neutrinos.

Models of many dark matter components had been proposed recently\cite{dienes}, these models provide the tantalizing possibility that the decay of one heavy component can seed the production of a lighter component. As we will discuss below, this possibility also arises if there is a hierarchy of heavy neutrinos.

 \vspace{5mm}

\vspace{5mm}

\textbf{Motivation and Goals:}
As discussed above the   study of sterile neutrinos with masses in a wide range, from few
$eV$ through $\mathrm{keV}$ up to several $GeV$ is motivated from astrophysical observations, cosmological simulations, terrestrial accelerators experiments and the new generation of experiments that will directly search for signals of sterile neutrinos. Most theoretical extensions beyond the Standard Model that provide mechanisms to generate neutrino masses invoke one or more generations of heavy ``sterile'' neutrinos. While the focus on sterile neutrinos as a dark matter candidate has been on the mass range of few $keV$ (largely motivated by the cusp-vs-core and related problems of structure formation,  and more recently by the possible detection of an X-ray line at $\simeq 7 \,\mathrm{keV}$), \emph{if extensions beyond the standard model allow for a hierarchy of heavy   neutrinos} these \emph{may} yield a mixture of warm, cold (and hot) dark matter. This possibility motivates us to explore possible mechanisms of production of heavier species of \emph{massive} neutrinos  and assess in particular cases their production, freeze-out and possible consequences of non-thermal distribution functions.

\emph{Mixed dark matter} described by several species of massive neutrinos with non-equilibrium distribution functions will certainly evade   Lyman-$\alpha$ constraints\cite{lymanboyar}.

Rather than proposing yet new models, the purpose of this work is twofold:
\begin{itemize}
\item \textbf{i):} to understand the production and freeze out of heavy sterile-like neutrinos \emph{from standard model charged and neutral current interaction vertices} under a \emph{minimal set of assumptions}, assessing their suitability as DM candidates. We argue that different processes with different kinematic channels are available in a wide range of temperatures and produce heavy neutrinos with different contributions to their distribution functions. The total distribution function is, therefore, a \emph{mixture} of the   contributions from the various different processes.

   \item \textbf{ii):}  to provide an explicit example within the context of the well understood effective field theory of weak interactions of pions including finite temperature corrections. Pion decay is a production mechanism that is available shortly after the QCD crossover and produces heavy neutrinos from different kinematic channels. Therefore furnish an explicit example of the \emph{mixed} nature of the distribution function.

\end{itemize}

The \emph{minimal set of assumptions} are the following:

\begin{itemize}
\item{ We assume that the massless flavor neutrinos fields of the standard model are related to the fields that create and annihilate the    mass eigenstates as
    \be \nu_{\alpha,L}(x) = \sum_{m=1,2,3} U_{\alpha m}\,  \nu_{m,L}(x) + \sum_{h=4,5\cdots} H_{\alpha h}\,\nu_{h,L}(x) \label{relanus} \ee where $m=1,2,3$ refer to the light mass eigenstates that explain atmospheric and solar oscillation observations, $h=4,5\cdots$ refer to heavy mass eigenstates. We assume a hierarchy of masses with $M_m \ll M_h$. The heavier mass eigenstates correspond to the various putative massive neutrinos with $M_h \simeq 1 \,\mathrm{eV}$ to explain the LSND/MiniBooNE anomalies, or the $M_h \simeq 50-80 \,\mathrm{MeV} $ proposed to explain these anomalies via radiative decay\cite{gninenko}, or $M_h \simeq 7 \,\mathrm{keV}$ that could explain the X-ray line, or even $M_h \simeq \,\textrm{few}\,\textrm{MeV}-\textrm{GeV}$ that could be a CDM candidate. We do not endorse nor adopt any particular extension beyond the Standard Model with particular mass generation mechanisms. We only assume the relationship (\ref{relanus}) and the existence of a hierarchy of very massive neutrinos with mass scales well separated from the atmospheric and solar ones.   }

    \item{We assume that $|U_{\alpha m}| \simeq \mathcal{O}(1)$ and $|H_{\alpha h}| \ll |U_{\alpha m}|$ although this is a feature of generic see-saw type mechanisms, we do \emph{do not invoke} a particular mechanism.  We then write the charged and neutral current weak interaction vertices in the mass basis and keep
        \emph{only the linear terms} in $H_{\alpha h}$. We \emph{only} consider the production of the heavy species with $M_h$ from these weak interaction vertices in the \emph{mass basis}. In this manner the production of heavy neutrinos is similar to the production of standard model neutrinos via charged and neutral current vertices \emph{if the kinematics allows for the particular process to produce a heavy neutrino mass eigenstate}.  }

        \item{Our last assumption is that the light mass eigenstates (active-like) $\nu_{1,2,3}$ along with all the other standard model particles are in thermal equilibrium. We will assume that the relevant processes occurred during the radiation dominated era with $T \geq 0.1 \,\textrm{MeV}$ at which standard model active neutrinos decouple.  }

\end{itemize}

This minimal set of assumptions allows us to implement  well understood standard methods from quantum kinetic theory to describe the production of heavy neutrinos from  weak interaction vertices and
asses their distribution function when the production freezes and discuss finite temperature corrections to the production processes.

Here we do \emph{not} consider the possibility that the masses of sterile neutrinos is a consequence of Yukawa coupling to the Standard Model Higgs, as this entails a particular model. Production of heavy neutrinos from scalar decay has been discussed in refs.\cite{kusenko1,petraki,boyan1,merle,twohiggs}.

As a definite example we  implement this program in  the particular case of production of   heavy neutrinos from pion decay after the QCD transition (crossover) within the well understood effective field theory of charged pion decay  including finite temperature corrections to the pion decay constant and mass. This production channel is one of the most ubiquitous sources of neutrinos in accelerator experiments. Furthermore, pion decay shortly after the hadronization transition (at $t\simeq 10\,\mu s$) in the Early Universe certainly leads to  neutrino production just as    in accelerator experiments.

  It is shown that this production mechanism yields heavy neutrinos from different channels ($\mu,e$) which freeze out with highly non-thermal distributions, and furnishes a definite example of ``kinematic entanglement'' with important  cosmological consequences on their clustering properties.

\textbf{Brief Summary of Results:}

\begin{itemize}

    \item  In the mass basis the same standard model processes that lead to the production of active-like (light) neutrinos yield heavy (``sterile-like'') neutrinos if   kinematically allowed. We identify several processes that lead to the cosmological production of heavy neutrinos during the radiation dominated era in a wide range of temperatures including $\gamma\nu_m \rightarrow \nu_h$ whose inverse process describes the X-ray telltale of sterile neutrinos as well as  the possibility of production from \emph{collective excitations in the medium}. To leading order in the mixing matrix elements $H_{\alpha h}$ we obtain the quantum kinetic equations and give their exact solution in terms of gain and loss rates that   obey detailed balance and can be calculated with standard model rules. We analyze the possibility of thermalization and argue that heavy neutrinos with lifetimes larger than $1/H_0$ will freeze-out with non-equilibrium distribution functions. We establish the bounds from abundance, coarsed grained phase space density (Tremaine-Gunn) and \emph{stability} for suitable DM candidates and discuss clustering properties of heavy neutrino species in terms of the distribution function after freeze-out.

    \item  We generalize the concept of \emph{mixed DM} to encompass not only different species of heavy neutrinos, but also  the different distributions functions of a single species of mass $M_h$ from different production channels. We argue that heavy neutrinos produced via standard model charged and neutral current interactions are \emph{kinematically entangled} with leptons produced in the same reaction. This kinematic entanglement leads to different distribution functions for different channels of production of a given species, some colder than others depending on the mass of the lepton, quantified by the equation of state in each case. The total distribution function is therefore a \emph{mixture} of all the different distribution functions from each channel. If the heavy neutrino does not thermalize (and those with lifetimes larger than $1/H_0$ do not), its distribution function at freeze-out exhibit memory of this kinematic entanglement. This memory is manifest in the equation of state, velocity dispersion, coarse grained phase space density and free streaming length.

    \item As a specific example, we study the production of heavy neutrinos after the QCD transition (crossover) from pion decay within the effective field theory description of weak interactions of pions  including finite temperature corrections to the pion decay constant and mass in a large kinematically allowed  window $M_h \lesssim 140\,\mathrm{MeV}$. Specifically we study in detail the distribution function from both available channels, $\pi \rightarrow \mu \nu_h~;~\pi \rightarrow e \nu_h$, which furnishes a clear example of the ``kinematic entanglement'' with the corresponding charged lepton. The distribution function from the $\mu$ channel is distinctly \emph{colder} than that from the $e$ channel and the total distribution function in this case is a sum (mixture) of both components, with concentrations depending on   the \emph{ratio} of mixing matrix elements $|H_{\alpha h}|^2$ for each channel. We assess heavy neutrinos produced via this mechanism as possible DM candidates by analyzing the allowed region of parameters that fulfills the abundance, phase space density and stability constraints, highlighting how the phenomenon of kinematic entanglement is manifest along the boundaries of the allowed regions. We also study the equation of state (velocity dispersion) and free streaming length (cutoff in the power spectrum) and show that both interpolate between the colder ($\mu$-channel) and warmer (e-channel) components as a function of $M_h$ and the \emph{ratio} $|H_{eh}|^2/|H_{\mu h}|^2$ showing explicitly the \emph{mixed} nature of the distribution function.

        \item We \emph{conjecture} that heavy neutrinos with lifetimes \emph{shorter} than $1/H_0$ produced during the radiation dominated era may decay in a \emph{cascade} into active-like neutrinos well after Big Bang Nucleosynthesis, providing a late injection of   neutrinos out of thermal equilibrium into the cosmic neutrino background, and possibly, into lighter (but still heavy) and stable(r) neutrinos that could be DM candidates after matter radiation equality.

\end{itemize}

\section{Production and freeze-out: quantum kinetics}\label{sec:qk}
Sterile neutrinos are $SU(2)$ weak singlets that do not interact via standard model weak interactions, they only couple to the massless, flavor active neutrinos via an off diagonal mass matrix. This is the general description of sterile neutrinos, different models propose different types of (see-saw like) mass matrices. For our purposes the only relevant aspect is that the diagonalization of the mass matrix leads to mass eigenstates, and these are ultimately the relevant degrees of freedom that describe dark matter in its various possible realizations, cold, warm or hot. Cold and warm dark matter would be described by heavier species, whereas the usual (light) active-like neutrinos could be a hot dark matter component depending on the absolute value of their masses.
As mentioned above in this article we focus on the production of heavy neutrinos from \emph{standard model} couplings: in the mass basis, heavy neutrinos couple via standard model charged and neutral current interactions albeit with very small mixing matrix elements. If kinematically allowed, the same processes that produce active neutrinos will produce heavy (sterile-like) neutrinos with much smaller branching ratios.

We consider the standard model augmented by neutrino masses under the assumption of a hierarchy of heavy neutrinos with masses much larger than those associated with the (light-active like) atmospheric and solar neutrinos. Upon diagonalization of the mass matrix, the active (massless) flavor neutrino fields  are related to the neutrino fields that create/annihilate mass eigenstates as
\be    \nu_{\alpha,L}(x) = \sum_{m=1,2,3} U_{\alpha m}\,  \nu_{m,L}(x) + \sum_{h=4,5\cdots} H_{\alpha h}\,\nu_{h,L}(x) \label{relanus2} \ee where $\alpha = e,\mu,\tau$, $m=1,2,3$ refer to the light (atmospheric, solar) mass eigenstates  with massess $M_m$  and  $h=4,5\cdots$ refer to heavy mass eigenstates of mass $M_h$. The form of the mixing matrix is taken in a way that is \emph{model independent} in that no assumptions are made as to the particular source of the neutrino masses and mixing. No specification of the number of Dirac or Majorana mass terms nor their source is made. In particular, we do not assume that any Dirac mass terms are generated through Yukawa couplings to the Standard Model Higgs as this constitutes a choice of one particular model. We focus \emph{solely} on production processes from  standard model charged and neutral interactions.

As discussed above our main assumptions are
\be |U_{\alpha m}| \simeq \mathcal{O}(1) \gg |H_{\alpha h}|~~;~~ M_h \gg M_m \, \label{assu}\ee and we assume that charged and neutral vector bosons, quarks, leptons and light $\nu_{1,2,3}$  neutrinos are all in thermal equilibrium. This assumption is justified with the usual arguments that the standard model interaction rates are much larger than the expansion rates down to $T \simeq 1\,\mathrm{MeV}$ when active-like neutrinos decouple from the plasma.  To present the arguments in the simplest terms, we will consider vanishing chemical potentials for all relevant species, \emph{a posteriori}, it is straightforward to include  lepton asymmetries with chemical potentials. Furthermore, we will not distinguish between Dirac and Majorana neutrinos as the difference typically results in factors $2$ in various transition probabilities. Lastly, we will not consider the possibility of CP-violation, implying that forward and backward (gain and loss) transition probabilities are the same. All of these assumptions can be relaxed with the concomitant complications which will be addressed elsewhere.

 In the mass basis the full Lagrangian density is
\be \mathcal{L} = \mathcal{L}_{mSM} + \sum_{h=4,5\cdots} \mathcal{L}_{0h} + \mathcal{L}_I \,, \label{totallag} \ee where $\mathcal{L}_{mSM}$ is the Standard Model Lagrangian augmented with a diagonal neutrino mass matrix for the active like neutrinos $\nu_m$ of masses $M_m$ ($ m=1,2,3$), in this Lagragian density the charged and neutral interaction vertices are written in terms of the neutrino mass eigenstates with the mixing matrix $U_{\alpha  m}$,  $\mathcal{L}_{0h}$ is the free field Lagrangian density for the heavier neutrinos $\nu_h$ of masses $M_h$ ($ h= 4,5\cdots$) and to linear order in $H_{\alpha h}$
\bea \mathcal{L}_I & = & -\frac{g_w}{\sqrt{2}} W^+_\mu \,\sum_{\alpha=e,\mu,\tau}\,\sum_{h=4,5\cdots} H^*_{h \alpha}\overline{\nu}_h(x)\gamma^\mu \mathcal{P}_L l_\alpha(x) + h.c \nonumber \\
& - & \frac{g_w}{ {2}\cos\theta_W} Z^0_\mu  \,\sum_{h=4,5\cdots}\sum_{m=1,2,3} \widetilde{H}^*_{h m}\overline{\nu}_h(x)\gamma^\mu \mathcal{P}_L \nu_m(x) + h.c. \,, \label{Lint} \eea  where
\be \widetilde{H}^*_{h m} = \sum_{\alpha=e,\mu,\tau} H^*_{h \alpha}U_{\alpha m}\,,\ee with $\mathcal{P}_L = (1-\gamma_5)/2$. From now onwards, we refer to heavy neutrinos instead of ``sterile'' neutrinos, because unlike the concept of a sterile neutrino, heavy neutrinos do interact with standard model degrees of freedom via charged and neutral current vertices. The new mass eigenstates will undoubtedly contribute to the tightly constrained ``invisible width'' of the $Z^0$ \cite{pdg}; however, the contribution from the heavier neutrinos is suppressed with respect to the light active-like neutrinos by  small branching ratios $Br \simeq |H_{l h}|^2/|U_{lm}|^2 \ll 1$ and the tight constraints on the number of neutrinos contributing to the width of the $Z^0$ may be evaded by very small matrix elements, which is, in fact, one of the underlying assumptions of this work.

The strategy is to pass to the Hamiltonian
\be H = H_{mSM}+H_0(\nu_h)+H_I \equiv H^{(0)}+ H_I \label{tothami} \ee where $H^{(0)}$ is the \emph{total } Hamiltonian of the Standard Model in the mass basis of the light (active-like) neutrinos plus the free field Hamiltonian of the heavy neutrinos $h= 4,5\cdots$ and $H_I$ is the interaction Hamiltonian  obtained from $\mathcal{L}_I$ in  (\ref{Lint}). We now pass to the interaction picture of $H^{(0)}$ wherein
\be H_I(t) = e^{iH^{(0)}t} H_I e^{-iH^{(0)}t} \,, \label{HIip}\ee  namely we will obtain the quantum kinetic equations to leading order in $g_w|H_{\alpha h}|\ll g_w$ but \emph{in principle  to all orders } in the weak interaction coupling $g_w$ (more on this issue below). In terms of the interaction vertices (\ref{Lint}) we can obtain the quantum kinetic equations for production of massive neutrinos to   order    $(|H_{\alpha h}|)^2$  from standard master (gain-loss) equations of the form
\be \frac{d n_h(q;t)}{dt} = \frac{d n_h(q;t)}{dt}\big|_{\textrm{gain}}-\frac{d n_h(q;t)}{dt}\big|_{\textrm{loss}}\,,\label{quakin}\ee where $n_h(q;t)$ is the distribution function of the heavy neutrino. This is calculated for each process: the gain term describes the increase in the population $n_h(q;t)$ by the creation of a heavy state and the loss by the annihilation, the best way to understand the calculational aspects is with a few examples.

\vspace{2mm}

\subsection{Setting the stage in Minkowski space-time.}

\vspace{2mm}

We begin by describing the main aspects in \textbf{Minkowski space time} to highlight  important concepts, and generalize the formulation to an expanding cosmology in the next subsection.

 To begin with consider a temperature $M_{W,Z} \lesssim T \ll T_{EW}$ where $T_{EW}\simeq 160 \textrm{GeV}$ is the temperature of the electroweak transition. In this temperature range the massive vector bosons (described as three physical degrees of freedom in unitary gauge) are populated in thermal equilibrium in the plasma with the Bose-Einstein distribution functions. If the mass of the heavy neutrino is $M_h + m_\alpha < M_{W,Z}$ the massive vector Bosons can \emph{decay} into the massive neutrino, thereby contributing to the gain term. For example, take $W^+\rightarrow l^+_\alpha \nu_h$; each $l_\alpha$ and each $\nu_h$ constitute different decay channels which lead to a \emph{gain} contribution whereas the \emph{loss} contribution is the inverse process $ l^+_\alpha \nu_h \rightarrow W^+$. The gain and loss terms are calculated by obtaining the total transition probability per unit time to a particular channel where an explicit calculation is detailed in appendix (\ref{kinetics}). For $W^{\pm} \rightleftharpoons l_{\alpha}^{\pm} \nu_h$, it is straightforward to find
\bea \frac{d n_h(q;t)}{dt}\big|_{\textrm{gain}} & = &  \frac{2\pi}{2E_h(q)} \int \frac{d^3k\,\overline{|\mathcal{M}_{fi}|^2}}{(2\pi)^3 2E_W(k)2E_\alpha(p)} \,n_W(k)(1-n_\alpha(p))(1-n_h(q;t)) \nonumber  \\ & \times & \,\delta(E_W(k)-E_\alpha(p)-E_h(q)) \label{gainW} \\
\frac{d n_h(q;t)}{dt}\big|_{\textrm{loss}} & = &  \frac{2\pi}{2E_h(q)} \int \frac{d^3k\,\overline{|\mathcal{M}_{fi}|^2}}{(2\pi)^3 2E_W(k)2E_\alpha(p)} \,(1+n_W(k))n_\alpha(p)n_h(q;t)\nonumber \\ & \times & \,\delta(E_W(k)-E_\alpha(p)-E_h(q)) \,,\label{lossW}\eea where $p= |\vec{k}-\vec{q}|$ and
\be n_W(k) = \frac{1}{e^{E_W(k)/T}-1} ~~;~~ n_\alpha(p) = \frac{1}{e^{E_\alpha(p)/T}+1} \,,\label{disLTE}\ee and
$\overline{|\mathcal{M}_{fi}|^2}$ is the usual transition matrix element for $W^+\rightarrow l^+_\alpha \nu_h$ summed over the three polarizations of the vector Boson, summed over the helicity of the charge lepton and summed over the helicity of the heavy neutrino; obviously $\overline{|\mathcal{M}_{fi}|^2}\propto g^2_w |H_{\alpha h}|^2$. Therefore the quantum kinetic equation (\ref{quakin}) becomes of the form
\be \frac{d n_h(q;t)}{dt} = \Gamma^<(q)(1-n_h(q;t)) - \Gamma^>(q) n_h(q;t) \label{gainminlos} \ee where the gain and loss \emph{rates} are
\be \Gamma^<(q) =  \frac{2\pi}{2E_h(q)} \int \frac{d^3k\,\overline{|\mathcal{M}_{fi}|^2}}{(2\pi)^3 2E_W(k)2E_\alpha(p)} \,n_W(k)(1-n_\alpha(p))\,\delta(E_W(k)-E_\alpha(p)-E_h(q)) \label{gainrate} \ee
\be \Gamma^>(q) =  \frac{2\pi}{2E_h(q)} \int \frac{d^3k\,\overline{|\mathcal{M}_{fi}|^2}}{(2\pi)^3 2E_W(k)2E_\alpha(p)} \,(1+n_W(k)) n_\alpha(p)\,\delta(E_W(k)-E_\alpha(p)-E_h(q)) \,.\label{lossrate} \ee Because the $W,l_\alpha$ are in thermal equilibrium the gain and loss rates obey the \emph{detailed balance condition}
\be \Gamma^<(q)\,e^{E_h(q)/T} = \Gamma^>(q)\,, \label{detailedbalance} \ee which can be confirmed straightforwardly from the explicit expressions (\ref{gainrate},\ref{lossrate})  using the energy conserving delta functions and the relations
\be 1+n_B(E) = e^{E/T} n_B(E)~~;~~ 1-n_F(E) = e^{E/T} n_F(E) \label{relas}\ee where $n_{B,F}$ are the Bose-Einstein and Fermi-Dirac distribution functions respectively.

   A similar exercise yields the quantum kinetic equation for the process $Z^0 \rightarrow \overline{\nu}_m \nu_h$ and the inverse process, it is of the same form as (\ref{gainminlos}) now with
\be \Gamma^<(q) =  \frac{2\pi}{2E_h(q)} \int \frac{d^3k\,\overline{|\mathcal{M}_{fi}|^2}}{(2\pi)^3 2E_Z(k)2E_m(p)} \,n_Z(k)(1-n_m(p)) \,\delta(E_Z(k)-E_m(p)-E_h(q))\label{gainrateZ} \ee
\be \Gamma^>(q) =  \frac{2\pi}{2E_h(q)} \int \frac{d^3k\,\overline{|\mathcal{M}_{fi}|^2}}{(2\pi)^3 2E_Z(k)2E_m(p)} \,(1+n_Z(k)) n_m(p) \delta(E_Z(k)-E_m(p)-E_h(q))\,,\label{lossrateZ} \ee where now $\overline{|\mathcal{M}_{fi}|^2}\propto g^2_w |\widetilde{H}_{hm}|^2$ is the transition matrix element for the process $Z^0 \rightarrow \overline{\nu}_m \nu_h$ from the charged current vertex in (\ref{Lint}). Again because the vector Boson and the active-like neutrino $\nu_m$ are in thermal equilibrium, the gain and loss rates obey the detailed balance condition (\ref{detailedbalance}).

As highlighted above, the interaction picture corresponds to considering the usual standard model vertices with light neutrinos $\nu_m$ to \emph{all} orders in charged and neutral current weak interactions, as well as, in principle to all orders in strong interactions. This feature, in fact, leads to many different production channels of heavy neutrinos, provided the kinematics allows the particular channels. In thermal equilibrium the typical energy of quarks or leptons whose masses are $\ll T$ is $\langle E \rangle \simeq 3.15 \, T$ so that for $M_h \lesssim T$ various higher order processes are available; for example, $\overline{f} f \rightarrow Z^0 \rightarrow \nu_m \nu_h$ and its inverse process where $f,\overline{f}$ refer to quark-antiquark or lepton-antilepton or  $\nu_m \overline{\nu}_m$ and similar charged current processes involving either quarks or charged leptons. If $T \simeq M_Z, M_W$ the intermediate $Z^0, W^\pm$ are on-shell, with $n_Z(E), n_W(E) \simeq 1$ for $E\simeq T$    and the process is ``resonantly'' enhanced, the gain and loss terms are of the form
\bea \frac{d n_h(q;t)}{dt}\big|_{\textrm{gain}} & = &  \frac{2\pi}{2E_h(q)} \int \frac{d^3k_1 d^3k_2\,\overline{|\mathcal{M}_{fi}|^2}}{(2\pi)^6 2E_f(k_1)2E_{\overline{f}}(k_2)2E_m(p)}\,n_f(k_1)n_{\overline{f}}(k_2)(1-n_m(p))(1-n_h(q;t))
\nonumber \\ & \times & \delta(E_f(k_1)+E_{\overline{f}}(k_2)-E_m(p)-E_h(q)) \label{gain3} \\ \frac{d n_h(q;t)}{dt}\big|_{\textrm{loss}} & = &  \frac{2\pi}{2E_h(q)} \int \frac{d^3k_1 d^3k_2\,\overline{|\mathcal{M}_{fi}|^2}}{(2\pi)^6 2E_f(k_1)2E_{\overline{f}}(k_2)2E_m(p)}\,(1-n_f(k_1))(1-n_{\overline{f}}(k_2))n_m(p)n_h(q;t)\nonumber \\ & \times &
\delta(E_f(k_1)+E_{\overline{f}}(k_2)-E_m(p)-E_h(q)) \label{loss3}\eea  where $p=|\vec{k}_1+\vec{k}_2-\vec{q}|$ and the contribution from the intermediate vector Boson is included in $\overline{|\mathcal{M}_{fi}|^2}$, wherein the $Z^0$ propagator must include the width (with the corresponding thermal contributions) because for $T \simeq M_Z$ the intermediate state goes ``on shell'' and is enhanced. This results in a region in the phase space integrals with a large ``resonant'' contribution but with width $\simeq \Gamma_Z$. A similar enhancement and treatment arises for processes of the form $\nu_m l_\alpha \rightarrow W \rightarrow l_\alpha \nu_h$ for $T \simeq M_W$ where the intermediate $W$ becomes resonant. Obviously the difference (gain-loss) looks just like a typical Boltzmann equation, however, this is an equation for the \emph{production} of the heavy neutrino $\nu_h$ as the distribution functions of $f,\overline{f},\nu_m$ are all in thermal equilibrium as per our previous assumption. For this contribution it follows that $\overline{|\mathcal{M}_{fi}|^2} \propto g^4_w |\widetilde{{H}}_{m h}|^2$, for temperatures $ T \ll M_Z$ the Fermi limit implies that the production rate is $\propto G^2_F T^5 \mathcal{F}_1[E_h/T]|\widetilde{H}_{m h}|^2$ and similar contributions with charged current exchange, of the form $  G^2_F T^5 \mathcal{F}_2[E_h/T]|{H}_{m h}|^2$ on dimensional grounds, with $\mathcal{F}_{1,2}$ dimensionless functions of their arguments. A similar quantum kinetic equation describes the (gain) process $f_1 \overline{f}_2 \rightarrow W \rightarrow l_\alpha \nu_h$ and its inverse (loss)  process where $f_1,f_2$ correspond to either up/down-type quarks or $l_\beta,\nu_m$ for charged current interactions with the concomitant change in $|\mathcal{M}_{fi}|^2 \propto g^4_w |H_{\alpha h}|^2$. It is clear that for these cases the general form of the quantum kinetic equation (\ref{gainminlos}) holds where again $\Gamma^<,\Gamma^>$ obey the detailed balance condition (\ref{detailedbalance}).

Another process that is important and relevant in cosmology is the production of $\nu_h$ via
$e^+ e^- \nu_m \rightarrow  \nu_h$, the inverse process corresponds to the decay $\nu_h \rightarrow e^+ e^- \nu_m$ and a similar process mediated by neutral currents $ \nu_m \nu_m  \overline{\nu}_n \rightarrow \nu_h$ and the corresponding decay inverse process. In both cases the gain and loss terms are of the form
\bea \frac{d n_h(q;t)}{dt}\big|_{\textrm{gain}} & = &  \frac{2\pi}{2E_h(q)} \int \frac{d^3k_1 d^3k_2\,\overline{|\mathcal{M}_{fi}|^2}}{(2\pi)^6 2E_{1}(k_1)2E_{2}(k_2)2E_{3}(p)}\,n_1(k_1)n_2(k_2) n_3(p) (1-n_h(q;t))
\nonumber \\ & \times & \delta(E_1(k_1)+E_2(k_2)+E_3(p)-E_h(q)) \label{gain4} \\
\frac{d n_h(q;t)}{dt}\big|_{\textrm{loss}} & = &  \frac{2\pi}{2E_h(q)} \int \frac{d^3k_1 d^3k_2\,\overline{|\mathcal{M}_{fi}|^2}}{(2\pi)^6 2E_1(k_1)2E_{2}(k_2)2E_3(p)}\,(1-n_1(k_1))(1-n_{2}(k_2))(1-n_3(p)) \nonumber \\ & \times & n_h(q;t)\,
\delta(E_1(k_1)+E_{2}(k_2)+E_3(p)-E_h(q)) \,,\label{loss4}\eea where we read off the gain and loss rates
\bea \Gamma^<(q) & = &  \frac{2\pi}{2E_h(q)} \int \frac{d^3k_1 d^3k_2\,\overline{|\mathcal{M}_{fi}|^2}}{(2\pi)^6 2E_{1}(k_1)2E_{2}(k_2)2E_{3}(p)}\,n_1(k_1)n_2(k_2) n_3(p)
\nonumber \\ & \times & \delta(E_1(k_1)+E_2(k_2)+E_3(p)-E_h(q)) \label{Ggain4} \\
\Gamma^>(q) & = &  \frac{2\pi}{2E_h(q)} \int \frac{d^3k_1 d^3k_2\,\overline{|\mathcal{M}_{fi}|^2}}{(2\pi)^6 2E_1(k_1)2E_{2}(k_2)2E_3(p)}\,(1-n_1(k_1))(1-n_{2}(k_2))(1-n_3(p)) \nonumber \\ & \times &
\delta(E_1(k_1)+E_{2}(k_2)+E_3(p)-E_h(q)) \,,\label{Gloss4}\eea

where $\vec{p} = \vec{q}-\vec{k}_1-\vec{k}_2$  and the labels $1,2,3$ refer to the respective fermions either $e^\pm$ or $\nu_m$ for example (not to be confused with the labels for the light neutrinos $\nu_m$). We notice that as the temperature diminishes, setting the occupation factors $n_j=0$ in the loss term one recovers the \emph{decay rate} of the heavy neutrino, this observation will be important in the discussion of thermalization and stability of the DM candidate below.

It is clear from this discussion that in the mass basis, standard model charged and neutral current interaction vertices lead to production processes for heavy neutrinos that are similar to those of active-like neutrinos constrained by the usual kinematics. For example at temperatures $T  \gtrsim 1\,\mathrm{GeV}$, tau-lepton decay may lead to the production of heavy neutrinos with masses up to $\simeq \mathrm{GeV}$. Heavy lepton decay is an available mechanism down to $T\simeq M_\mu \simeq 100 \mathrm{MeV}$ and are processes that have not yet been studied in detail and clearly merit attention.

\begin{figure}[ht!]
\begin{center}
\includegraphics[height=3.5in,width=3.5in,keepaspectratio=true]{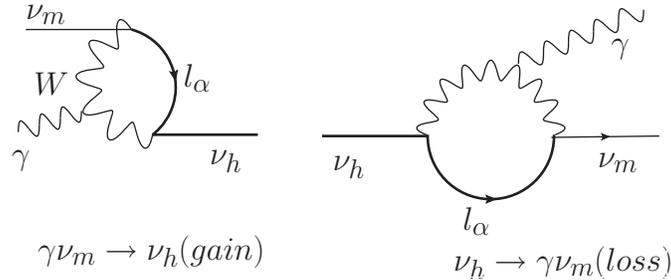}
 \caption{  $\gamma \nu_m \rightarrow \nu_h$ (gain) and the inverse process $\nu_h \rightarrow \nu_m \gamma$  }
\label{fig:xray}
\end{center}
\end{figure}

Finally, at temperatures $T \gtrsim M_h$ the production (gain) process $\gamma \nu_m \rightarrow \nu_h$  is kinematically allowed, the inverse process $\nu_h \rightarrow \gamma \nu_m$ is precisely the process conjectured to yield the X-ray line as a telltale of $keV$ neutrinos. The corresponding Feynman diagrams for the gain and loss terms are depicted  in
fig. (\ref{fig:xray})
 For these processes we find
 \be  \Gamma^<(q)   =    \frac{2\pi}{2E_h(q)} \int \frac{d^3k\,\overline{|\mathcal{M}_{fi}|^2}}{(2\pi)^3 2E_\gamma(k)2E_m(p)} \,n_\gamma(k) n_m(p)  \,\delta(E_\gamma(k)+E_m(p)-E_h(q))\label{gainratex} \ee
  \be \Gamma^>(q)    =     \frac{2\pi}{2E_h(q)} \int \frac{d^3k\,\overline{|\mathcal{M}_{fi}|^2}}{(2\pi)^3 2E_\gamma(k)2E_m(p)} \,(1+n_\gamma(k)) (1- n_m(p)) \delta(E_\gamma(k)+E_m(p)-E_h(q))\,,\label{lossratex} \ee  where the $W-l_\alpha$ loop is included in the $\overline{|\mathcal{M}_{fi}|^2}$. How large can the production rate be?, at $T \ll M_W$ on dimensional grounds we expect (see discussion in section (\ref{sec:stable}))
  \be \Gamma^{<}_{\gamma \nu_m \rightarrow \nu_h} \propto \alpha_{em} G^2_F T^5 \mathcal{F} \,\sum_{l} |H_{h l}U_{l m}|^2 \label{invgam}\ee with $\mathcal{F}$ a dimensionless function of the ratios $M_h/T, M_l/T$ with a finite limit for $T \gg M_{h,l}$. For $T \simeq \mathrm{GeV}$ this contribution to the production rate \emph{could} be of the same order as that for non-resonant production (DW) at $T \simeq 150\,\mathrm{MeV}$\cite{dodwid,abazajian3,asaka}, clearly motivating a deeper assessment of these processes.

   At $T=0$ only the loss term (\ref{lossratex}) survives, and the corresponding decay rate has been obtained in ref.\cite{palwolf,palbook}, this will be an important aspect discussed below. The gain process is actually kinematically allowed at temperatures $T \lesssim M_h$ because of the tail in the fermionic and blackbody distributions, although suppressed at lower temperatures.

The important aspect of these latter processes is that whereas the gain rate $\Gamma^<(q) \rightarrow 0 $ as the lepton and photon  populations vanish at small temperature, the loss rate $\Gamma^>(q)$ \emph{does not} vanish when the lepton populations vanish, the reason for this is that any population of $\nu_h$ \emph{decays} into the lighter leptons. These particular processes will be important in the discussion within the cosmological context. The radiative decay $\nu_h \rightarrow \nu_m \gamma$ is conjectured to be a telltale of the presence of ``sterile'' (  heavy)  neutrinos. We then see that the inverse process \emph{produces} the heavy (sterile-like) neutrinos at high enough temperature.

\subsection{Finite temperature corrections:} 
\begin{figure}[ht!]
\begin{center}
\includegraphics[height=3.5in,width=3.2in,keepaspectratio=true]{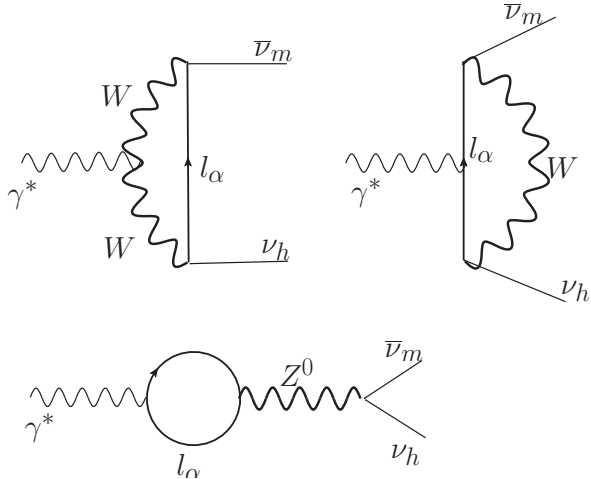}
 \caption{Gain processes: $\gamma^*\rightarrow \overline{\nu}_m \nu_h$ }
\label{fig:rare}
\end{center}
\end{figure}

There are important loop corrections at finite temperature,  self-energy corrections to the incoming and outgoing external ``legs'' as well as vertex corrections. There are also finite temperature corrections to the mixing angles arising from self-energy loops, these tend to suppress the mixing matrix elements\cite{medium,dolgovreview,boywu} therefore in medium the matrix elements $H_{hl}\rightarrow H_{hl,eff}(T)$ and typically in absence of MSW resonances $H_{hl,eff}(T) < H_{hl}$. An explicit example is given in section (\ref{distfunction}) below (see eqn. (\ref{mixanT})).  At high temperature there are   hard thermal loop corrections to the self-energy of fermions and vector bosons\cite{weldon,htl,pisarskihtl,frenkel,lebellac} that lead to novel collective excitations  with masses $\propto g T$ where $g$ is the gauge coupling. Photons and fermion-antifermion pairs form plasmon collective excitations with mass $\propto e T$. For $T > T_{EW}$ the $W,Z$ vector bosons \emph{do not} acquire a mass via the Higgs mechanism because the ensemble average of the Higgs field vanishes, but they acquire \emph{thermal} masses $\propto g_w T$ akin to the plasmon collective excitations. Thermal masses for collective excitations of $W,Z$ \emph{could} open   up kinematic windows for decay into $\nu_h$ \emph{above} the electroweak transition.  Plasmon collective  excitations from photons in the medium can also produce $\nu_h$ via  the \emph{electromagnetic} process  $\gamma^* \rightarrow \overline{\nu}_m \nu_h$ via the Feynman diagrams displayed in fig. (\ref{fig:rare}). These processes are similar to the mechanism of energy loss by plasmon decay into neutrinos in highly evolved massive stars\cite{raffeltbook,braaten,esposito} such as red giants and are also available in the Early Universe.

 Although  \emph{a priori} these processes are subleading being suppressed by higher orders in the couplings,  photons are populated all throughout the thermal history of the Universe, so these processes \emph{may} contribute to production during a longer time scale as the leading processes described above. A similar possibility \emph{may be} associated with plasmon collective excitations for $Z^0$ above $T_{EW}$ resulting in the production of the heavier species at $T>T_{EW}$, this possibility merits deeper understanding, which is beyond the realm of this article.  A follow up study will be reported elsewhere.

The main point of this discussion is to highlight that in the mass basis, standard model interactions provide a wide variety of mechanisms to produce heavy neutrinos which are available at high temperatures in the early Universe. The final distribution function of a particular heavy neutrino species after freeze-out is a \emph{mixture} of the distribution functions arising from the various production processes. A full assessment of a particular species as a DM candidate thus requires a thorough understanding of the various physical processes that lead to its production.

\subsection{Thermalization in Minkowski space-time} An important aspect of the general form of the quantum kinetic equation (\ref{gainminlos}) is the linear dependence on the (time dependent) distribution function $n_h(q;t)$. This linearity is a consequence of keeping the lowest order term in $|H_{\alpha h}|;|\widetilde{H}_{\alpha m}|$ and along with detailed balance  has profound consequences: in Minkowski space time the quantum kinetic equation (\ref{gainminlos}) leads to thermal equilibration of the heavy neutrino as the following argument shows.

The solution of (\ref{gainminlos}) is
\be  n_h(q;t)  =    n_h(q;0)\,e^{-\gamma(q)t} + e^{-\gamma(q)t} \int_0^t \Gamma^<(q) e^{\gamma(q) t'}\,dt' ~~;~~ \gamma(q) = \Gamma^<(q)+\Gamma^>(q)\,, \label{sol1} \ee carrying out the time integral and using the detailed balance result (\ref{detailedbalance}) one finds
\be n_h(q;t) = n_{eq}(q)+ \big[n_h(q;0)-n_{eq}(q)\big]\,e^{-\gamma(q) t} \,,\label{equili} \ee where $n_{eq}(q)$ is the equilibrium (Fermi-Dirac) distribution function. Obviously $\gamma(q)$ is the rate of relaxation towards equilibrium, for $t\gg 1/\gamma(q)$ the distribution function is that of thermal equilibrium.
Since $\gamma \propto g^2_w |H|^2$ one would be tempted to neglect the exponential terms in (\ref{sol1}), however the exponent is \emph{secular}, growing with time and becoming non-perturbative on a time scale $t \simeq 1/\gamma$. In Minkowski space time where the gain and loss rates are constant in time the distribution function will always reach thermal equilibrium at long time $t \gg 1/\gamma$. This is an important observation: detailed balance between the gain and loss term guarantees that at asymptotically long time $t \gg 1/\gamma$ the heavy neutrino thermalizes and its distribution function is $n_{eq}$.

This is a general result \emph{in Minkowski space time} which will be revisited below within the context of an expanding cosmology, where the relevant time scales are determined by the Hubble time scale.

\subsection{Production, freeze-out, LTE  and decay with cosmological expansion:}
In an expanding cosmology described by an isotropic and homogeneous Friedmann-Robertson-Walker metric, and during a radiation dominated epoch the temperature varies with time $T(t) \propto 1/a(t)$\cite{kolbturner} where $a(t)$ is the scale factor. The energy of a particle species of mass $M$ measured locally by an observer is
\be E(t) = \sqrt{p^2_f(t)+M^2} ~~;~~ p_f(t) = \frac{p_c}{a(t)} \,, \label{eandp} \ee where $p_f(t),p_c$ are the physical and comoving momenta respectively. The distribution function of a particle species in local thermodynamic equilibrium (LTE) is (in absence of chemical potentials)
\be n^\pm (E(t);t) = \frac{1}{e^{E(t)/T(t)} \pm 1}\,, \label{befd} \ee for Bose-Einstein and Fermi-Dirac respectively. The ratio
\be \frac{p_f(t)}{T(t)} = y = \frac{p_c}{T_0}\,,\label{ydef} \ee is constant during a radiation dominated era and $T_0$ would be the temperature of the plasma today, related to the temperature of the cosmic radiation by accounting of the reheating of the photon gas whenever the number of the relativistic degrees of freedom changes\cite{kolbturner}. As the temperature drops during the expansion when $M/T(t) \gg 1$ the population of the massive species is strongly thermally suppressed (by setting vanishing chemical potential we assume that there is no conservation of this species). In LTE the calculation of the gain and loss rates carry over from Minkowski space-time by replacing the energies and momenta by $E(t),p_f(t)$ (\ref{eandp}) and the distribution functions of the species in LTE  by (\ref{befd}). Although it is typical to separate the explicit time dependence in the rate equations from the time dependence of $p_f(t)$, we will consider the rates and distribution functions as functions of $p_c$ and the explicit time dependence of $a(t)$. Thus the quantum kinetic equation in the cosmological setting becomes
\be \frac{d n_h(q_c;t)}{dt} = \Gamma^<(q_c;t)(1-n_h(q_c;t)) - \Gamma^>(q_c;t) n_h(q_c;t) \,,\label{qkcosmo} \ee where the gain and loss rates are obtained as in Minkowski space time replacing the momenta and energies by the local time dependent counterparts in the expanding cosmology and the (LTE) distribution functions for the various bosonic or fermionic species (with vanishing chemical potentials).

As a consequence of the linearity of the quantum kinetic equation (\ref{qkcosmo}), which, in turn is a consequence of keeping only terms of $\mathcal{O}(|H_{\alpha h}|^2), \mathcal{O}(|\widetilde{H}_{m h}|^2)$, the full quantum kinetic equation is a simple sum over all possible channels with total gain and loss rates
\be  \Gamma^<_{tot}(q_c;t) = \sum_{all~ channels} \Gamma^<(q_c;t)~~;~~ \Gamma^>_{tot}(q_c;t) = \sum_{all~  channels} \Gamma^>(q_c;t) \,,\label{gamatots}\ee therefore the gain and loss rates in the quantum kinetic equation (\ref{qkcosmo}) are the total rates (\ref{gamatots}).

Because for each channel the rates  $\Gamma^{<,>}$ are calculated with distribution functions in (LTE) and obey the detailed balance condition this translates into
\be \Gamma^<_{tot}(q_c;t)\,e^{E_h(t)/T(t)} = \Gamma^>_{tot}(q_c;t)\,.\label{detbalcosmo} \ee
The general solution of (\ref{qkcosmo}) with the total gain and loss rates is
\be n_h(q_c;t) = n_h(q_c;t_0)\,e^{-\int^t_{t_0}\gamma(q_c;t')dt'}+ e^{-\int^t_{t_0}\gamma(q_c;t')dt'}\,\int^t_{t_0}\Gamma^<_{tot}(q_c;t')\,e^{ \int^{t'}_{t_0}\gamma(q_c;t'')dt''}\,dt' \,,\label{solcosmo} \ee where the \emph{relaxation rate}
\be \gamma(q_c;t) = \Gamma^<_{tot}(q_c;t)+\Gamma^>_{tot}(q_c;t) = \frac{\Gamma^<_{tot}(q_c;t)}{n_{eq}(t)} \, , \label{gammacosmo} \ee
 with
 \be n_{eq}(t)= \frac{1}{ e^{E_h(t)/T(t)}+1 }\,,\label{neq} \ee
 and we used the detailed balance result (\ref{detbalcosmo}).

 A  fixed point solution with $\dot{n}_{h}(q_c;t) = 0$ corresponds to \emph{freeze out}. Although $\Gamma^<$ and $\Gamma^>$ are related by detailed balance, the LTE distribution function $n_{eq}$  is not stationary, therefore is not a fixed point of the quantum kinetic equation\footnote{This is also the case for the Maxwell-Boltzmann distribution in a radiation dominated cosmology\cite{bernstein}.}.    However, the more relevant criterion for freeze-out is  that the distribution function varies much more slowly than the expansion time scale, namely\cite{bernstein,kolbturner},
\be \frac{\dot{n}_h(q_c;t)}{n_h(q_c;t)} \ll H(t) \,. \label{frees}\ee

As discussed above, the gain rate $\Gamma^<$ depends on the population of the particles whose decay or combination results in the production of the heavy neutrino, as a result the production rate diminishes during cosmological expansion eventually vanishing exponentially because of thermal suppression of the respective LTE distribution functions. Therefore \emph{if}
\be \int^{\infty}_{t_{0}} \gamma(q_c,t')\, dt' \ll 1 \label{freezecond} \ee we can neglect the exponential terms in the solution (\ref{solcosmo}). The remaining integral is finite at asymptotically long time because the gain rate vanishes exponentially, leading to a \emph{non-equilibrium frozen  distribution}
\be n_h(q_c;\infty) \simeq n_h(q_c;t_0)+ \int^{\infty}_{t_0} \Gamma^<_{tot}(q_c;t')\,dt'\,.  \label{Nhfroz}\ee The freeze-out condition is clearly achieved since (\ref{frees}) is fulfilled with $\dot{n}_h(q_c;t) = \Gamma^<_{tot}(q_c;t)$ vanishing exponentially at sufficiently long time, whereas $H(t)\propto 1/t$, and  $n_h(q_c;\infty)\neq 0$.

 When the condition (\ref{freezecond}) is valid and the distribution function is \emph{linearly} related to the production rate as in (\ref{Nhfroz}) we can separate the contribution from the different production channels which will, generally, lead to \emph{different distribution functions}, the total distribution function being a simple sum over all the different production channels. Because in each production channel the heavy neutrino is ``kinematically entangled'' with other leptons the distribution function at freeze-out for each channel will reflect the different kinematics, which can be interpreted as a ``memory'' of the particular production process.  We will study a relevant example below.

  Although the gain and loss rates satisfy the detailed balance condition (\ref{detbalcosmo}) the LTE distribution function is \emph{not} a solution of the quantum kinetic equation
(\ref{qkcosmo}) because $\dot{n}_{eq}(t) \neq 0$. We can assess if and when a distribution function is very  nearly in  LTE by exploiting the relations (\ref{detbalcosmo},\ref{gammacosmo}) to cast (\ref{qkcosmo}) in the  form
\be \frac{d n_h(q_c;t)}{dt} = - \gamma(q_c;t)\Big[n_h(q_c;t)-n_{eq}(t)\Big]\,, \label{qkcosmonew}\ee  writing
\be n_h(q_c;t)  =   n_{eq}(t)+ \delta n_h(q_c;t)\,, \label{delN}\ee and inserting into (\ref{qkcosmonew}) one finds
\be \delta n_h(q_c;t)=  \delta n_h(q_c;t_0)e^{-\int^t_{t_0}\gamma(q_c;t')dt'}- e^{-\int^t_{t_0}\gamma(q_c;t')dt'}\,\int^t_{t_0}\dot{n}_{eq}(q_c;t')\,e^{ \int^{t'}_{t_0}\gamma(q_c;t'')dt''}\,dt' \label{soludelNh} \ee with
\be \dot{n}_{eq}(t)= -n_{eq}(t)(1- n_{eq}(t))~ \frac{M^2_h\,H(t)}{T(t)E_h(t)}  \,, \label{dotneq}\ee and \be H(t) = 1.66 \, g^{\frac{1}{2}}_{eff}(T)\,\frac{T^2(t)}{M_{Pl}} \simeq 2\times 10^5 \,  g^{\frac{1}{2}}_{eff}(T) \Big( \frac{T(t)}{\mathrm{GeV}}\Big)^2\,s^{-1}\label{Hoft} \ee during radiation domination.  As the temperature drops during expansion $n_{eq}$ diminishes rapidly and  we expect that the integral with $\dot{n}_{eq}(t)$ in
(\ref{soludelNh}) will be finite at long time as the exponential suppression from $n_{eq}(t)$ will overwhelm the perturbative growth of the integral of $\gamma$. Hence  the condition that the distribution function becomes nearly LTE, becomes
 \be \int^{t}_{t_{0}} \gamma(q_c,t') dt' \gg 1 \,.\label{nelte} \ee at long time $t \lesssim  1/H_0$. If this condition is fulfilled the full solution (\ref{solcosmo}) must be considered as the integrals of $\gamma(q_c,t)$ cannot be neglected. In this case the distribution function at long time   may be described by (nearly) LTE, and the ``memory'' of the production process and the kinematic ``entanglement'' characteristic of each production channel is erased. Therefore, it is important to assess under what circumstances the condition (\ref{nelte}) could be satisfied, since under such circumstances the heavy neutrino can \emph{thermalize} with the rest of the standard model particles.

As the temperature diminishes throughout the expansion history, the gain term is strongly suppressed by the thermal population factors because these processes entail the annihilation of particles in the initial state which are thermally populated (by assumption) (see the examples described above). The loss rate will also vanish  exponentially  by thermal suppression  at long time if it involves the annihilation of \emph{any thermal species}, therefore for these processes  the condition (\ref{freezecond}) is expected to be fulfilled and the distribution function is expected to freeze out of LTE.

 However for processes in which the heavy neutrino \emph{decays} into the final products the loss term does \emph{not} vanish at low temperatures and the distribution function eventually decays, an example of this case is the loss term (\ref{Gloss4}) describing the decay processes $\nu_h \rightarrow e^+ e^- \nu_m$ or $\nu_h\rightarrow \nu_{m_1}\nu_{m_2}\nu_{m_3}$. This has important implications: consider the quantum kinetic equation in the form (\ref{qkcosmonew}) (note that $\gamma(q_c;t) >0$) and an initial condition in the far past with $n_h(q_c;t_0) =0$. At early times $n_h$ \emph{grows} $\dot{n}_h > 0$ as the gain terms dominate and the heavy neutrinos are being produced. \emph{If} at late times the population \emph{decays}, namely $\dot{n}_h <0$ this means that at some time the distribution function has reached LTE at which $\dot{n}_h =0$, as the cosmological expansion continues the loss term dominates and the population decays. This simple analysis leads to the conclusion that if the lifetime of the heavy neutrino is smaller than the age of the Universe, meaning that it is now decaying, at some point in the past its distribution function \emph{reached LTE}. Conversely, if the lifetime is much longer than the Hubble time $1/H_0$ then the distribution function has not reached LTE and the heavy neutrino is non-thermal. In other words, a heavy neutrino that is stable during the age of the Universe $\sim 1/H_0$ and is a suitable DM candidate \emph{must be out of LTE}.

The analysis above indicates that when the lifetime of the heavy neutrino is much smaller than the age of the Universe, its distribution function must have passed through  LTE during the evolution. The LTE solution $n_{eq}(t)$ is \emph{not} a true fixed point of the kinetic equation (\ref{qkcosmo}) because $\dot{n}_{eq}(E_h(t)) \neq 0$, however, we can ask what is the condition that ensures that the distribution function remains \emph{nearly} in LTE if and when it reaches LTE. To answer this question we write
\be n_h(q_c;t) = n_{eq}(t)+ \delta n^{(1)}_h(t) + \cdots \label{nearLTE} \ee where  $\delta n^{(1)}_h(t) \ll n_{eq}(t))$ etc,  the expansion is in a small parameter to be quantified \emph{a posteriori}. Introducing this ansatz into (\ref{qkcosmo}) we find
\be \frac{\delta n^{(1)}_h(t)}{n_{eq}(t)} = (1- n_{eq}(t))~ \frac{M^2_h}{T(t)E_h(t)}\,\frac{H(t)}{\gamma(q_c;t)}\,,\label{chapens}\ee
in terms of  the relaxation time
\be \tau(q_c;t) = \frac{1}{\gamma(q_c;t)} \,, \label{reltime}\ee
we find that  once LTE is attained, the distribution function will linger near LTE whenever
\be \tau(q_c;t) H(t)  \ll \frac{T(t)}{M_h} \,.\label{relLTE} \ee
  As the temperature drops, the contribution from $\Gamma^<$ (gain) to $\gamma = \tau^{-1}$ is suppressed and the relaxation time becomes the lifetime of the heavy neutrino (loss via decay). This can be seen for example, from the loss term for the process $\nu_h \rightarrow \nu_{m}\nu_{m'}\nu_{m''}$ with $\Gamma^<$ given by (\ref{Gloss4}) with $1,2,3=m,m',m''$ and the discussion following it.  Therefore the  condition (\ref{relLTE}) is actually equivalent to the statement of a lifetime much shorter than the Hubble time, consistently with the discussion above. If the heavy neutrino is a \emph{stable} DM candidate its lifetime must be $\tau > H^{-1}(t)$, implying that its distribution function will not be in LTE.

 \subsection{Stability and lifetime:}\label{sec:stable}

 To be a suitable DM candidate a heavy neutrino must feature a lifetime $\tau \geq 1/H_0$, thus it remains to understand the decay channels for a firmer assessment  of the lifetime of heavy neutrinos, their suitability as DM candidates  and the conditions for non-LTE distribution functions.

 The decay channels of a heavy neutrino were studied in refs.\cite{shrock,barger,palbook}, for example   the purely leptonic and radiative channels:  neutral current process $\nu_h \rightarrow \nu_m \nu_{m'}  \nu_{m''}$, or the charged current process $\nu_h \rightarrow e^+ e^- \nu_{m}  $  and $\nu_h \rightarrow \nu_{h'} \gamma~;~ \nu_h \rightarrow \nu_{m} \gamma$ these are the inverse processes associated with the production processes $  \nu_m \nu_{m'}  \nu_{m''} \rightarrow \nu_h $,
$e^+ e^- \nu_{m}  \rightarrow \nu_h$  and $\nu_{h',m} \gamma \rightarrow \nu_h$ respectively.
At $T=0$ the decay rates for  $\nu_h \rightarrow \nu_m \nu_{m'}  \nu_{m''}~;~\nu_h \rightarrow e^+ e^- \nu_{m} $ have been obtained in ref.\cite{shrock,barger},   these are given by
\be
\Gamma(\nu_h \rightarrow e^+ e^- \nu_m) \simeq  {3.5}\times {10^{-5}}  {|H_{e h}|^2}  \left(\frac{M_{h}}{MeV}\right)^5  ~K\Bigg[\frac{m^2_e}{M^2_h}\Bigg]\,\frac{1}{s}\label{nuhdec1}
\ee where\cite{shrock}
\be K\big[x\big] = (1-4x)^{1/2}\,(1-14x-2x^2-12x^3)+ 24x^2(1-x^2)\ln\Bigg[ \frac{1+(1-4x)^{1/2}}{1-(1-4x)^{1/2}}\Bigg]\,, \label{kofx}\ee  with $K \rightarrow 0$ for $M_h \rightarrow 2 m_e$ and $K \rightarrow 1$ for $M_h \gg m_e$.
 Therefore   for this process the ratio of the lifetime to the Hubble time today $1/H_0$ is given by
\be  H_0 \tau(\nu_h \rightarrow e^+ e^- \nu_m) \simeq \frac{10^{-13}}{|H_{e h}|^2K\Big[\frac{m^2_e}{M^2_h}\Big]}\,\left(\frac{MeV}{M_{h}}\right)^5 \,.\label{tauho}\ee

 The  decay rate  into active-like neutrinos mediated by neutral currents (not GIM (Glashow-Iliopoulos-Maiani) suppressed with sterile-like heavy neutrinos) is given by   (see \cite{barger})
  \be
\Gamma(\nu_h \rightarrow \nu_m \nu_m' \nu_m'') =    {3.5}\times {10^{-5}}  \sum_{\alpha = e,\mu,\tau}{|H_{\alpha h}|^2}  \left(\frac{M_{h}}{MeV}\right)^5  \,\frac{1}{s}\label{nuh3nus}
\ee therefore for this ``invisible'' channel we find
\be  H_0 \tau(\nu_h \rightarrow \nu_m \nu_m' \nu_m'') \simeq \frac{10^{-13}}{ \sum_{\alpha = e,\mu,\tau}|H_{\alpha h}|^2}\,\left(\frac{MeV}{M_{h}}\right)^5 \,.\label{tauho3nus}\ee

    The radiative decay $\nu_h \rightarrow \gamma \nu_m$ has been studied in ref.\cite{palwolf,barger,palbook}, for heavy (sterile-like) neutrinos  it is not GIM   suppressed but is suppressed by one power of $\alpha_{em}$ with respect to the leptonic channels above and  given by
\be \Gamma(\nu_h \rightarrow \gamma \nu_m) \simeq   10^{-7} \Big(\frac{M_h}{\mathrm{MeV}}\Big)^5 \, \sum_{\alpha} |H_{h\alpha}U_{\alpha m}|^2 \,\frac{1}{s}\label{gamahfot}\ee   namely
\be H_0 \tau(\nu_h \rightarrow \gamma \nu_m) \simeq   \frac{2\times  10^{-11}}{\sum_{\alpha}|H_{h\alpha}U_{\alpha m}|^2}\,\left(\frac{MeV}{M_{h}}\right)^5\,. \label{tauhogam}\ee For a Majorana neutrino there is an extra factor $2$ multiplying the rates for the expressions above. For the heavy neutrino to be stable during the lifetime of the Universe and be a suitable DM candidate it must be that $H_0 \tau(\nu_h \rightarrow e^+ e^- \nu_m)~;~ H_0 \tau(\nu_h \rightarrow \nu_m \nu_m' \nu_m'')  ~;~ H_0 \tau(\nu_h \rightarrow \gamma \nu_m) \geq 1$.

Assuming that the $|H_{\alpha h}|$ are all of the same order and $|U_{\alpha m}|\simeq  \mathcal{O}(1)$ these estimates suggest that heavy neutrinos with
\be |H_{h \alpha }|^2 (M_h/MeV)^5 \lesssim 10^{-13} \label{longlive} \ee feature lifetimes   $ \gtrsim 1/H_0$ and may be acceptable DM candidates. Interestingly, for $M_h \leq 1\,MeV$ the ``visible'' leptonic decay channel $\nu_h \rightarrow e^+ e^- \nu_m$ shuts off and the lifetime is dominated by the ``invisible'' channel $\nu_h \rightarrow \nu_m \nu_m' \nu_m'' $ therefore evading potential bounds from the ``visible'' decays into $l^+l^-$ pairs.

As per the discussion above the constrain that the heavy neutrino features a lifetime $\tau > 1/H_0$ also implies that its distribution function is out of LTE. We will discuss possible interesting consequences of a heavy sterile neutrino with a lifetime much smaller than the Hubble time in a later section (see section{\ref{sec:cascade}).

\subsection{Comparisons and caveats}

A formulation of the production rates of sterile neutrinos firmly based on the quantum field theory of neutrino mixing was introduced in ref.\cite{asaka}. It relies on a see-saw type mass matrix with vanishing masses for the active neutrinos, large (Majorana) masses for sterile neutrinos on the diagonal and small off-diagonal matrix elements that mix the sterile and active neutrinos. This small (compared to the large Majorana masses) off-diagonal mixing sub-matrix is treated in a perturbative expansion. The authors in ref.\cite{asaka} obtain the quantum density matrix  by tracing over the degrees of freedom of the standard model (assumed in thermal equilibrium) up to second order in the off-diagonal mixing, and, in principle,  to all orders in the strong and weak interactions. The quantum master equation that describes the time evolution of the reduced density matrix is completely determined by   correlation functions  of the active neutrino including self-energy corrections, which is written in terms of spectral representations. The production rate of sterile neutrinos is obtained from the imaginary part of this self energy evaluated on the mass-shell of the sterile neutrinos, in principle to all orders in standard model couplings. The authors focus on temperature scales $\ll M_{W,Z}$, and argue that sterile neutrinos with masses in the $keV$ range are primarily produced in the temperature range $T \sim 150 \,\mathrm{MeV}$. Furthermore, they only consider a ``gain'' term in this temperature regime neglecting the loss term.

  There are several differences   between the approach of ref.\cite{asaka} and the framework presented here: \textbf{1)} we describe the production process with standard quantum kinetic equations for the \emph{mass eigenstates}, without resorting to any particular model for the mixing mass matrix, however, similarly to  \cite{asaka}  our results are in principle valid to all orders in standard model couplings, but to second order in the (small) mixing matrix elements between the light (active-like) and heavy (``sterile-like'') neutrinos. \textbf{2)} By going to the mass basis, we recognize that standard model processes that produce active neutrinos via charged and neutral current vertices  lead to the production of heavy neutrinos provided the kinematics is favorable and identify various processes available in a wide temperature region. \textbf{3)} we consider both the gain and loss terms, giving an exact result for the non-equilibrium evolution of the distribution function, this is the result (\ref{solcosmo}). We discuss the important issue of thermalization \emph{vis a vis} the constraints of stability of the DM candidate.

    The authors of ref.\cite{asaka} identified the ``one-loop'' contributions to the active neutrino self energy, whose imaginary parts are precisely the contributions described by the quantum kinetic equations (\ref{gainrate},\ref{lossrate},\ref{gainrateZ},\ref{lossrateZ}) but neglected them by restricting their study to $T \ll M_{W,Z}$, whereas we argue that these contributions may be  important (obviously a statement on their impact requires a detailed calculation, beyond the scope of this article). The production processes mediated by W,Z exchange in an intermediate state with gain and loss terms given by (\ref{gain3}-\ref{loss4}) (and equivalent for charged current processes) correspond to the ``two loops'' contributions in ref.\cite{asaka}, although we pointed out that at $T\sim M_W, M_Z$ these processes could be resonantly enhanced as the intermediate $W,Z$ propagators feature ``on-shell'' thermal contributions which are not suppressed at these temperatures. \textbf{4)} We have identified production processes from ``collective excitations'' in the medium, suggesting that finite temperature corrections, leading to plasmon masses for the photon and for the W,Z bosons \emph{above} $T_{EW}$ \emph{may} yield important production mechanisms. \textbf{5)} we will argue below that different production channels with different kinematics yield different contributions to the distribution function. This is a result of ``kinematic entanglement'' between the heavy neutrino and the lepton produced in the reaction leading to distribution functions that may be colder for some channels and warmer for others depending on the mass of the lepton and the kinematics.
     In other words,  the final distribution is a \emph{mixture} of several components even for the same species.   This point will be elaborated upon in more detail within a specific example in the next sections.

  \vspace{1mm}

\textbf{Caveats:} As we discussed above at high temperature there are important self-energy corrections that must be assessed for a more reliable understanding of the gain and loss processes. Furthermore, our results are valid up to $\mathcal{O}(|H_{\alpha h}|^2)$ under the assumption that $|H_{\alpha h}|^2 \ll |U_{\alpha m}|^2$, (and in principle to all orders in weak coupling), however this approximation will break down if there are Mikheyev-Smirnov-Wolfenstein (MSW)\cite{msw,medium,boywu} resonances in the medium, because near the resonance the effective mixing angle reaches $\pi/4$. This is also a caveat in the approach of ref.\cite{asaka} because in this reference the quantum master equation has been obtained up to second order in the sterile-active mixing angle.

Including the possibility of MSW resonances in the medium requires adopting a different framework that does not rely on $|H_{\alpha h}| \ll 1$ but that would be firmly based on quantum field theory out of equilibrium. Such approach \emph{could be} based on   effective field theory out of equilibrium as advocated in ref.\cite{boyneqft}. We will report on this approach in a future study.

\section{Cosmological consequences and constraints.}\label{sec:cosmocons}
If there is a hierarchy of \emph{stable}    heavy neutrinos produced by the various mechanisms discussed above (we will comment later on unstable heavy neutrinos), each species with $M_h \gg 1\,\mathrm{eV}$ will contribute as a non-relativistic DM component after matter radiation equality. Once the distribution function at freeze out is obtained, various cosmological consequences of the dark matter species can be assessed. In this section we gather together various quantities of cosmological relevance which are determined by basic properties of the dark matter species: mass, number of intrinsic degrees of freedom,  distribution function and freeze-out (or decoupling) temperature with the purpose of assessing particular species of heavy neutrinos as DM candidates once their distribution function is obtained.

 Let us define the \emph{total} asymptotic ``frozen'' distribution function of a species $\nu_h$ of mass $M_h$ as
\be f_h(q_c) = n_h(q_c;\infty)\,, \label{frozdf} \ee where by $t\rightarrow \infty$ we mean a time sufficiently long that the distribution function satisfies the freeze-out condition (\ref{frees}).

As discussed above after freeze-out  the distribution function depends on the physical momentum $q_f(t)=q_c/a(t)$ and temperature $T(t)$ through the combination $y = q_f(t)/T(t) = q_c/T_0$ where $T_0$ is the temperature at which the species decoupled from  the plasma (freeze-out)  redshifted to  today. Through the usual argument of entropy conservation it is related to the temperature of the cosmic microwave background today $T_{\gamma,0}$ by
\be \frac{T_0}{T_{\gamma,0}}  = \Big(\frac{2}{g_{dh}} \Big)^{1/3}\,, \label{T0today} \ee where $g_{dh}$ is the number of ultrarelativistic degrees of freedom at the time when the particular species $\nu_h$ decoupled (freezes). Then the number density $\mathcal{N}_{\nu_h}$, energy density $\rho_{\nu_h}$ and \emph{partial pressure} $\mathcal{P}_{\nu_h}$  of   species $\nu_h$ is given by
\be
\mathcal{N}_{\nu_h} (t)   =   g_{\nu_h} \int \frac{d^3q_f}{(2\pi)^3} f_h(q_c) = \frac{g_{\nu_h}}{2 \pi^2} \Big(\frac{T_0}{a(t)}\Big)^3   \int^{\infty}_0 dy \, y^2  f_h(y) \label{numh} \ee
\be \rho_{\nu_h} (t)   =     g_{\nu_h} \int^{\infty}_0 \frac{d^3q_f}{(2\pi)^3} \sqrt{q^2_f + M^2_h} f_h(q_c) = \frac{g_{\nu_h} M_h}{2 \pi^2} \Big(\frac{T_0}{a(t)}\Big)^3  \int^{\infty}_0 dy \, y^2 \sqrt{1+\frac{y^2}{x^2_h(t)}} f_h(y) \label{rhoh}\ee
\be \mathcal{P}_{\nu_h} (t)    =  \frac{g_{\nu_h}}{3} \int^{\infty}_0 \frac{d^3q_f}{(2\pi)^3} \frac{|\vq_f|^2}{\sqrt{q_f^2 +M^2_h}} f_h(q_c) = \frac{g_{\nu_h}}{6 \pi^2 M_h} \Big(\frac{T_0}{a(t)}\Big)^5  \int^{\infty}_0 dy \, \frac{y^4\,f_h(y)}{ \sqrt{1+\frac{y^2}{x^2_h(t)}}}     \label{pressh}
\ee where $g_{\nu_h}$ is the number of internal degrees of freedom of the   species $\nu_h$, and
\be x_h(t)=M_h/T(t)\,.\label{xdef} \ee
The contribution of each non-relativistic species $\nu_h$ to $\Omega_{DM}$ today with $T_0/M_h \ll 1$    is given by
\be \Omega_{\nu_h} h^2 = \frac{g_{\nu_h} M_h}{2 \pi^2 \,\rho_c}   {T^3_0}  \,h^2 \, \int^{\infty}_0 dy \, y^2   f_h(y) = \Big( \frac{n_\gamma \,h^2}{\rho_c}\Big) \, \frac{g_{\nu_h} M_h}{ 4\, \xi(3)} \Big(\frac{T_0}{T_{\gamma,0}}\Big)^3 \int^{\infty}_0 dy \, y^2   f_h(y) \,. \label{Omegah} \ee Using the relation (\ref{T0today}) and $ n_\gamma h^2/\rho_c = 1/25.67 \, eV$ we find
 \be \Omega_{\nu_h} h^2 = \frac{M_h}{61.7\, eV}\,\Big(\frac{g_{\nu_h}}{g_{dh}}\Big)  \int^{\infty}_0 dy \, y^2   f_h(y) \,. \label{omegah2}\ee Therefore, assuming that all of DM is in the form of various species of heavy neutrinos, it follows that
 \be \Omega_{DM}h^2 = \sum_{h=4,5\cdots} \Big(\frac{M_h}{61.7\, eV}\Big) \,\Big(\frac{g_{\nu_h}}{g_{dh}}\Big)  \int^{\infty}_0 dy \, y^2   f_h(y)\,. \label{totdm}\ee
 The \emph{fraction} $\mathcal{F}_{\nu_h}$ of $\Omega_{DM}$ contributed by the species $\nu_h$ is given by
 \be \mathcal{F}_{\nu_h}\equiv \frac{\Omega_{\nu_h} h^2}{\Omega_{DM}h^2} = \frac{M_h}{7.4\, eV}\,\Big(\frac{g_{\nu_h}}{g_{dh}}\Big)  \int^{\infty}_0 dy \, y^2   f_h(y) \,, \label{fractionh}\ee where we used $\Omega_{DM} h^2 = 0.1199$\cite{pdg}.

 The equation of state for each species is

\be \label{eos}
w_{\nu_h}(T(t))) = \frac{\mathcal{P}_{\nu_h}(t)}{\rho_{\nu_h}(t)} = \frac{1}{3}  \frac{\int^\infty_0 dy \, \frac{y^4\,f_h(y) }{\sqrt{  y^2 +\frac{M^2_h}{T^2(t)}}}}{\int^\infty_0 dy \, y^2 \sqrt{y^2+\frac{M^2_h}{T^2(t)}}\, f_h(y)}    \,.
\ee The equation of state yields information on how ``cold'' is the particular species and provides a benchmark to compare the equilibrium and non-equilibrium distribution functions. In particular $\sqrt{w_{\nu_h}(T)}$ yields a generalization of the effective  adiabatic ``speed of sound'' for collisionless DM, namely,
$\mathcal{P}_{\nu_h}(t)   = c^2_{h}(T(t)) \,\rho_{\nu_h}(t) ~~;~~  c^2_{h}(T(t)) \equiv w_{\nu_h}(T(t))$.

In the non-relativistic limit $M_h \gg T(t)~~;~~x(t) \rightarrow \infty$, the average primordial velocity dispersion for a   species $\nu_h$ is given by
\be
\big\langle \vec{V}^2_{\nu_h}(t) \big\rangle = \frac{\langle {\vq}^{\,2}_f \rangle}{M^2_h} = \frac{T(t)^2}{M^2_h}\frac{\int dy \, y^4 f_h(y) }{\int dy \, y^2 f_h(y)} = \frac{3 \mathcal{P}_{\nu_h}}{\rho_{\nu_h}} \label{veldisp}
\ee which leads to the \emph{primordial} velocity dispersion relation

\be
\mathcal{P}_{\nu_h} = \sigma^2_{\nu_h} \, \rho_{\nu_h} ~~;~~ \sigma_{\nu_h} = \sqrt{\frac{\big\langle \vec{V}^2_{\nu_h}(t) \big\rangle}{3}}  = \frac{T(t)}{M_h}\sqrt{\frac{\int dy \, y^4 f_h(y) }{3 \int dy \, y^2 f_h(y)}} \,, \label{sigvel}
\ee therefore for a collisionless non-relativistic species $\nu_h$   the dispersion $ \sigma_{\nu_h}=w^{1/2}_{\nu_h}$ plays the role of the adiabatic sound speed. The equation of state (\ref{eos}) will be   important in assessing the consequences of kinematic entanglement of distribution functions obtained by different production channels since it yields information on the ``coldness'' of the distribution.

In analogy with a fluid description and following ref.\cite{wuboy} we introduce the \emph{comoving} free streaming wave vector for the species $\nu_h$ similarly to the comoving Jeans wavevector, namely
\be k^2_{fs,\nu_h}(t) = \frac{4\pi\,G\,\rho_m(t)}{ { \big\langle \vec{V}^2_{\nu_h}(t) \big\rangle }}\,a^2(t) \label{kfscom}\ee as discussed in ref.\cite{wuboy} $k_{fs}(t_{eq})$  describes the cutoff in the power spectrum for linear density perturbations, where $t_{eq}$ is the time of matter-radiation equality. Since for non-relativistic particles
\be \big\langle \vec{V}^2_{\nu_h}(t) \big\rangle = \frac{\big\langle \vec{V}^2_{\nu_h}(0) \big\rangle }{a^2(t)} \label{v2}\ee where $\big\langle \vec{V}^2_{\nu_h}(0)\big \rangle$ is the velocity dispersion today, it follows that
\be  k_{fs,\nu_h}(t_{eq}) = k_{fs,\nu_h}(0)~  \sqrt{a(t_{eq})}  \label{kaifes}\ee where
\be k_{fs,\nu_h}(0) =\Bigg[\frac{3\Omega_{M}h^2}{2\overline{ \vec{V}^2_{\nu_h}(0)}} \Bigg]^{1/2} H_0 \label{kfstod}\ee is the free streaming wavevector \emph{today}. The free streaming scale that determines the length scale associated with the cutoff in the power spectrum is
\be \lambda_{fs,\nu_h}(t_{eq}) = \frac{2\pi}{k_{fs,\nu_h}(t_{eq})}= \lambda_{fs,\nu_h}(0)\,(1+z_{eq})^{1/2} \label{lamfs}\ee    with $\Omega_M h^2 = \Omega_b h^2+ \Omega_{DM} h^2 = 0.14$ we find\footnote{The discrepancy with the result in ref.\cite{lellolightsterile} is due to the difference between $\Omega_{DM}h^2 = 0.12$ and $\Omega_M h^2 =0.14$\cite{pdg}.}
\be
\lambda_{fs,\nu_h}(0) = 9.72 \,  \left(\frac{keV}{M_h}\right) \left(\frac{2}{g_{dh}}\right)^{1/3} \sqrt{\frac{ \int dy \, y^4 f_h(y)}{ \int dy \, y^2 f_h(y)} } ~ \mathrm{kpc} \label{lamfes}
\ee and $(1+z_{eq})^{1/2}\simeq 57$.

Up to constants of $\mathcal{O}(1)$)  $ \lambda_{fs,\nu_h}(t_{eq})$ is equivalent to the comoving distance traveled by a free streaming particle with average velocity $\big\langle \vec{V}^2_{\nu_h}(t) \big\rangle$ from the time of matter-radiation equality until today\cite{wuboy}, as can be seen from the following argument. Consider a particle moving with velocity $v(t) = \sqrt{\big\langle \vec{V}^2_{\nu_h}(t) \big\rangle}$ the comoving distance traveled between $t_{eq}$ and today at time $t^* \gg t_{eq}$ is
\be l(t_{eq},t^*) = \int^{t^*}_{t_{eq}} \frac{v(t')}{a(t')}dt' \label{dist} \ee during a matter dominated cosmology when density perturbations grow
\be a(t) = \big[\Omega_M H^2_0 \big]^{1/3}\,t^{2/3} \label{aoft} \ee and using (\ref{v2}) for a non-relativistic species we find
\be l(t_{eq},t^*)\simeq 3 \Bigg[ \frac{\big\langle \vec{V}^2_{\nu_h}(0) \big\rangle}{\Omega_M H^2_0}\Bigg]^{1/2} (1+z_{eq})^{1/2} \,, \label{lfs2}\ee from which it follows that
\be \lambda_{fs,\nu_h}(t_{eq}) \simeq 1.7 \, l(t_{eq},t^*)\,. \label{compar}\ee

 Phase-space density (Tremaine-Gunn) constraints are based on the result that the coarse grained phase space density may only decrease from its primordial value during ``violent relaxation'' in the process of galaxy formation and evolution\cite{tremainegunn,violentrelax,galacticdynamics,dalcantonhogan,destri}. The coarse grained phase space density for the species $\nu_h$ is defined as\cite{dalcantonhogan}
 \be \mathcal{D}_h = \frac{\mathcal{N}_h(t)}{   \big[\langle q^2_f(t) \rangle_{h}\big]^{3/2}} \,, \label{phspdens}\ee where the average is with the distribution function $f_h(q_c)$ and $\mathcal{N}_h(t)$ is given by (\ref{numh}). Therefore the primordial coarse grained phase space density is given by
 \be
\mathcal{D}_h  = \frac{g_{\nu_h}}{2 \pi^2} \frac{\left[\int^\infty_0 dy \, y^2 f_h(y)\right]^{5/2} }{\left[\int^\infty_0  dy \, y^4 f_h(y)\right]^{3/2}}\,. \label{Dhfini}
\ee

When the DM  particles become non relativistic  it becomes

\be
\mathcal{D}_{h}   = \frac{\rho_{\nu_h}}{M^4_h \big\langle \vec{V}^2_{\nu_h}\big\rangle^{ 3/2}} = \frac{1}{3^{3/2} M^4_h} \frac{\rho_{\nu_h}}{\sigma^3_{\nu_h}}
\ee For a primordial phase space density, $\mathcal{D}_{h}$, imposing the bound $\mathcal{D}_{today} \leq \mathcal{D}_{h}$\cite{dalcantonhogan} gives us the constraint

\be
\mathcal{D}_{h} \ge \frac{1}{3^{3/2} M^4_{h}}\, \Bigg[\frac{\rho_{DM}}{\sigma^3_{DM}} \Bigg]_{today} \label{tg1}
\ee  If there is only \emph{one species} the values of $\rho_{\nu_h}$ and $\sigma_{\nu_h}$  can be inferred from the kinematics of dwarf spheroidal galaxies\cite{dalcantonhogan} and this constraint can be used to obtain a bound which complements those from CMB observations.

\vspace{2mm}

\subsection{Non-LTE freeze-out from multiple production channels:}
The distribution function $f_h(y)$ is a result of \emph{all} the possible production channels that are kinematically allowed as the total gain and loss terms determine the solution of the cosmological quantum kinetic equation.  Furthermore, each production process of a heavy neutrino species $\nu_h$ may actually be the result of the decay of a heavy standard model particle (such as $W,Z,\tau, \mu,\cdots$) into different channels and each channel may yield a different distribution function because of the kinematics, we refer to this as ``kinematic entanglement''.   Thus even for a single species $\nu_h$ of heavy neutrino, its frozen distribution function may be a \emph{mixture} of several contributions some colder than others as a consequence of the kinematics.

  While the general solution (\ref{solcosmo}) is \emph{non-linear} in the gain and loss terms of each channel because of the exponential terms in the relaxation rate $\gamma$, \emph{if} the condition (\ref{freezecond}) is fulfilled,  these exponentials can be neglected and the result is given by (\ref{Nhfroz}) which is linear in the gain rates allowing for an identification of the distribution functions associated with \emph{each production channel}. In each production channel the heavy neutrino is \emph{kinematically entangled} with a lepton and the non-LTE distribution function at freeze out will reveal this ``entanglement'' for example in the form of production thresholds. This aspect is another manifestation of ``mixed DM'', in the sense that even for a given heavy species $\nu_h$, its distribution function at freeze-out is a result of different contributions from different channels each with different kinematics. The concentration of each component depends, among other factors, of the ratios of mixing angles for the different channels. If freeze-out occurs out of LTE, the distribution functions will maintain memory of the processes that led to the production and the kinematic entanglement, in LTE this memory is erased as the distribution function becomes the Fermi-Dirac distribution regardless of the production process.

Separating the contribution from the different channels will also allow an assessment of the ``coldness'' of the heavy neutrino as a result of the particular production channel. An explicit example of these phenomena will be studied in detail within the context of heavy neutrinos produced from pion decay in the next section.

\vspace{2mm}

\subsection{Summary of cosmological constraints:}  We now summarize the main cosmological constraints in terms of the decoupled distribution function, degrees of freedom and mass of a particular heavy species $\nu_h$. Once the distribution function has been obtained from the solution of the quantum kinetic equations these constraints inform the feasibility of such species as a suitable DM candidate. In the discussion below $f_h(y)$ is the \emph{total distribution function} solution of (\ref{solcosmo}) after freeze-out.

\begin{itemize}
\item{ \textbf{Abundance:} the fraction of DM from a particular species $\nu_h$ must obey, $\mathcal{F}_{\nu_h} \leq 1$ leading to the abundance constraint
    \be  \frac{M_h}{7.4\, eV}\,\Big(\frac{g_{\nu_h}}{g_{dh}}\Big)  \int^{\infty}_0 dy \, y^2   f_h(y) \leq 1 \label{abucons} \,,\ee  }

\item{ \textbf{Phase space density (Tremaine-Gunn):} the result that the phase space density only diminishes during ``violent relaxation'' in gravitational collapse\cite{tremainegunn,violentrelax,galacticdynamics,dalcantonhogan} leading to (\ref{tg1}) yields the constraint
    \be \,  0.26\, g_{\nu_h}\, \Bigg[\frac{M_h}{keV}\Bigg]^4 \, \frac{\left[\int^\infty_0 dy \, y^2 f_h(y)\right]^{5/2} }{\left[\int^\infty_0  dy \, y^4 f_h(y)\right]^{3/2}} \,> \Bigg[\frac{\rho_{DM}}{\sigma^3_{DM}\,(keV)^4 } \Bigg]_{today} \,,\label{tg2} \ee
    or alternatively\footnote{$1 \,keV^4 = 1.27 \times 10^8 \frac{M_\odot}{kpc^3} \,(\frac{km}{s})^{-3}$.}
    \be  0.335 \times 10^8 \, g_{\nu_h}\, \Bigg[\frac{M_h}{keV}\Bigg]^4 \, \frac{\left[\int^\infty_0 dy \, y^2 f_h(y)\right]^{5/2} }{\left[\int^\infty_0  dy \, y^4 f_h(y)\right]^{3/2}} \,> \Bigg[\frac{\rho_{DM}}{M_\odot/kpc^3} \Bigg]_{today} \,\times \Bigg[\frac{(km/s)^3}{\sigma^3_{DM}} \Bigg]_{today}\,.\label{tg3} \ee The most DM dominated dwarf spheroidal galaxies feature phase space densities within a wide range (see \cite{destri} and references therein)
    \be 10^{-4}\, (keV)^4 \lesssim  \frac{\rho}{\sigma^3}   \lesssim 1\,(keV)^4 \,.\label{dsph} \ee }

    \item{\textbf{Stability:} To be a DM candidate the candidate particle must feature a lifetime larger than the Hubble time $1/H_0$, namely $\tau H_0 > 1 $. Heavy neutrinos feature various leptonic and radiative decay channels analyzed in section (\ref{sec:qk}). A conservative bound on the lifetime from the dominant leptonic decay is (see eqns. (\ref{nuhdec1}-\ref{longlive}))
        \be |H_{h \alpha }|^2 (M_h/MeV)^5 \lesssim 10^{-13} \,,\label{longlive2} \ee  in particular for heavy neutrinos with $M_h < 1\,\mathrm{MeV}$ the decay channel with the largest branching ratio corresponds to the ``invisible'' decay into three light active-like neutrinos. This decay mode provides  interesting possibilities \emph{even when the heavier neutrino is unstable}, this will be  discussed in section (\ref{sec:cascade}) below.   }

\end{itemize}

\vspace{2mm}

\textbf{Caveats:}
If indeed DM is composed of several species including a hierarchy of different masses, the individual free streaming lengths and phase space densities of a species $\nu_h$ of mass $M_h$ may \emph{not} be the proper characterization. Although the different species do not
interact directly with each other (or do so extremely feebly), they interact indirectly via the common gravitational potential which is sourced by all   species. While a previous study addressed the gravitational clustering properties of mixed dark matter\cite{boyanfreestream} that study did not include cosmological expansion and should be deemed, at best, as of preliminary guidance. As far as we know, there has not yet been a consistent study of free streaming and phase space dynamics for mixed DM including cosmological expansion during the different stages.  In particular on the important question of what is the correct cut-off scale in the linear power spectrum in the case of various components. This entails the study of the linearized collisionless Boltzmann equation including cosmological expansion. While a numerical study implementing the public Boltzmann codes may yield insight, to the best of our knowledge an analytical study for arbitrary concentrations of the different components clarifying the contributions of each species to the effective cutoff scale is still lacking. Until such study emerges, we will consider the free-streaming and phase space density constraints of the individual species as \emph{indicative}.

\section{Heavy  Neutrino production from  Pion decay. } \label{distfunction}

The  previous section  discusses in general terms the quantum kinetic approach to production and freeze-out of heavy neutrino species and the various processes that may produce them in the early Universe. It is clear from the discussion in section (\ref{sec:qk}) that standard model interaction vertices that lead to the production of active neutrinos will also lead to the production of the heavy neutrinos as long as the processes are kinematically allowed. This entails a far broader range of production mechanisms than those that had been the focus in the literature and suggests that a firm assessment of a particular candidate, such as a $\mathrm{keV}$ ``sterile'' neutrino requires a deeper understanding of \emph{all} the standard model processes that may lead to their production during the various cosmological stages.

We recognized the necessity to include finite temperature corrections to masses and interaction vertices to obtain a reliable description.
In this section we implement this program with a definite example: the production of heavy neutrinos from pion decay shortly after the QCD phase transition (crossover) into the confined phase.

 In ref.\cite{asaka} the authors proposed to study hadronic contributions to the production of sterile neutrinos by considering the self-energy corrections to the active (massless) neutrinos in terms of correlators of vector and axial-vector currents. While this is correct in principle, it is an impractical program: the confined phase of QCD is strongly coupled and the description in terms of nearly free quarks is at best an uncontrolled simplification, casting doubts on the reliability of the conclusions in this reference.

Instead, in this section we study the production from pion decay by relying on the well understood effective field theory description of weak interactions of pions, enhanced by the results of a systematic program that studied finite temperature corrections to the pion decay constant and mass implementing non-perturbative techniques from chiral perturbation theory, linear and non-linear sigma models and lattice gauge theory \cite{boyan5,rajawilc,tytgat1,jeon,nicola1,dominfete,harada,nicola2,hotqcd}. There are at least three important reasons that motivate this study: \textbf{1)} it is a clear and relevant example of the quantum kinetic equation approach to production and freeze out that includes consistently finite temperature corrections. In fact this case is similar to the production via the decay of $W,Z$ vector bosons the main difference being the three polarizations of the latter and kinematic factors. \textbf{2)} Pion decay surely contributed to the production of heavy neutrinos in the early Universe if such species of neutrinos do exist: the QCD transition to its confined phase undoubtedly happened, and the consequent hadronization resulted in  baryons and low lying mesons, pions being the lightest, are the most populated meson degrees of freedom near the QCD scale. \textbf{3)} The two leptonic decay channels $\pi\rightarrow \mu \nu_h; \pi \rightarrow e \nu_h$ feature different kinematics and thresholds, in particular the (V-A) vector-axial coupling results in chiral suppression of the $e$ channel for vanishingly small $M_h$. Since the heavy neutrino produced in the decay is \emph{kinematically entangled} with the emitted lepton\cite{lello1}, we expect that the distribution functions from each channel will display differences as a manifestation of this kinematic entanglement, thus providing an explicit example of memory effects as a consequence of ``kinematic entanglement'' in the non-equilibrium distribution function and dependence on the particular production channel.

 At temperatures larger than the QCD Phase transition $T_{QCD} \sim 155 MeV$\cite{hotqcd}, quarks and gluons are asymptotically free. Below this temperature, QCD bound states form on strong interaction time scales, the lightest   being the pion. Recent lattice gauge theory calculations \cite{hotqcd} suggest that the confinement-deconfinement transition is not a  sharp phase transition  but a smooth yet  rapid crossover at a temperature $T_c \sim 155 MeV$  within a temperature range $\Delta T \pm 10 MeV$. This occurs in the radiation dominated epoch at $t \simeq 10 \,\mu secs$ within a time range $\Delta t \sim 2-3\, \mu secs$. This is \emph{much} larger than the typical strong and electromagnetic interaction time scales $\simeq 10^{-22}\,secs$ implying that pions that form shortly after the confining cross-over are brought to LTE via strong, electromagnetic and weak interactions on time scales much shorter than $\Delta t$. After pions are produced they  reach LTE via  $\pi-\pi$ scattering on strong interaction time scales, they decay into leptons and maintain LTE via detailed balance with the inverse process since charged leptons and active-like neutrinos  are also in LTE.  In conclusion, for $T \lesssim 155 MeV$, pions are present in the plasma in thermal/chemical equilibrium due to pions interacting on strong interaction time scales ($10^{-22}s$) while the crossover  transition occurs on the order of $2-3\,\mu secs$.

 If the pions (slowly) decay into heavy neutrinos $\nu_h$  with very small mixing angles, detailed balance for $\pi \rightleftharpoons l \nu_h, h=4,5\cdots$ will \emph{not} be maintained if the heavy neutrinos are \emph{not} in LTE. As the pion is the lowest lying bound state of QCD, it is a reasonable assumption that during the QCD confinement-deconfinemet crossover pions will be the most dominantly produced mesonic bound state and, during this time,   pions will remain in LTE with the light \emph{active-like} neutrinos and charged leptons by detailed balance $\pi \rightleftharpoons l_\alpha \nu_{\alpha}$. We focus on heavy neutrino $\nu_h$ production from $\pi \rightarrow l \nu_h$ which is \emph{suppressed} by $|H_{\alpha h}|^2 \ll 1$ with respect to the active neutrinos and does not maintain detailed balance.

 A full study of sterile neutrino production through $\pi$ decay  in the early universe requires various finite temperature corrections and a preliminary study which focused on the production of    neutrinos in the $\mathrm{keV}$  mass range\cite{lellolightsterile} has implemented the first step. This study yielded a suitable warm dark matter candidate with free streaming lengths on the order of several kpc whereas it is expected that   heavier neutrinos will yield a  colder dark matter species.

 The issue of a heavy   neutrino production ($M_h \gtrsim MeV$) through $\pi$ decay has not been addressed and is the main focus of this section.  Through the two possible channels $\pi \rightarrow \mu \nu_\mu; \pi \rightarrow e \nu_e$ pion decay offers a wide kinematic window to the production of heavy neutrinos and provides  a natural mechanism to produce a \emph{mixed dark matter} scenario provided that a hierarchy of  heavy neutrino species exist and their production is kinematically allowed\cite{mixedsterile,boyanfreestream}. Furthermore, ``kinematic entanglement'' will yield different distribution functions for the different channels, thus providing an example of a \emph{mixed} distribution for the same species.

Whereas  pion decay $\pi \rightarrow \mu \nu$, is one of the most ubiquitous mechanisms to produce neutrinos in many terrestrial experiments, including short baseline experiments, this  production mechanism has not been fully addressed within the cosmological setting. While we \emph{do not} claim that the processes studied below are more or less important than the processes described in the previous section or discussed elsewhere in the literature, this study leads to a clear example of the concepts and methods described in the previous section, including the finite temperature corrections. Furthermore, this production mechanism \emph{may} yield a substantial (if not the dominant) contribution to DM below the QCD scale.

  In studying the production of heavy neutrinos, we make the following assumptions for the quantum kinetic equation   that governs the $\nu_h$ population build up:

\begin{itemize}
\item A substantial body of work has investigated finite temperature corrections to the pion decay constant \cite{tytgat1,nicola1,nicola2,jeon,harada}. To leading order in chiral perturbation theory,    the correction is given by

\be
f^2_{\pi} \rightarrow f^2_{\pi}(t) = f^2_{\pi}(0)\left(1-\frac{T(t)^2}{6 f_{\pi}(0)^2}\right) ~~;~~ f_{\pi}(0) = 93 MeV\,. \label{fpiofT}
\ee This result will be needed for the quantum kinetics equation as we consider production beginning at $T_{QCD} \sim 155 MeV$. These results are well established but this subfield remains an active area of investigation \cite{newlattice} and, for precision studies,  these results will need refinement. The assumption is that there are no pions present in the plasma above $T_{QCD} \sim 155 MeV$ and that hadronization happens instantaneously in comparison to the neutrino production time scales.

\item Finite temperature corrections to the masses of both the pions and charged leptons occur in the plasma. The   pion mass including electromagnetic corrections is shown in figure 2 of \cite{nicola1}. It is shown that between the temperatures of 10 and 150 MeV, the pion mass varies between  140  and 144 MeV. This change is relatively small and will be neglected for the calculation and we will use an average $M_{\pi} = 142 MeV$. Finite temperature corrections to the charged lepton mass are of  $\mathcal{O}(e T) $\cite{htl,pisarskihtl,lebellac,weldon} and for muons  this correction is only a fraction of its mass so it can be neglected. However, mass corrections to the electron may be substantial, but in this case we are interested in a kinematic window of large mass for $\nu_h$.   For the purposes of this work, the charged lepton mass will be taken as a constant for both muons and electrons. The effects of charged lepton mass corrections will be investigated elsewhere.

\item   It is argued in \cite{rafelski} that the chemical potentials (including pions) are on the order of $\sim 10^{-6} eV$ for the temperature range we consider here. We  assume  that the lepton asymmetry is small and will  be neglected in our calculations consistently with the neglect of chemical potentials in the quantum kinetic equations of the previous section.

   \item   The mixing angle between active-sterile neutrinos in the presence of a matter potential will develop a temperature and lepton asymmetry dependence as shown in refs. \cite{medium,dolgovreview,dolgovenqvist,boywu}. For vanishing lepton asymmetries (chemical potentials) and $T/M_W \ll 1$ there are no in medium MSW resonances\cite{medium,dolgovreview,boywu}. In absence of lepton asymmetry the finite temperature in-medium potential  lead to the following modification of the mixing angles (we consider mixing between one active-like and a heavy neutrino only)

    \be
    \sin(2\theta_m) = \frac{   \sin 2\theta}{ \left[\sin^2 2\theta + ( \cos 2\theta + V^{th}/\Delta)^2\right]^{1/2}} \label{mixanT}
    \ee where $\Delta = \delta M^2/2E = (M_h^2 - M_a^2)/2E$,   and $V^{th} \sim 10^2 G_F^2 E T^4$\cite{medium,dolgovreview}. We will be concerned with the temperature range $E\sim T \sim \, 100\, MeV$,   we find
    \be V^{th}/\Delta \sim 10^{-2} \,\Big(\frac{T}{100\,\mathrm{MeV}}\Big)^6\,\Big(\frac{\mathrm{keV}}{M_h} \Big)^2 \label{Vth}\ee

      The focus of this study is mainly on neutrinos with $M_h \gtrsim \mathrm{few} \,keV$ and $\sin(2\theta) \ll 1$ at temperatures below $T \sim 150 MeV$ so the temperature dependence of the mixing can be ignored. In any event, the temperature correction to the mixing angles may, at most, lead to a slight quantitative change but not to a major modification of our main results, in particular the contributions from different channels with  kinematic entanglement is a robust feature,  as it will become clear from the analysis  below.

\end{itemize}

The effective low energy field theory that describes pion decay appended by finite temperature corrections to the $\pi^\pm$ mass and decay constant yields the following interaction Hamiltonian (in the interaction picture),

\be
H_I = \sum_{\alpha=e,\mu} \sqrt{2} \,G_F \,V_{ud}\, f_{\pi}(T) \,\int d^3x \left[ \bar{\nu}_{\alpha}(\vec{x},t) \gamma^{\sigma} \mathcal{P}_L \Psi_{\alpha} (\vec{x},t) J^{\pi}_{\sigma} (\vx,t) + h.c. \right] \label{piHeff}\,
\ee where $ J^{\pi}_{\sigma} = i \partial_{\sigma} \pi (x,t)$ is the pseudoscalar pion current, $G_F$ is the Fermi constant, $V_{ud}$ is the CKM matrix element, $f_{\pi}$(T) is the pion decay constant with finite temperature corrections given by (\ref{fpiofT}) and the ``flavor'' neutrino fields $\nu_\alpha$ are related to the fields that create-annihilate neutrino mass eigenstates by the relation (\ref{relanus2}).

Pions   decay into heavy neutrinos through the channels $\pi^{+} \rightarrow \mu^{+} \nu_h \, ~;~\pi^{+} \rightarrow e^{+} \nu_h$ (and the charge conjugate). Because of the kinematics  the heavy neutrino $\nu_h$ is ``entangled'' with the charged lepton $l$ in the sense that the gain and loss rates depend on the particular channel.  Therefore for each channel we label the gain and loss rates and the distribution function with the labels $l,h$. As discussed in detail in section (\ref{sec:qk}) the total rates are the sum over all channels  and the quantum kinetic equation inputs the total rates as per eqn. (\ref{gamatots})

 The quantum kinetic equation for $\nu_h$ production is given by (\ref{qkcosmo}) with the total gain and loss rates summed over the kinematically allowed channels labeled by the corresponding charged lepton $l$. The results for these rates are given in detail in appendix (\ref{kinetics}),
\bea   \Gamma^<_{lh}(q)  & = & \frac{ |H_{l h} |^2 |V_{u d}|^2 G_F^2 f_{\pi}^2(T) }{8 \pi} \frac{M^2_{\pi}(M^2_{l} + M^2_{h} ) - (M^2_{l} -M^2_{h})^2}{q E_{h}(q)}
 \nonumber \\ & \times  & \int^{p_+}_{p_-} \frac{dp \, p}{\sqrt{p^2 + M_{\pi}^2 } } \Big[ n_{\pi}(p) (1 - n_{\bar{l}}(|\vp-\vq|)) \Big] \label{gamalesfi}\eea and
 \be \Gamma^>_{lh}(q) = e^{E_h(t)/T(t)}\,\Gamma^<_{lh}(q)\,, \label{gamapisrel}\ee where momenta and energies are replaced by their local expressions in the expanding cosmology (\ref{eandp}) and
 \be
p_{\pm}(t) = \left| \frac{E_{h}(q,t)}{2M_h^2} [ (M_{\pi}^2 - M^2_{l} + M_h^2)^2 -4M^2_{\pi}M^2_{h} ]^{1/2} \pm \frac{q_f(t) (M^2_{\pi} - M^2_{l} + M^2_{h} )}{2M^2_{h}} \right|  \,. \label{limits2}
\ee and the constraint from the energy conserving delta function
\be E_\pi(p,t) = E_l(k,t)+E_h(q,t)~~;~~k = |\vp-\vq| \,. \label{eners}\ee As argued previously, the pions and charged leptons $\pi^{\pm},l^{\pm}$ are assumed to be in LTE so that their distributions will take the standard bosonic and fermionic forms

\be
n_{\pi^{\pm}} = \frac{1}{e^{E_{\pi}(p,t)/T} - 1} ~~;~~ n_{\bar{l}/l} = \frac{1}{e^{E_{l}(k,t)/T} + 1} ~~;~~   E(q,t) = \sqrt{\frac{q_c^2}{a(t)^2} + M^2}
\ee where $q_c$ is the comoving momentum and we neglect  chemical potentials in this study.

Inserting the distributions into eqn. (\ref{gamalesfi})  and using $E_{l}(k) = E_{\pi}(p) -E_{h}(q)$, the gain rate becomes

\bea
\Gamma^<_{lh}(q) & = & \frac{ |H_{l s} |^2 |V_{u d}|^2 f_{\pi}(T(t))^2 }{16 \pi} \frac{\Big[M^2_{\pi}(M^2_{l} + M^2_{h} ) - (M^2_{l} -M^2_{h})^2\Big]}{q\sqrt{q^2+M_h^2}}  \\
\nonumber & * & \int^{p_+(t)}_{p_-(t)} \frac{dp \, p}{\sqrt{p^2 + M_{\pi}^2 } } \left[ \frac{e^{-E_{\nu}(q)/T} e^{E_{\pi}(p)/T}}{(e^{E_{\pi}(p)/T}-1) (e^{-E_{\nu}(q)/T} e^{E_{\pi}(p)/T}+1)} \right] \,, \label{Glespi}
\eea where $p^{\pm}$ are the solutions to the phase space constraints of eqn. (\ref{limits}) and the time dependence of the physical momentum has been suppressed. This integral is readily solved by a substitution with the following result

\bea \label{rate}
\Gamma^<_{lh}(q) & = & \frac{ |H_{l h} |^2 |V_{u d}|^2 f_{\pi}^2(t) }{16 \pi } \frac{\Big[M^2_{\pi}(M^2_{l} + M^2_{h} ) - (M^2_{l} -M^2_{h})^2\Big]}{q(t) E_{h}(q,t) (e^{E_{h}(q,t)/T(t)}+1)}\, T(t) \\
 \nonumber & * & \ln \left( \frac{ e^{E_\pi/T(t)} -1  }{ e^{-E_{h}(q,t)/T(t)} e^{E_\pi/T(t)}+ 1  } \right) \Bigg|^{E_\pi=E^+_\pi(q_f(t))}_{E_\pi=E^-_\pi(q_f(t))} \,.
\eea  where
\be E^\pm_\pi(q_f(t)) = \frac{1}{2M^2_h}\Bigg[E_h(q_f(t))\Big(M^2_\pi+M^2_h-M^2_l) \Big)\pm q_f(t) \lambda(M_\pi,M_l,M_h)  \Bigg] \, \label{Eplmin}\ee and the threshold function is defined as
\be \lambda(M_\pi,M_l,M_h) = \Bigg[M^4_\pi+M^4_l+M^2_h-2M^2_\pi M^2_l-2 M^2_\pi M^2_h - 2M^2_h M^2_l \Bigg]^{1/2}\,. \label{threshfun}\ee
In the limit $M_h \rightarrow 0$ the bracket in the first line of (\ref{rate}) yields the usual factor $M^2_l(M^2_\pi - M^2_l)$, which is the hallmark of pion decay vanishing in the $M_l \rightarrow 0$ limit.

In eqn. (\ref{rate}), it is readily seen that $\Gamma^{<}$ depends on the ratio   $y = \frac{q_f(t)}{T(t)} = \frac{q_c}{T_0}$ where $T_0$ the temperature of the plasma \emph{today} with the scale factor normalized today  at ($t=t_*$) with $a(t_*)=1$    and $q_c$ is the comoving momentum.

  The threshold function is one of the signatures of ``kinematic entanglement''; for fixed $M_l$, the threshold function vanishes at $M_\nu = M_\pi-M_l$, for this value of $M_h$ the two roots $E^\pm_\pi$ coalesce and the rate vanishes. This is important because for $M_h \lesssim 36\,\mathrm{MeV}$ there are \emph{two} production channels of heavy neutrinos, $\pi \rightarrow \mu \nu_h; \pi \rightarrow e \nu_h$, whereas for $M_h > 36\,\mathrm{MeV}$ only $\pi \rightarrow e \nu_h$ is kinematically available.

 Restricting our study to the production process $\pi \rightarrow l \nu_h$ and its inverse, we anticipate that the condition for production and freeze-out out of LTE (\ref{freezecond}) will be fulfilled, this will be proven self-consistently below. In this case the solution of the quantum kinetic equation is given by (\ref{Nhfroz}), namely the quantum kinetic equation (\ref{qkcosmo}) simplifies to
 \be \frac{d n_h(q_c;t)}{dt} = \Gamma^<_{tot}(q_c;t)  \,,\label{qkcosmopitot} \ee where
 \be \Gamma^<_{tot}(q_c;t) = \sum_{l=\mu, e} \Gamma^<_{lh}(q_c;t) \label{sumonl}\ee and the sum is over the kinematically allowed channels. In this approximation the linearity of (\ref{qkcosmopitot}) allows us to introduce a distribution function for \emph{each channel} $n_{lh}(q_c;t)$ that obeys
  \be \frac{d n_{lh}(q_c;t)}{dt} = \Gamma^<_{lh}(q_c;t)  \,, \label{qkcosmopil} \ee which will prove useful   as it allows the opportunity to   understand how the ``kinematic entanglement'' associated with each channel is manifest in the frozen distribution function.

   In this approximation the total distribution function is
  \be n_h(q_c;t) = \sum_{l=\mu, e}n_{lh}(q_c;t)\,. \label{totaln}\ee and after freeze-out
  \be f_h(q_c) = n_h(q_c;t=\infty) = \sum_{l=\mu, e}f_{lh}(q_c;t=\infty)\,, \label{totalfh}\ee in other words the total distribution function after freeze-out is a \emph{mixture} of the contributions from the different channels.

We emphasize that whereas the above definition is useful to learn the  effects of the ``kinematic entanglement'' in the frozen distribution function, the cosmologically relevant quantities, such as coarse-grained phase space densities, free streaming length etc, are \emph{non-linear} functions of the distribution function as they involve moments, therefore for these quantities only the total distribution function $n_h$ (\ref{totaln}) is relevant.

 It proves convenient to introduce  a dimensionless parameter $\tau$ that will play the role of  time along with with a  change of variables that introduces manifestly a factor of the Hubble factor (during radiation domination):

\be \label{hubble}
y=\frac{q_f(t)}{T(t)}=\frac{q_c}{T_0} ~~;~~ \tau = \frac{M_{\pi}}{T(t)} ~~;~~ \frac{d\tau}{dt} = \tau H(t) ~~;~~ H(t) = 1.66 \, g^{\frac{1}{2}}_{eff}(T)\,\frac{  T^2(t)}{M_{Pl}}\,,
\ee furthermore take the overall scale to be $M_\pi$ and define the following dimensionless ratios
\be m_h \equiv \frac{M_h}{M_\pi}~~;~~m_l \equiv \frac{M_l}{M_\pi}~~;~~ l= \mu, e \,. \label{ratiosm} \ee

 As argued previously, production from pion decay begins after the QCD crossover   during the radiation dominated epoch. We will argue that freeze out occurs at temperatures $T \gtrsim 10 MeV$ so that the Hubble factor is given by eq \ref{hubble} for the entire period of production. With this change of variables and factoring the constants out of Eq. (\ref{rate}) leads to the definition of the overall scale of the problem:

\be \label{scale}
\Lambda_{lh} = \frac{|H_{l h}|^2}{\sqrt{g(t)}} \left[ \frac{ |V_{u d}|^2 f^2_{\pi}(0) G^2_F}{8 \pi*1.66}  M_{pl}M_\pi  \right] \left(m_l^2 +m_{h}^2 - (m_l^2-m_{h}^2)^2  \right) \,,
\ee  where the dimensionless ratios $m_{l,h}$ are defined in (\ref{ratiosm}).
After the temperature falls below the hadronization temperature ($T\sim 155MeV$), the relativistic degrees of freedom ($g(t)$) remain constant until the temperature cools to below the $\mu$ mass ($T\sim 106 MeV$) and again remains constant until cooling to below the electron mass. For $155 MeV \gtrsim T \gtrsim 106 MeV$ the relativistic degrees of freedom are $g(t) \sim  14.25$ and for $106 MeV \gtrsim T \gtrsim 0.5 MeV$ the degrees of freedom are $g(t) \sim 10.75$ \cite{pdg}. As will be argued, $\nu_h$ production is complete by $10 MeV$ and since $ {g(t)}$ has a small variation in the temperature range of production, $ {g(t)}$ is replaced by its average value $\bar{g} = 12.5$. With this, the time variation of the overall scale  $\Lambda_h$  can be neglected, and using $|V_{ud}|= 0.974$\cite{pdg} we find
\be \Lambda_{lh} \simeq 0.13 \, \Bigg(\frac{|H_{l h}|^2}{10^{-5}}\Bigg) \Big[m_l^2 +m_{h}^2 - (m_l^2-m_{h}^2)^2  \Big]\,. \label{numlambda} \ee
 The function
 \be C[m_l,m_h] = \Big[m_l^2 +m_{h}^2 - (m_l^2-m_{h}^2)^2  \Big] \label{chiral} \ee reveals the usual  \emph{chiral suppression} in the case of negligible neutrino masses as a consequence of the (V-A) coupling,  being much larger for the $\mu$-channel than the $e$-channel with $C[m_e,m_h]/C[m_\mu,m_h] \simeq 10^{-4} $ for $m_h \rightarrow 0$.

  The compilation of bounds on mixing angles of heavy neutrinos in ref.\cite{kusbounds} in combination with bounds from X-ray astrophysical data\cite{bulbul,boyarsky} suggests that $|H_{lh}|^2 \ll 10^{-5}$ for $ \mathrm{few}\, \mathrm{keV} < M_h < 140 \,\mathrm{MeV}$ consequently  $\Lambda_{lh} \ll 1$ for $M_h$ in this range.

Production from pion decay begins shortly after the QCD crossover at $T_{QCD} \simeq 155 \,\mathrm{MeV}$ when pions form, therefore  $M_\pi/T_{QCD} = \tau_0 \leq \tau < \infty$ where $\tau_0 \simeq 0.92$.

Upon substituting the dimensionless variables, Hubble factor and overall scale the kinetic equation may be written in the following form:

\be \label{exactrate}
\frac{1}{\Lambda_{lh}}\frac{dn_{lh}}{d\tau}(y,\tau) = \frac{\tau^2\Big(1-\frac{M^2_\pi}{6\tau^2 f^2_\pi(0)}\Big)}{ \sqrt{y^2+ m^2_h \tau^2}\,\Big(e^{E_h(q,t)/T(t)}+1 \Big) }\,\ln \left( \frac{  e^{E_\pi(p,t)/T(t)} -1  }{ e^{-E_{h}(q,t)/T(t)} e^{E_\pi(p,t)/T(t)}+ 1  } \right) \Bigg|^{p=p_+(t)}_{p=p_-(t)} \,.
\ee Upon evaluating the pion energy at the solutions of Eq. (\ref{limits}), in terms of the dimensionless variables defined above (\ref{hubble},\ref{ratiosm}) we obtain\footnote{We do not include a label $l$ in the result for $E_\pi$ to avoid cluttering of notation.}

\be
 \frac{E_{\pi}(p_{\pm}(t),t)}{T(t)} = \frac{1}{2m^2_h}\Bigg[\Delta_{lh} \Big[y^2 + m^2_h \tau^2\Big]^{1/2}   \pm     y\,\delta_{lh}\Bigg]    \label{epipm} \,.
\ee where we introduced
\bea \Delta_{lh} & = &  {1 - m_l^2 +m_h^2} \label{delgran}\\ \delta_{lh} & = & \Big[1+m^4_l+m^4_h-2 m^2_l-2 m^2_h - 2m^2_h m^2_l \Big]^{1/2} = \Big[ \Delta^2_{lh}-4\,m^2_h \Big]^{1/2}\label{delsmal}\eea

From this equation, the rate can be integrated numerically to study the distribution function after freezeout and obtain the   cosmological quantities discussed in section (\ref{sec:cosmocons}). However, before we proceed to a numerical integration of (\ref{exactrate})  we recognize important features of this equation that anticipate the behavior of the solution:

\begin{itemize}

\item The prefactor, proportional to $\tau^2$ is small initially but   the argument of the logarithm attains its largest value as the initial  temperature $T(t_0) \geq M_{\pi}$, however, as $\tau$ increases, the temperature decreases and the   logarithmic term decreases eventually exponentially beating the growth of the prefactor. This analysis indicates that the production rate peaks as a function of time $\tau$ and falls off fast. We confirm this behavior numerically in fig. (\ref{fig:evkinetics}). This figure shows the fast rise and eventual fall off of the production \emph{rate} which becomes exponentially suppressed for $\tau \gtrsim 10 $. This entails a freeze out of the distribution function: while the rate is exponentially suppressed, the total integral will remain finite, thereby fulfilling the freeze-out condition (\ref{frees}). In ref. \cite{lellolightsterile} a similar behavior was noticed for lighter neutrinos and we confirmed numerically that a similar pattern holds for the whole kinematic range of $M_h$ in each lepton channel.

 \begin{figure}[h!]
\begin{center}
\includegraphics[height=3.5in,width=3.2in,keepaspectratio=true]{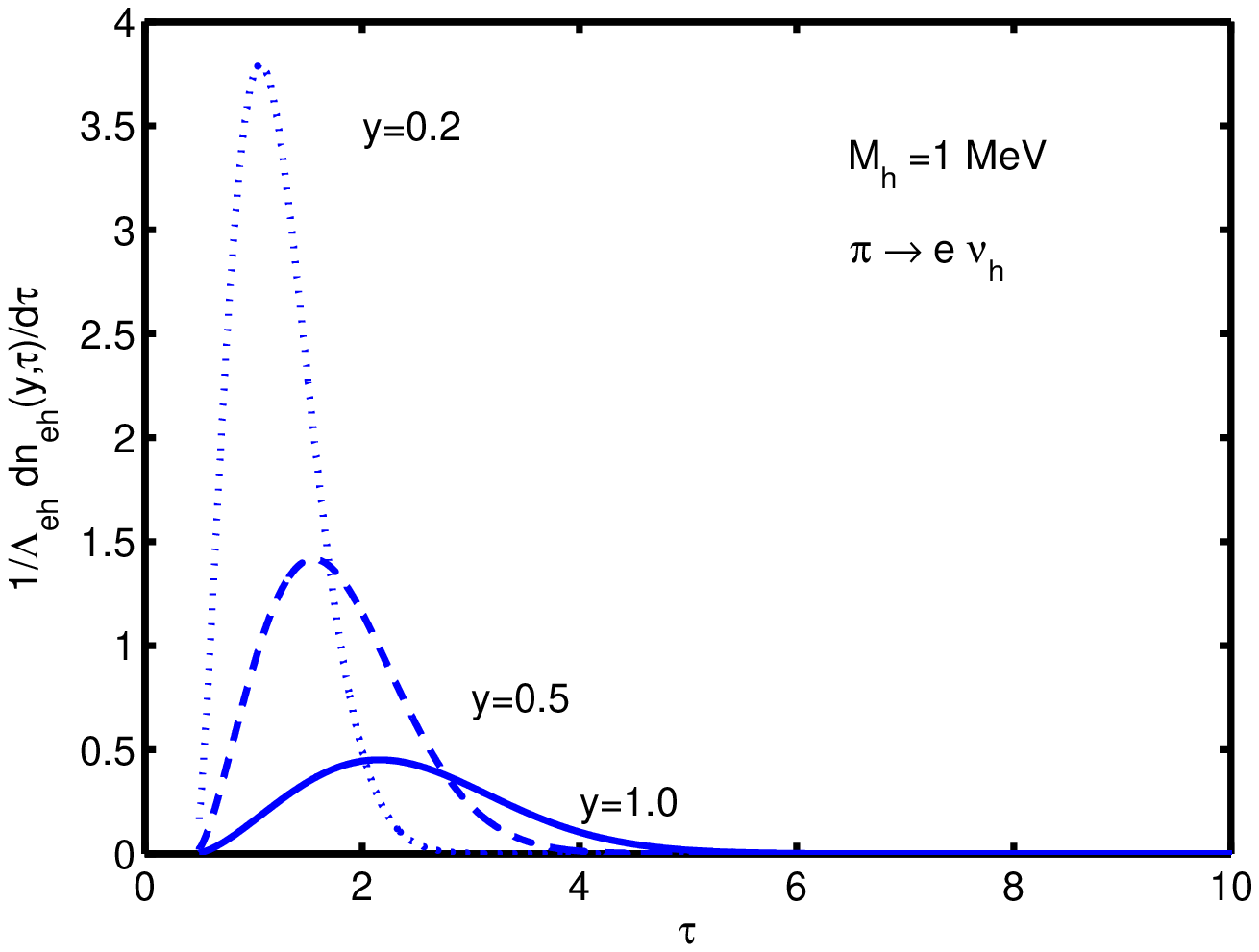}
\includegraphics[height=3.5in,width=3.2in,keepaspectratio=true]{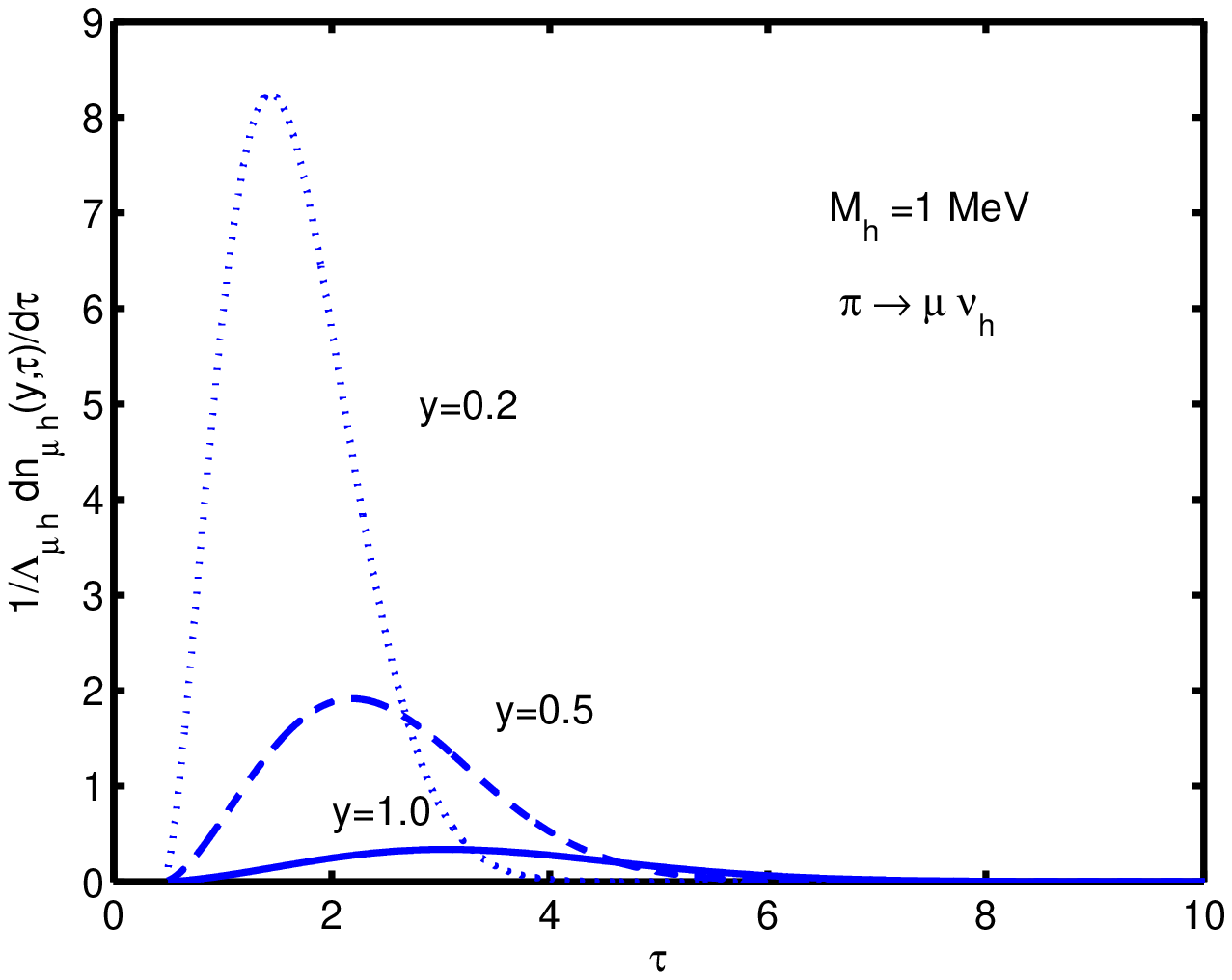}
\caption{Production rate of a heavy  $\nu_h$   from $\pi \rightarrow l \nu_s$ (eqn. (\ref{exactrate}) with $l=\mu,e$  for $M_{h} = 1 MeV$.}
\label{fig:evkinetics}
\end{center}
\end{figure}

\item The expression (\ref{epipm}) with (\ref{delgran},\ref{delsmal}) which is a consequence of the kinematics has very important implications on the ``kinematic entanglement'' of the heavy neutrino. For fixed $y,m_h$ the ratio (\ref{epipm}) is \emph{smaller} in the $\mu$ channel than in the $e$ channel because both $\Delta_{lh}$ and $\delta_{lh}$ are smaller, this implies that the rate is \emph{larger} in the $\mu$ channel. This feature is also displayed in fig. (\ref{fig:evkinetics}).  Therefore, we expect that $n_{\mu h}(y,\tau)/\Lambda_{\mu h} > n_{e h}(y,\tau)/\Lambda_{e h}$ and the distributions at freeze out to display this difference. Furthermore, the function $C[m_l,m_h]$ (\ref{chiral}) in $\Lambda_{lh}$ (\ref{numlambda}) is \emph{also} larger for the $\mu$-channel than in the $e$-channel for small $m_h$. This is a consequence of the chiral suppression as a consequence of $V-A$ and is displayed in fig.(\ref{fig:coef}).

     \begin{figure}[h!]
\begin{center}
\includegraphics[height=3.5in,width=3.2in,keepaspectratio=true]{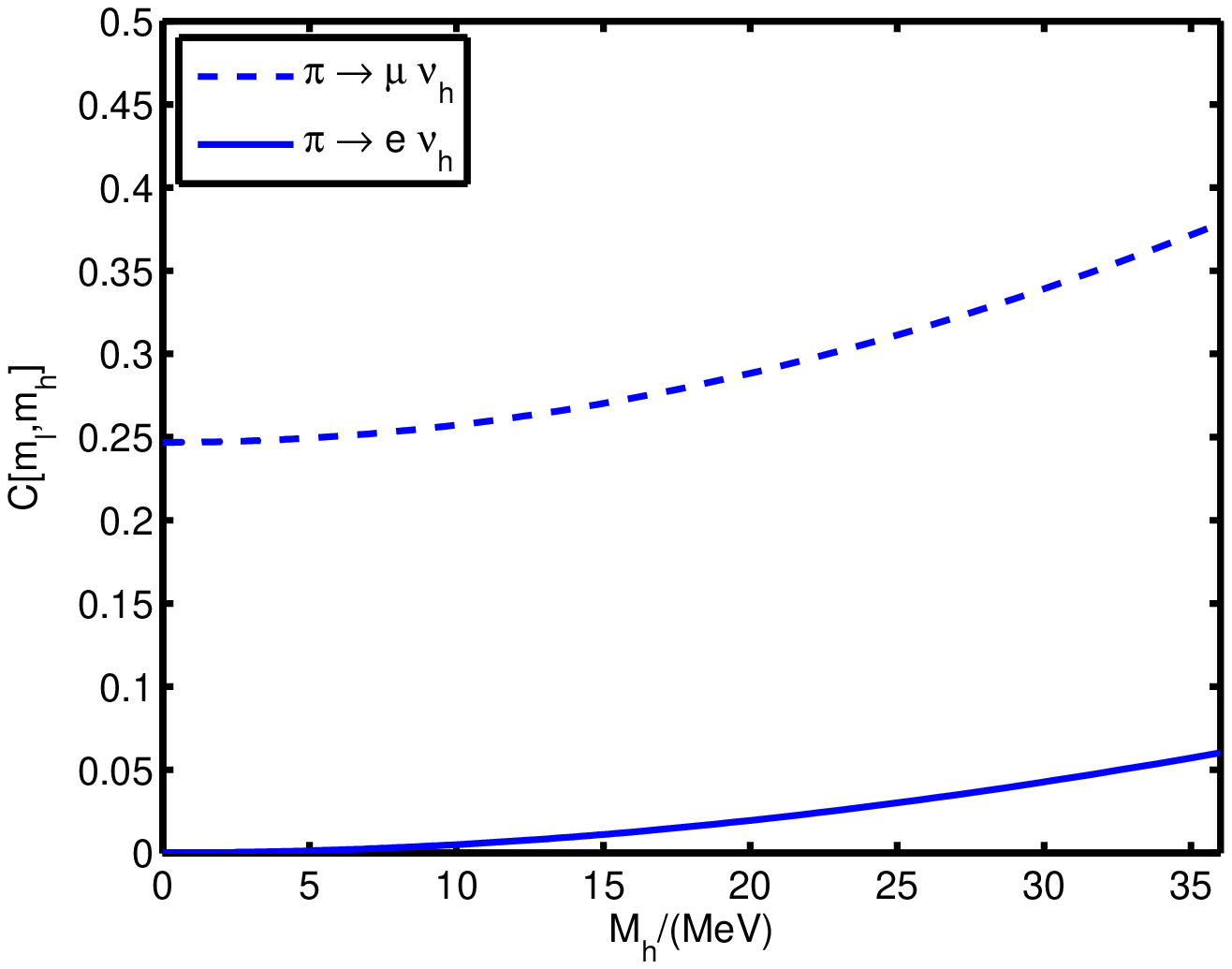}
\includegraphics[height=3.5in,width=3.2in,keepaspectratio=true]{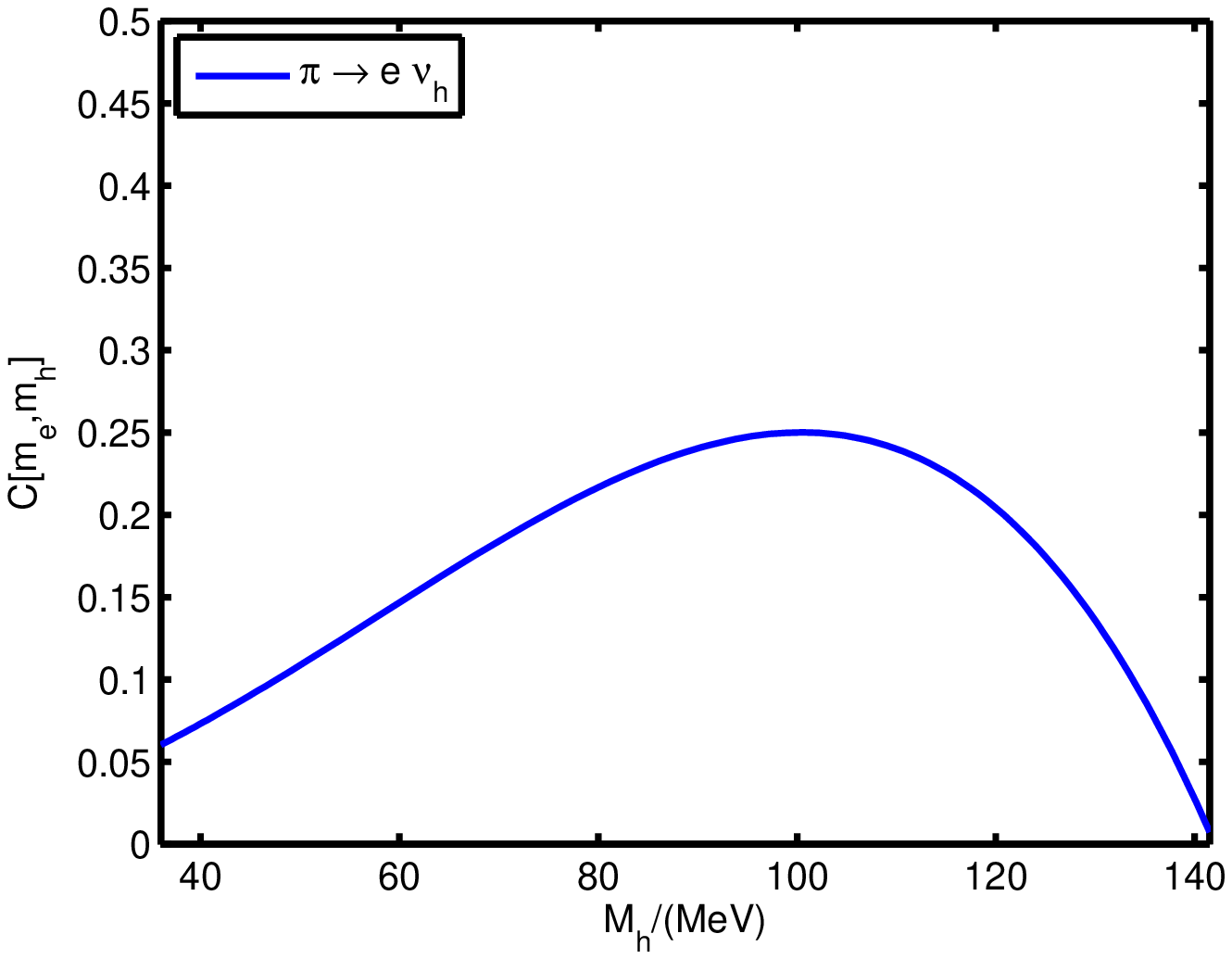}
\caption{  $C[m_\mu,m_h]$ for both channels in the kinematically allowed region of $M_h$. }
\label{fig:coef}
\end{center}
\end{figure}

\end{itemize}

The loss rate $\Gamma^>$ is obtained from the gain rate (\ref{exactrate}) from the detailed balance condition (\ref{gamalesfi}) and we can now proceed to integrate the quantum kinetic equation. However, we first confirm that the gain-loss processes solely from pion decay (and recombination) lead to freeze-out out of LTE, namely we first confirm that the condition (\ref{freezecond}) is fulfilled. This is shown explicitly in figs. (\ref{fig:gammas}). These figures confirm that the condition (\ref{freezecond}) is fulfilled provided $\Lambda_{lh} \ll 1$. The result (\ref{numlambda}) indicates that this is the case for $|H_{lh}|^2 \ll 10^{-5}$. Cosmological bounds from X-ray data\cite{bulbul,boyarsky,casey} and a compilation of accelerator bounds\cite{kusbounds} suggest that $|H_{lh}|^2 \ll 10^{-5}$ for $\mathrm{few}\,\mathrm{keV} \leq M_h \leq 140 \,\mathrm{MeV}$, therefore the condition $\Lambda_{lh} \ll 1 $  is   satisfied guaranteeing self-consistently that the condition (\ref{freezecond}) is fulfilled.

 \begin{figure}[h!]
\begin{center}
\includegraphics[height=3.5in,width=3.2in,keepaspectratio=true]{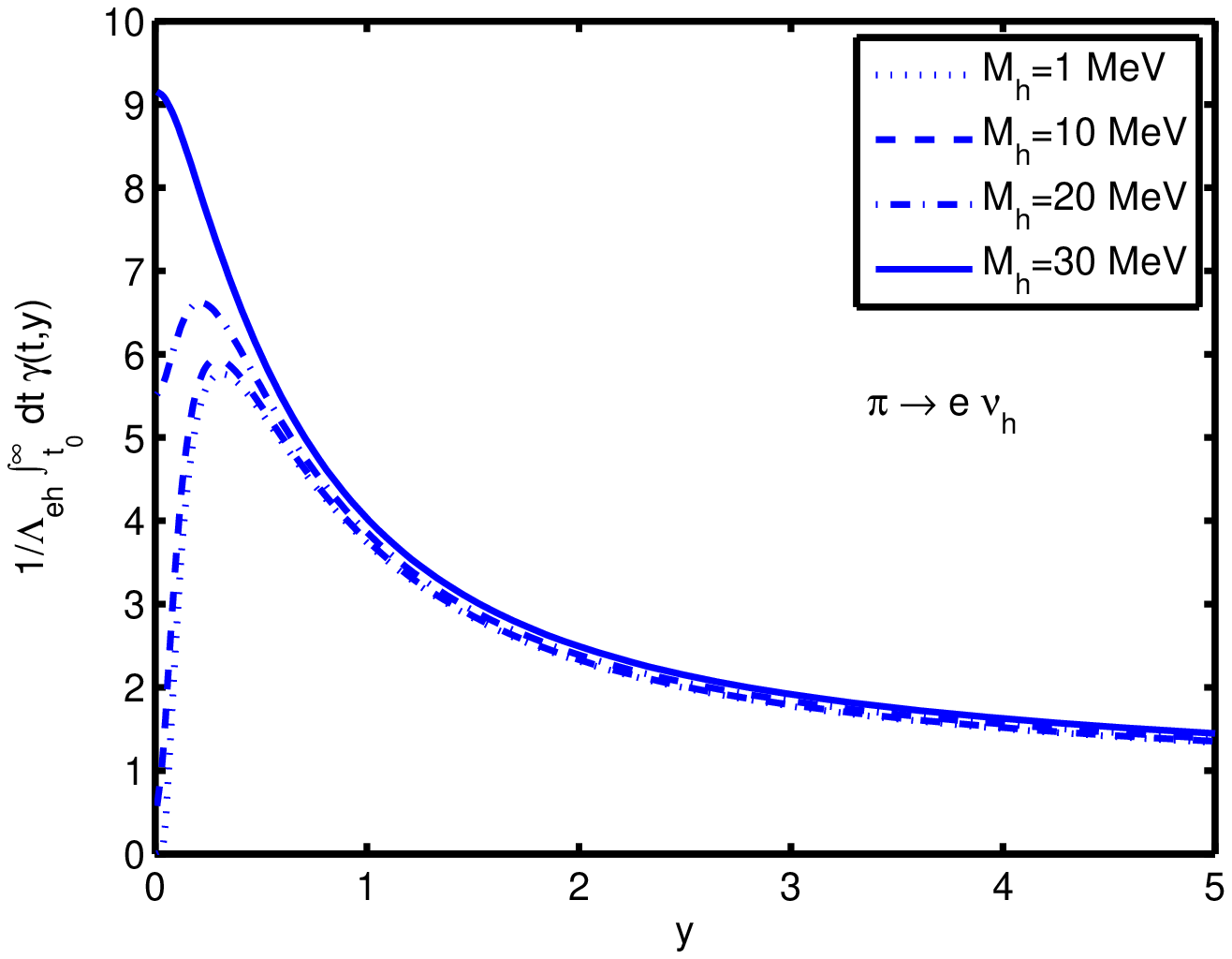}
\includegraphics[height=3.5in,width=3.2in,keepaspectratio=true]{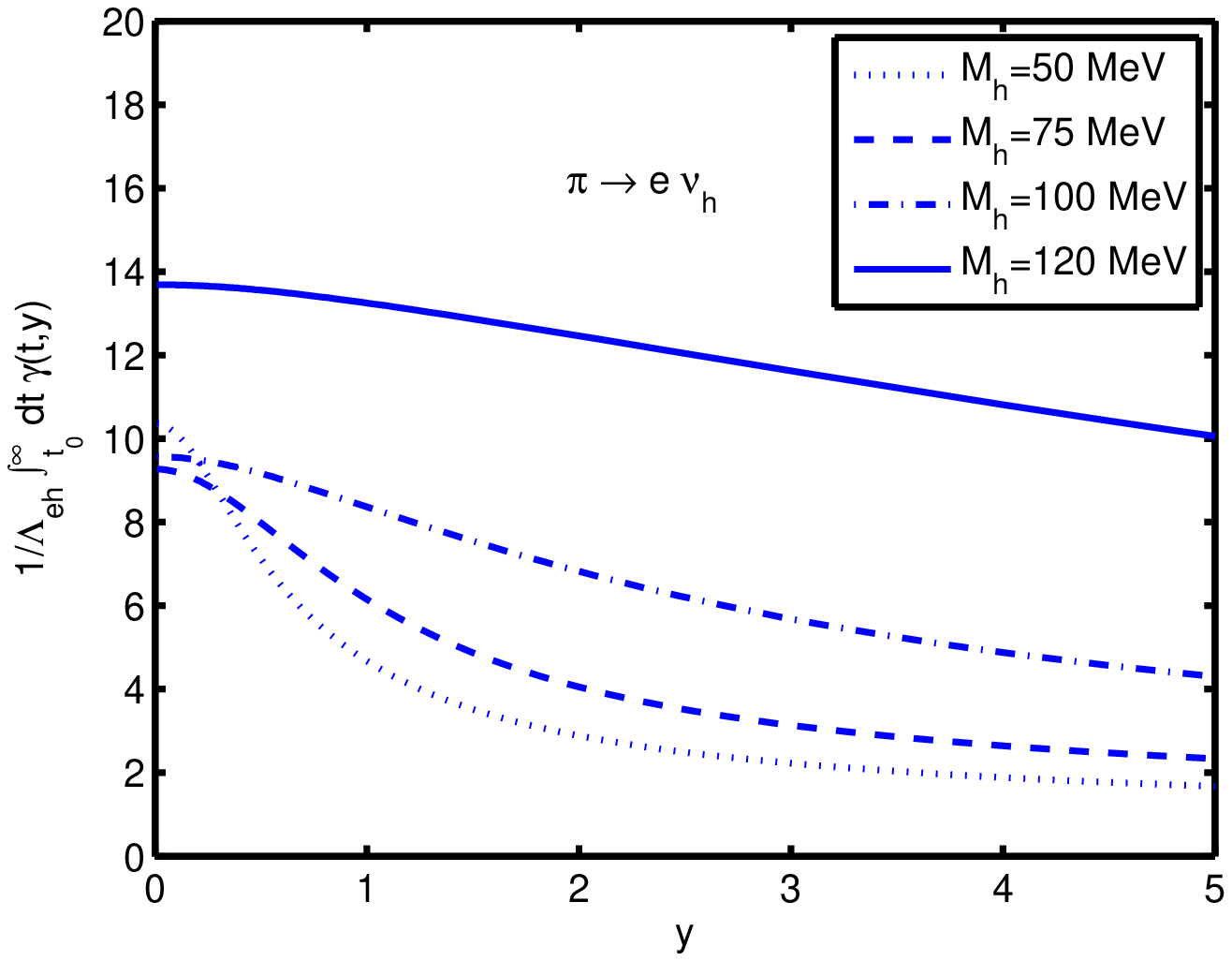}
\includegraphics[height=3.5in,width=3.2in,keepaspectratio=true]{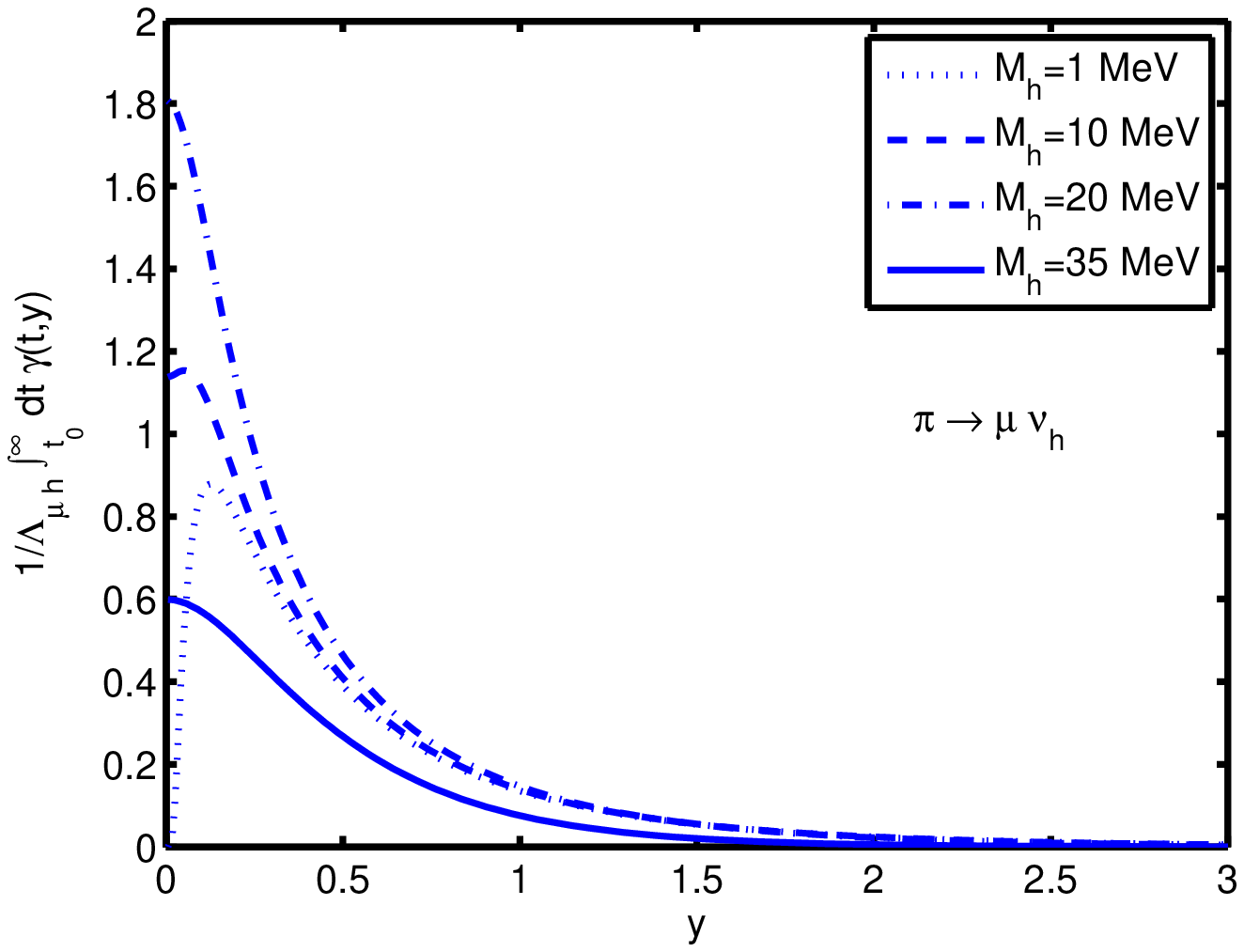}
\caption{$(1/\Lambda_{lh})\, \int_{\tau_0}^\infty \gamma_h(y,\tau')d\tau'$ for the two channels .}
\label{fig:gammas}
\end{center}
\end{figure}

In this case  the  population build up is obtained by   integration of the rate eqn. (\ref{exactrate}) as per the general result (\ref{Nhfroz}).
In order to exhibit clearly the contribution from $\pi$-decay we assume vanishing initial population, in this approximation any initial population must be added. For each channel we obtain the distribution function from direct integration of the gain rate
\be
n_{lh}(y,\tau) = \int^{\tau}_{\tau_0}   \frac{dn_{lh}}{d\tau}(y,\tau')\,d\tau' \label{numberh}\,.
\ee The results are shown in figs (\ref{fig:lightedist},\ref{fig:heavyedist},\ref{fig:mudist}). Each figure shows the distribution at different values of time ($\tau$) and, in all cases, the distribution has been frozen out at $\tau \simeq \tau_{fr} \simeq  10$  at which  $T(\tau_{fr})\lesssim 14 MeV$, this is because the rate is being suppressed through the pion thermal distribution as displayed in figs.(\ref{fig:evkinetics}).

\begin{figure}[h!]
\begin{center}
\includegraphics[height=3.5in,width=3.2in,keepaspectratio=true]{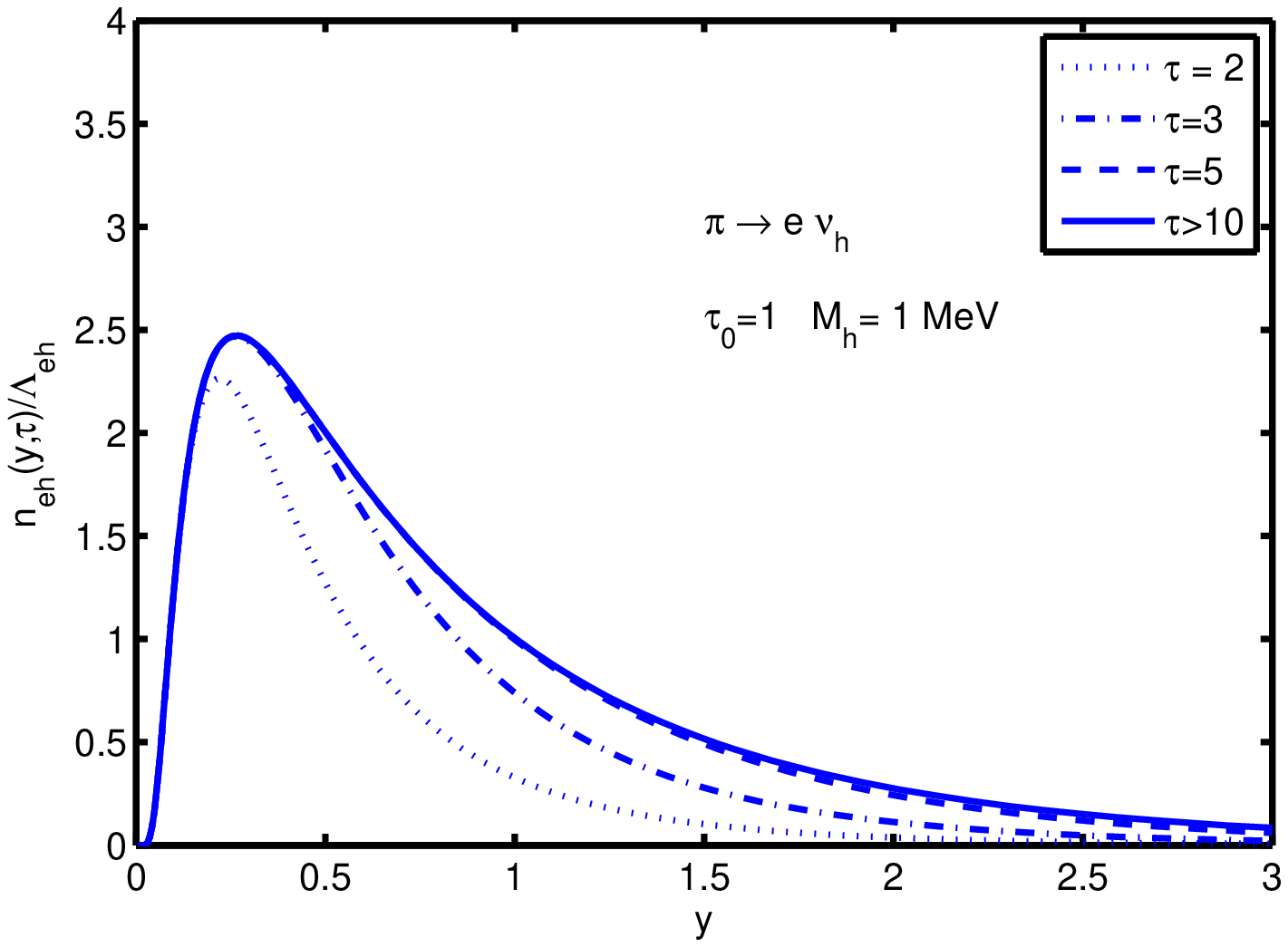}
\includegraphics[height=3.5in,width=3.2in,keepaspectratio=true]{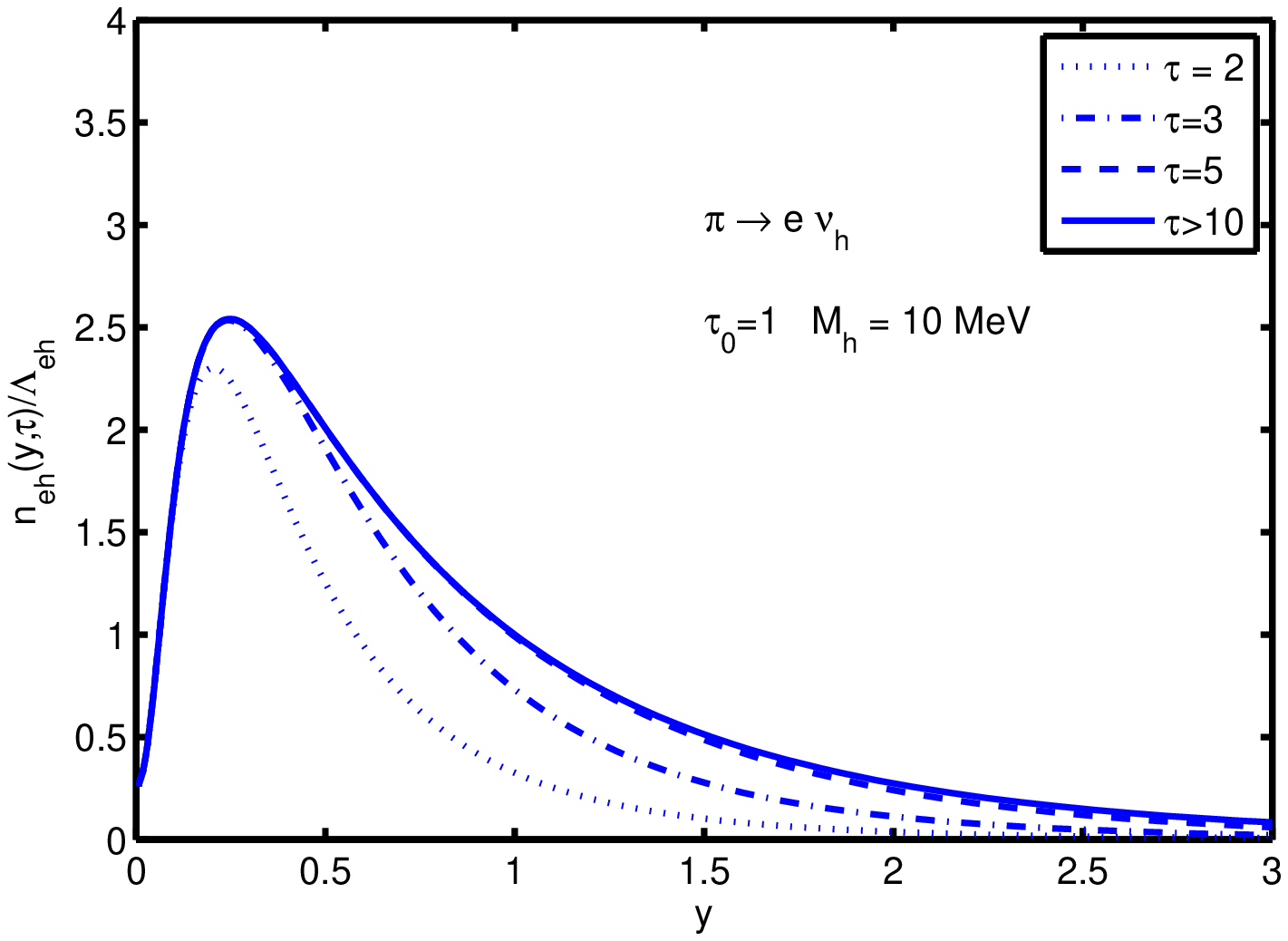}
\includegraphics[height=3.5in,width=3.2in,keepaspectratio=true]{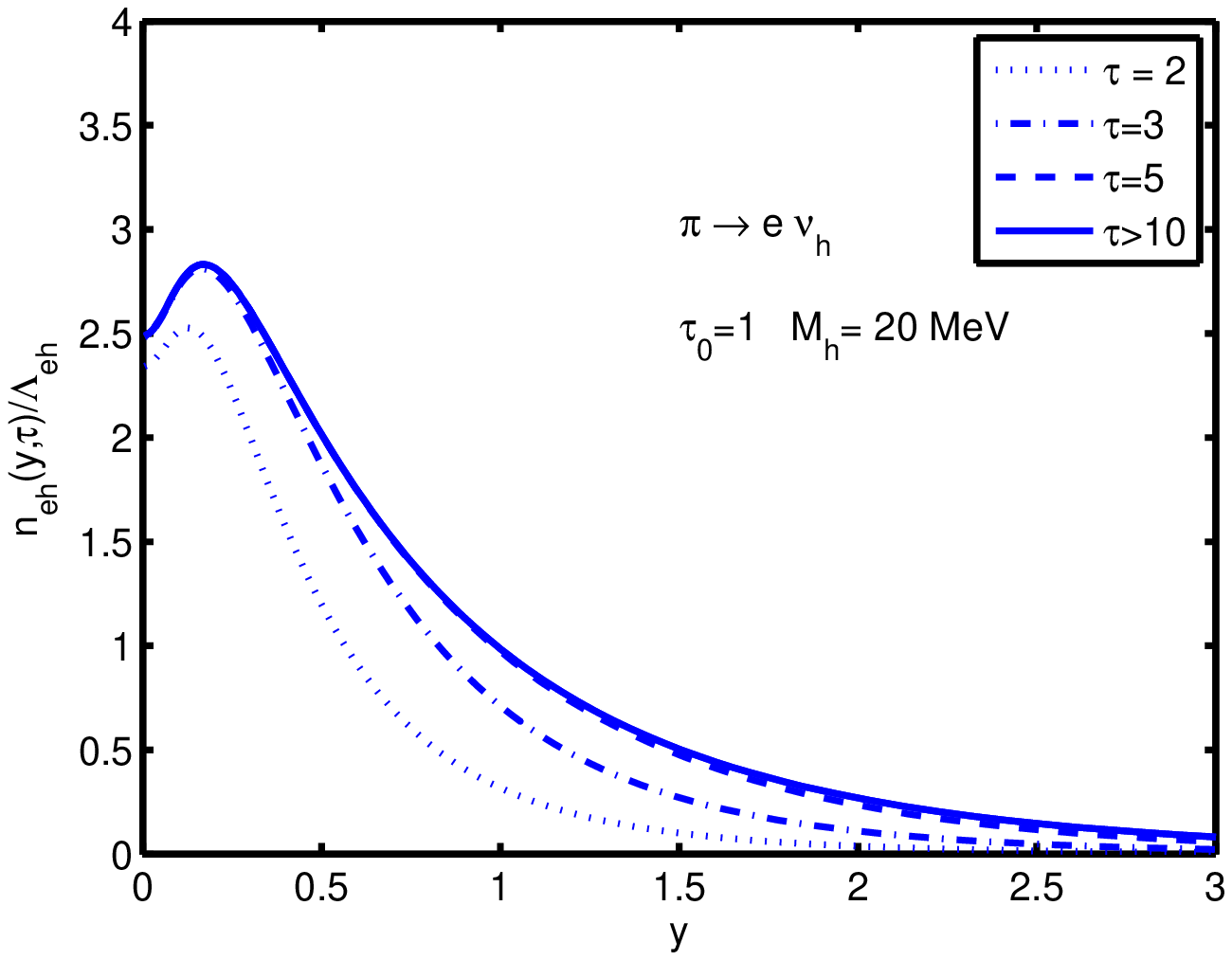}
\includegraphics[height=3.5in,width=3.2in,keepaspectratio=true]{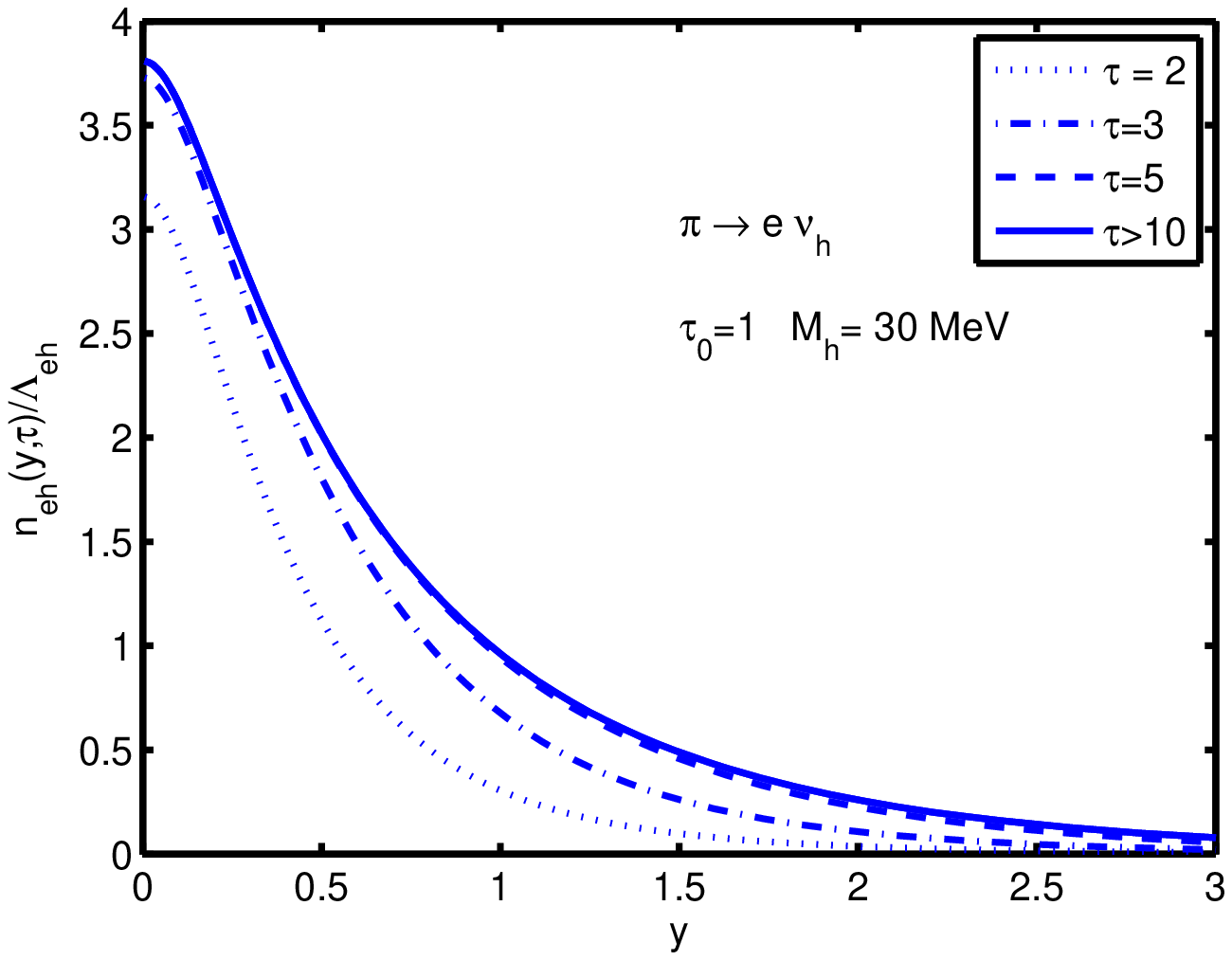}
\caption{Distribution function    $n_{eh}(y,\tau)$   from   $\pi \rightarrow e \nu_h$ with $M_h = 1,10,20,30 MeV$ and vanishing chemical potentials. The solid line is the asymptotic frozen distribution. }
\label{fig:lightedist}
\end{center}
\end{figure}

\begin{figure}[h!]
\begin{center}
\includegraphics[height=3.5in,width=3.2in,keepaspectratio=true]{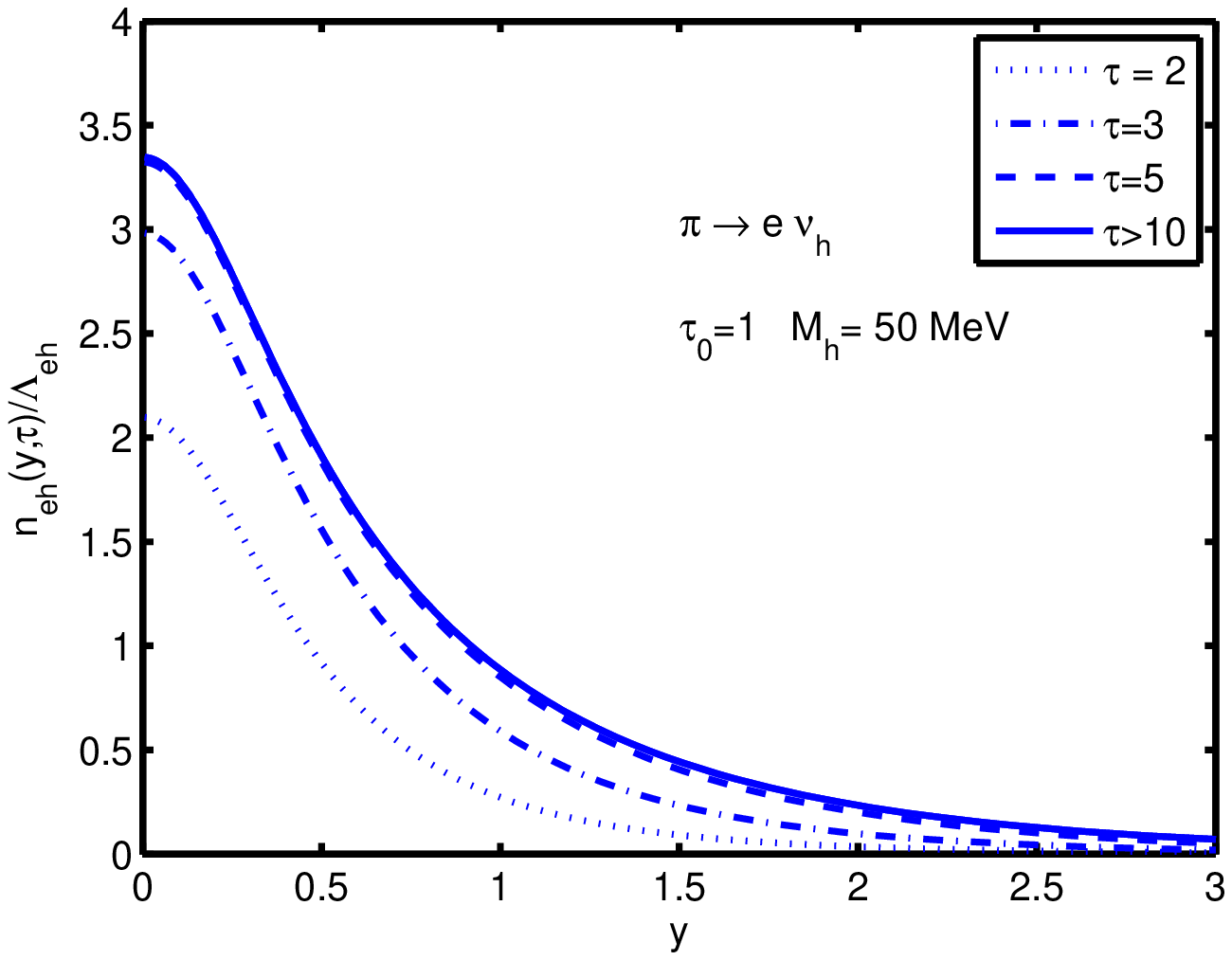}
\includegraphics[height=3.5in,width=3.2in,keepaspectratio=true]{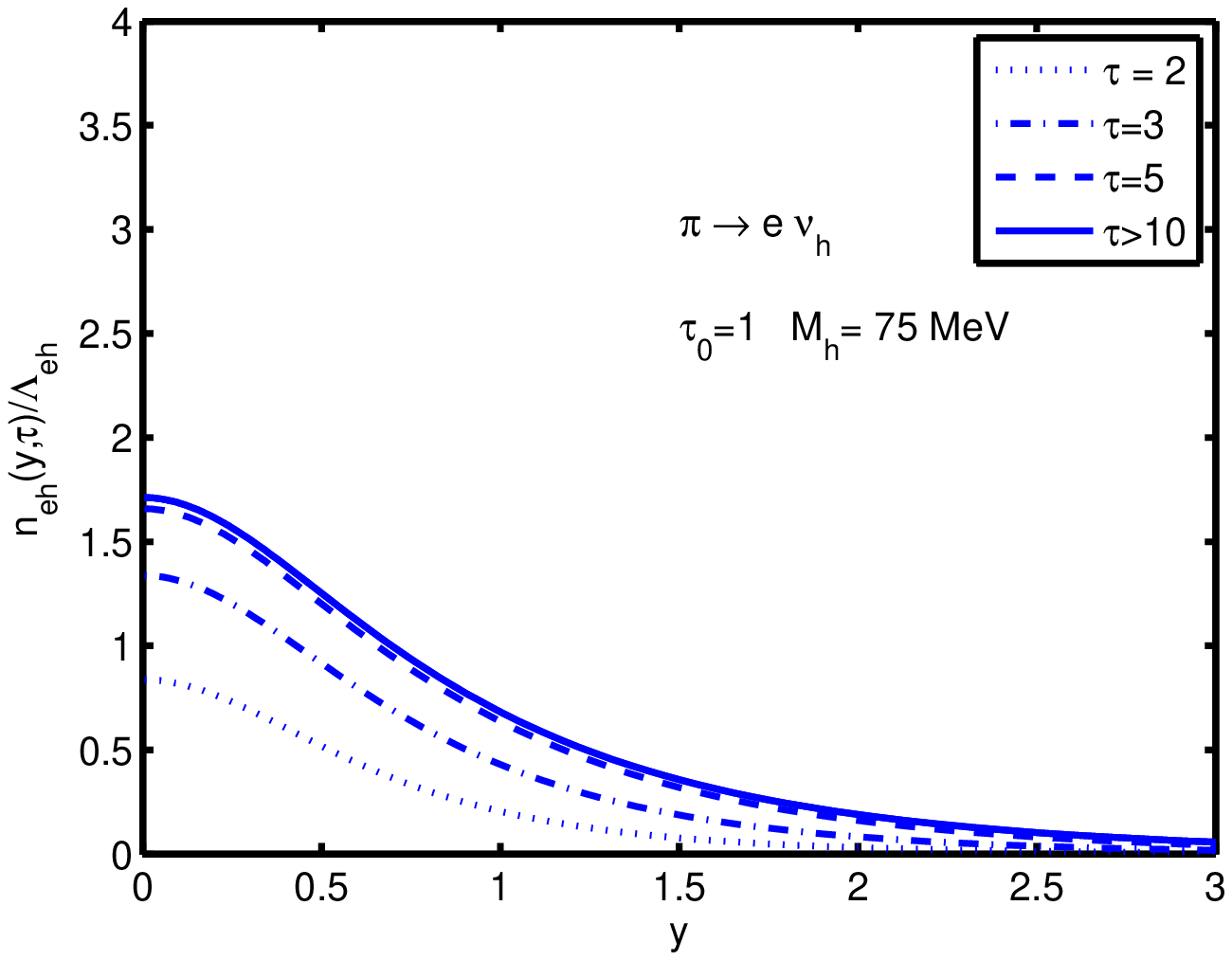}
\includegraphics[height=3.5in,width=3.2in,keepaspectratio=true]{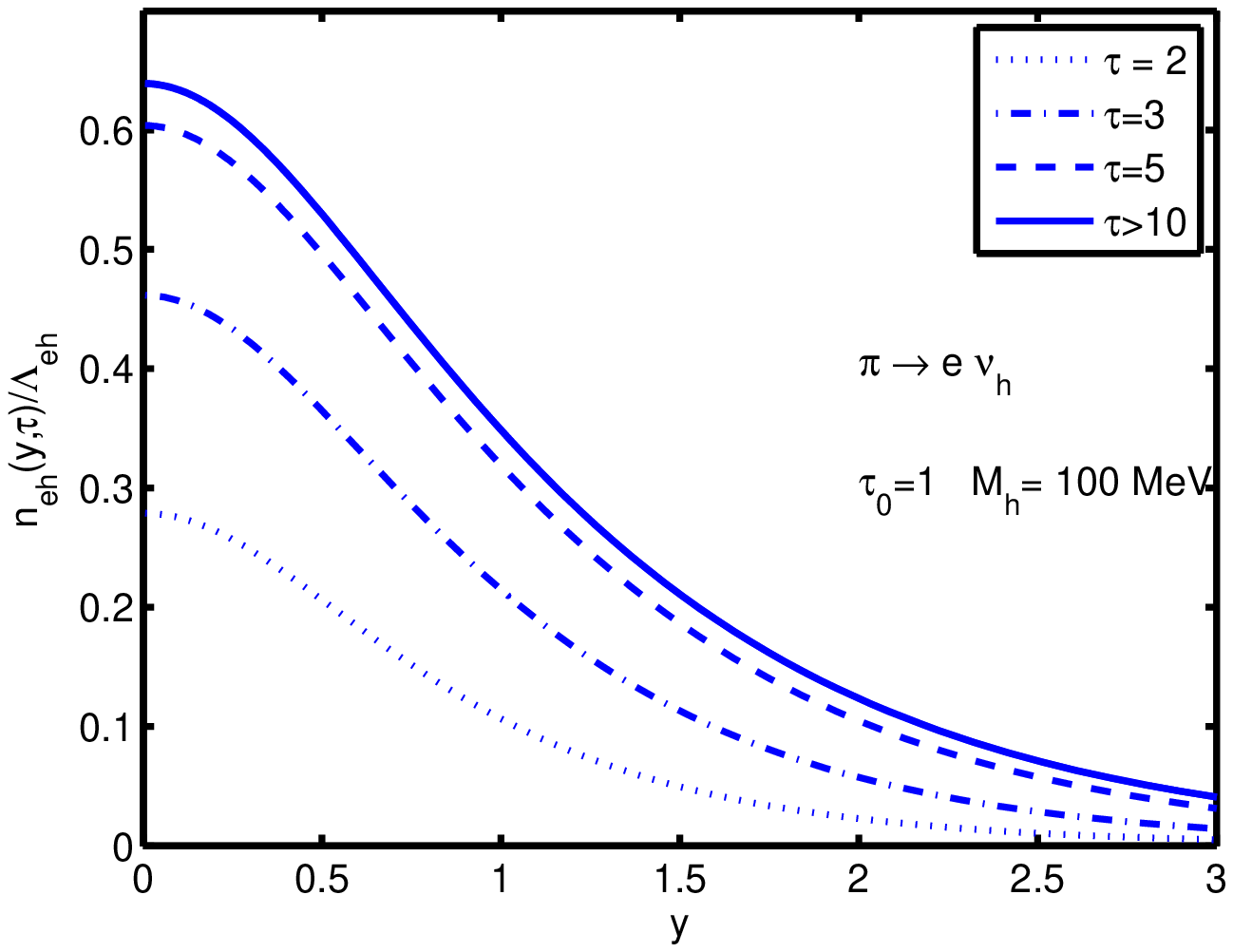}
\includegraphics[height=3.5in,width=3.2in,keepaspectratio=true]{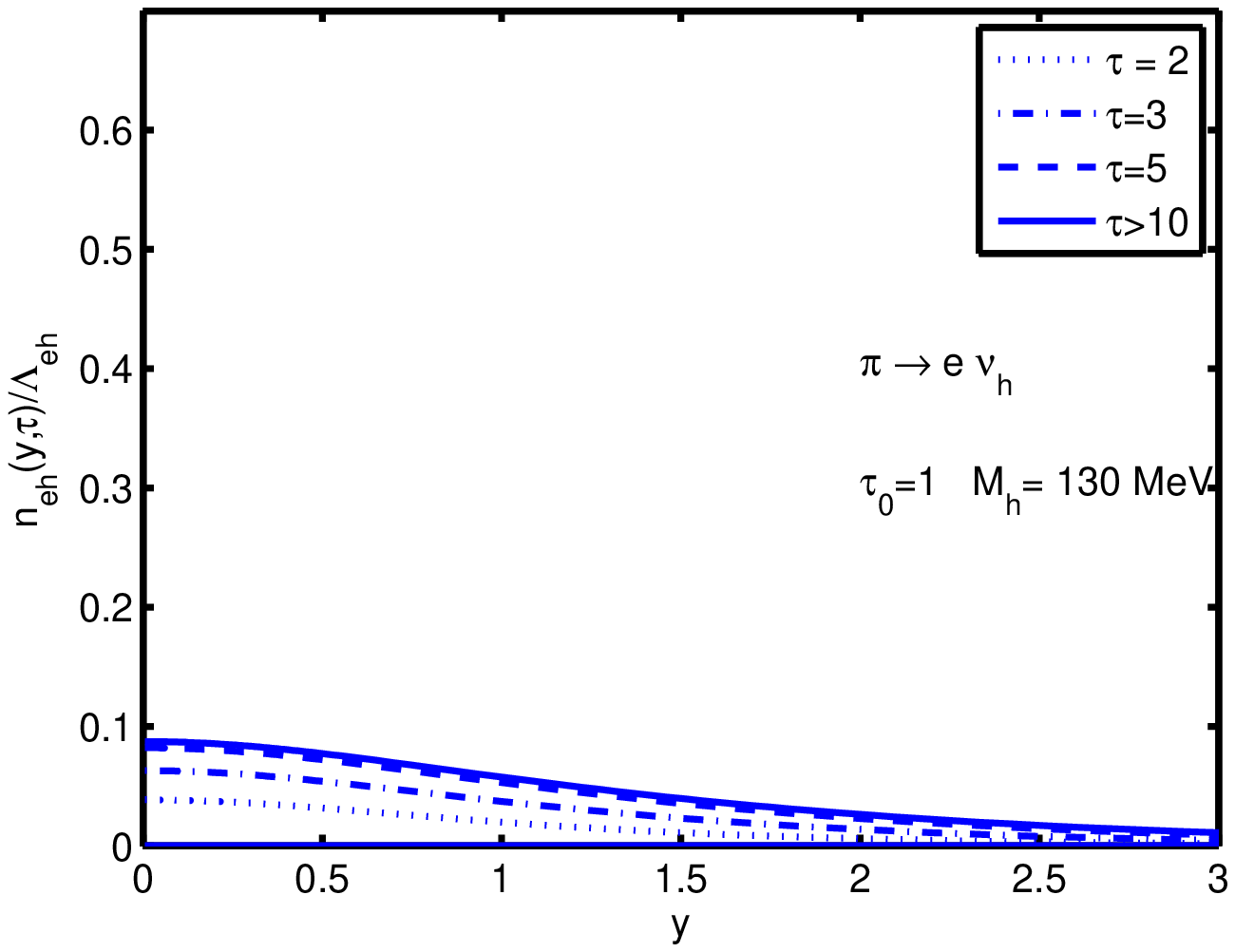}
\caption{Distribution function    $n_{eh}(y,\tau)$   from   $\pi \rightarrow e \nu_h$ with $M_{h} = 50,75,100,130 MeV$ and vanishing chemical potentials. The solid line is the asymptotic frozen distribution.}
\label{fig:heavyedist}
\end{center}
\end{figure}

\begin{figure}[h!]
\begin{center}
\includegraphics[height=3.5in,width=3.2in,keepaspectratio=true]{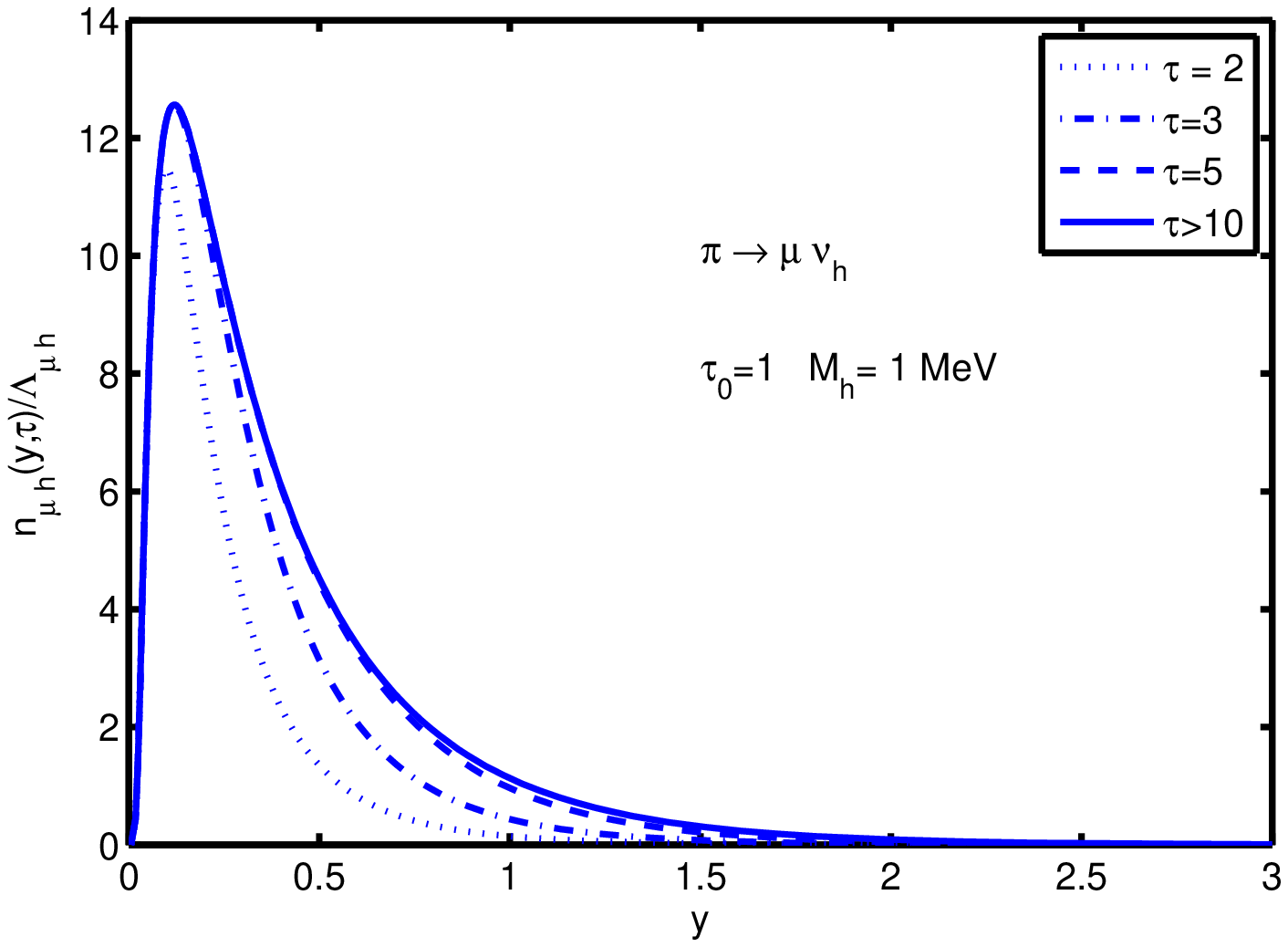}
\includegraphics[height=3.5in,width=3.2in,keepaspectratio=true]{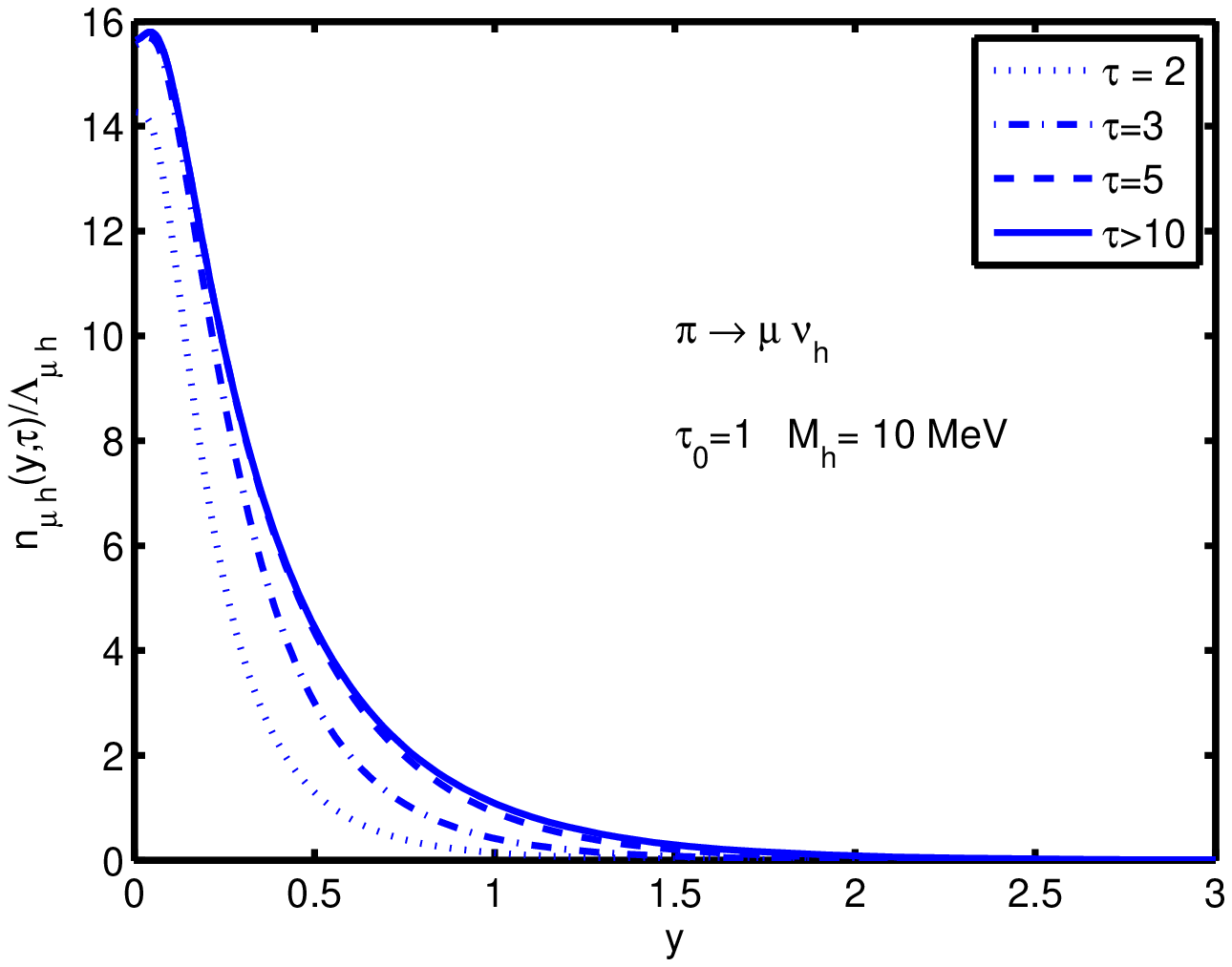}
\includegraphics[height=3.5in,width=3.2in,keepaspectratio=true]{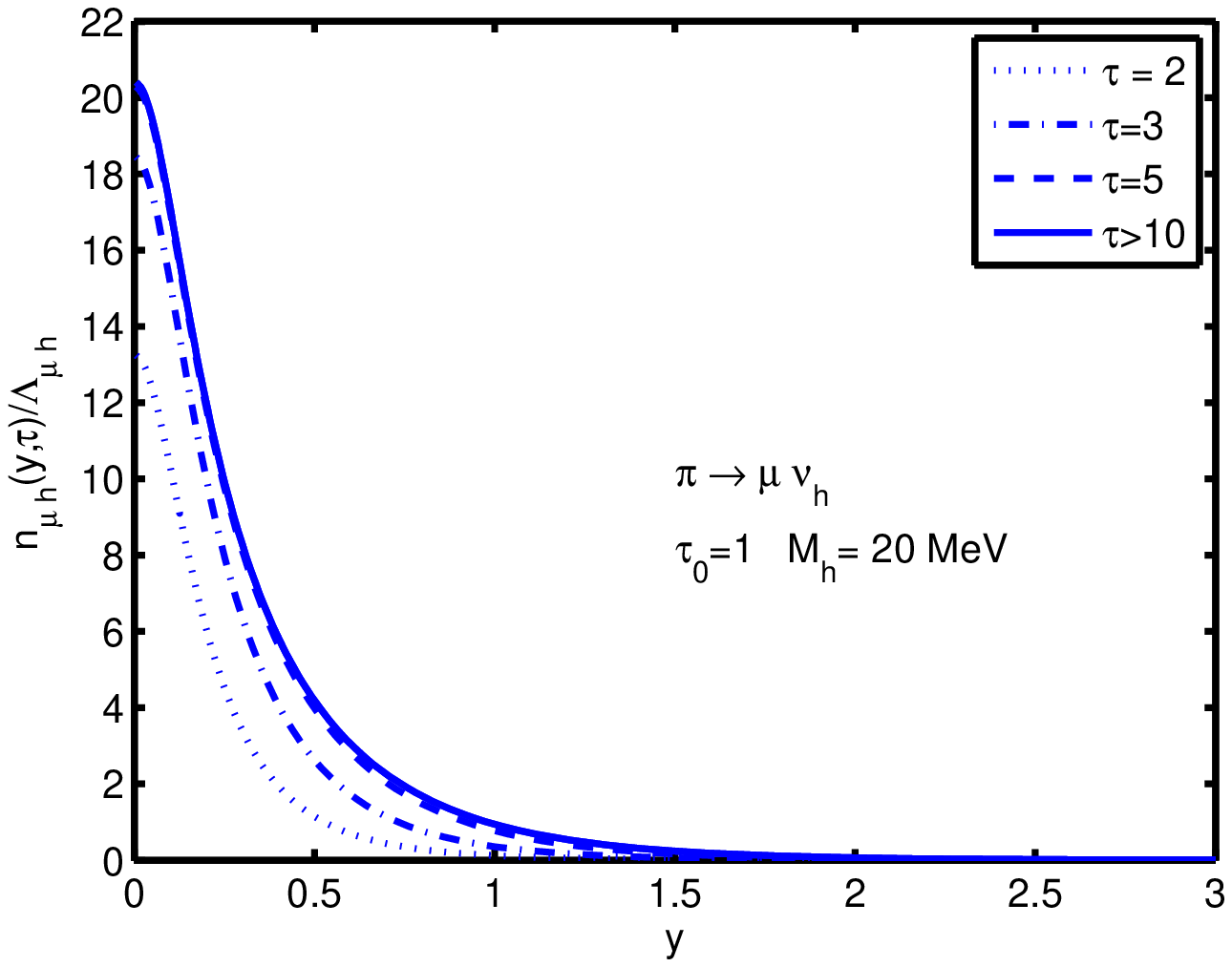}
\includegraphics[height=3.5in,width=3.2in,keepaspectratio=true]{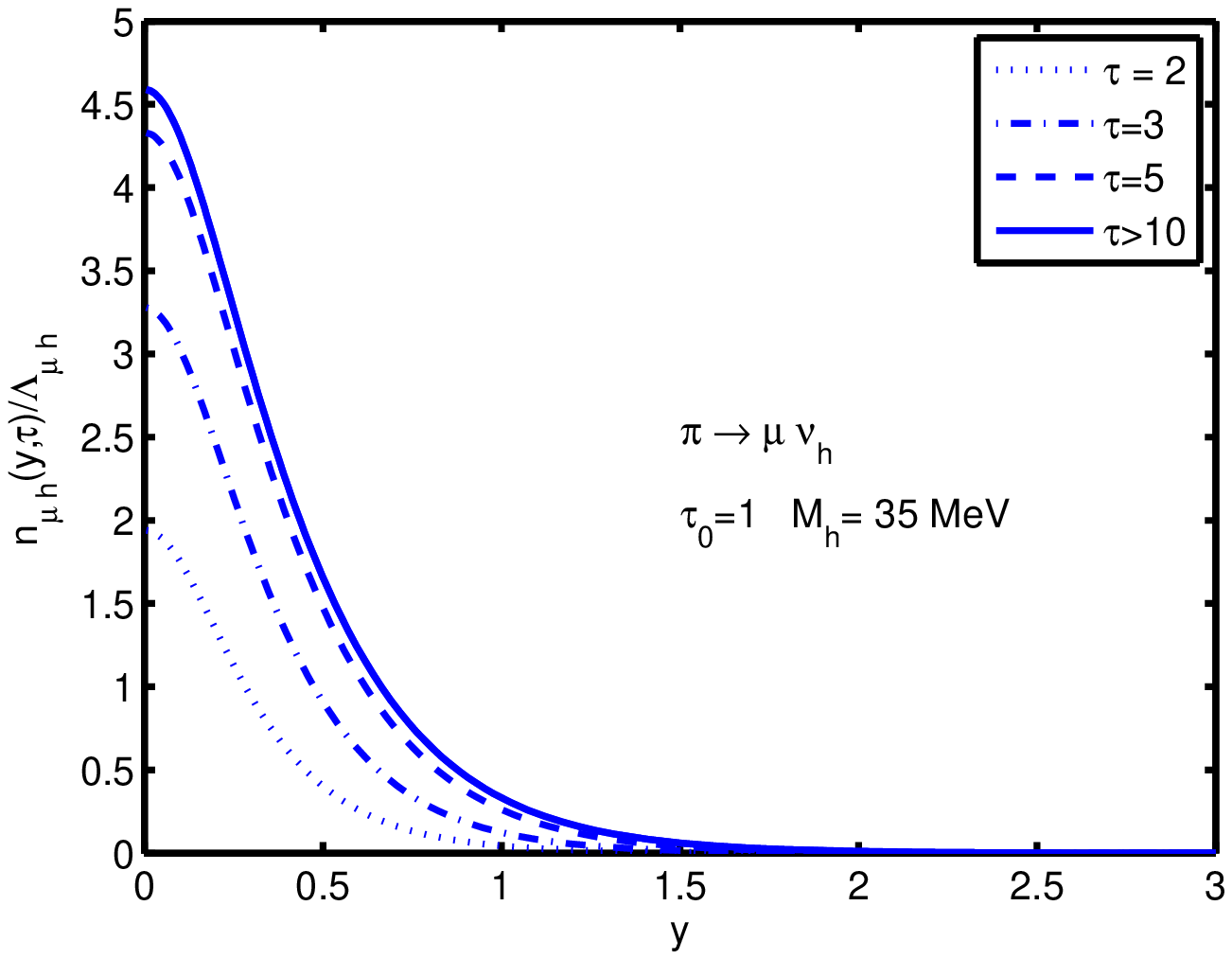}
\caption{Distribution function  $n_{\mu h}(y,\tau)$    from $\pi \rightarrow \mu \nu_h$ with $M_{h} = 1,10,20,35 MeV$ and vanishing chemical potentials. The solid line is the asymptotic frozen distribution.}
\label{fig:mudist}
\end{center}
\end{figure}

There are several features to observe from these plots,  the first of which is that, for low mass neutrinos ($M_h\lesssim 1 MeV$), the distributions observed in \cite{lellolightsterile} are recovered. For heavier   neutrinos ($M_h \gtrsim 30 MeV$) a very different behavior from the light species is observed. The light species features a vanishing distribution for small momentum ($y \rightarrow 0$), peaks at a particular momentum and falls off at large momentum whereas the heavier species has     non-vanishing support at zero momentum  ($y=0$) and monotonically decreases as momentum increases. An intermediate behavior can be observed in the electron channel for a heavy neutrino with $M_h = 20 MeV$; the crossover in behavior occurs for $M_h \simeq T(\tau_{fr}) \simeq 10-14\,\mathrm{MeV}$ for $\tau_{fr} \simeq 10$ is the time scale at which the distribution functions freeze out, the freeze out time $\tau \simeq 10$ is  nearly independently of the value of $M_h$.

Furthermore, comparison of figs. (\ref{fig:lightedist}) with those of figs. (\ref{fig:mudist}) is revealing. These two sets display the distributions from the electron and muon channels within the kinematically allowed window available for \emph{both} channels, $0< M_h \leq 36 \, \mathrm{MeV}$, and it is in this comparison that the relevance of the distribution functions $n_{lh}$ for $ l=\mu,e$ become manifest. Two important aspects stand out: a) for small $y$ the distribution function from the $\mu$ channel is systematically \emph{larger} than that for  the $e$ channel, this is a consequence of the kinematic factors $\Delta_{lh},\delta_{lh}$ in (\ref{epipm}),   both are much smaller in the $\mu$ channel than in the $e$ channel, this fact makes the energies $E_\pi$ in (\ref{epipm}) \emph{smaller} in the $\mu$ channel, therefore yield a \emph{larger} contribution since these are less thermally suppressed. This confirms the analysis presented above.  b) The distribution function for the e-channel has larger support for $y > 1$ than that for the $\mu$ channel, which is larger for $y<1$ but us strongly suppressed for $y>1$ as compared to that of the e-channel. Both these features lead to the conclusion that the distribution function from the $\mu$ channel yields a \emph{colder} component than that of the $e$ channel in the sense that the velocity dispersion obtained from $n_{\mu h}$ is smaller than that obtained from $n_{eh}$, this is a manifestation of the kinematic entanglement and its consequences will be analyzed in detail below.

The total distribution function is thus a \emph{mixture} of a colder and a warmer component.    Defining the distribution function at freeze-out for each channel as
\be f_{lh}(y) = n_{lh}(y,\infty) \label{frizf} \ee where by $\tau =\infty$ we mean $\tau > \tau_{fr} \simeq 10 $, the \emph{total} distribution function is given by
\be f_h(y) = \Lambda_{\mu h} \Bigg[ \frac{f_{\mu h}(y)}{\Lambda_{\mu h}}\Bigg] \Theta(36\,\mathrm{MeV}-M_h)+ \Lambda_{e h} \Bigg[ \frac{f_{e h}(y)}{\Lambda_{e h}}\Bigg] \Theta(141\,\mathrm{MeV}-M_h) \,.\label{totfrizf} \ee The $\Theta$ functions describe the thresholds in each channel, the brackets   $\Big[f_{lh}/\Lambda_{lh}\Big]$  are actually the result of the numerical integrations as per eqn. (\ref{exactrate}) and are given by the solid lines ($\tau > 10$) in figs. (\ref{fig:lightedist}-\ref{fig:mudist}).

 For a given range of masses $M_h$ the only unknowns are the mixing matrix elements $H_{lh}$. Because the total distribution function is a result of the combination of several production channels, each one with possibly a different mixing matrix element $H_{lh}$, it is convenient to factor out one of these and rewrite the total distribution function  (\ref{totfrizf}) in terms of ratios, namely
\bea f_h(y) & \equiv &   0.13\,\Big( \frac{|H_{\mu h}|^2}{10^{-5}}\Big) \, \widetilde{f}_h(y) \label{tilf} \\ \widetilde{f}_h(y) & = &  C[m_\mu,m_h] \Bigg[ \frac{f_{\mu h}(y)}{\Lambda_{\mu h}}\Bigg] \Theta(36\,\mathrm{MeV}-M_h)  \nonumber \\ & + &    \Bigg( \frac{|H_{e h}|^2}{|H_{\mu h}|^2} \Bigg)\,C[m_e,m_h]\Bigg[ \frac{f_{e h}(y)}{\Lambda_{e h}}\Bigg] \Theta(141\,\mathrm{MeV}-M_h)  \,.\label{totfrizf2} \eea where the coefficients $C[m_l,m_h]$ are given by (\ref{chiral}) and the brackets $[f/\Lambda]$ are the results obtained numerically and displayed   by the solid lines in figs. (\ref{fig:lightedist}-\ref{fig:mudist}).   In this manner the various constraints can be phrased in terms of an overall mixing matrix element and ratios, such as  $  |H_{e h}|^2/|H_{\mu h}|^2$ which along with the kinematic factors $C[m_l,m_h]$ determine the \emph{concentration} of the warmer species (from the electron channel) in the \emph{mixture} of colder and warmer distributions from both channels.

\subsection{ A Tale of Two Distributions:}

Closer examination of figs. \ref{fig:lightedist},\ref{fig:heavyedist},\ref{fig:mudist} reveal a transition in the shapes of distributions as a function of neutrino mass. In the case of relatively light neutrinos with $M_h \lesssim 1-10 MeV$, the distribution function vanishes at small momentum, reaches a maximum, and falls off with a long tail at high momentum. The situation for heavier neutrinos, $M_h \gtrsim 30 MeV$, produces a distribution function which features a non-vanishing plateau at low momentum which steadily falls off with increasing momentum. This latter distribution obviously produces a \emph{colder} dark matter candidate as the distribution function has considerably more support at small momentum. The transition between these two types of distributions is controlled by increasing neutrino mass and an intermediate distribution can be seen in the intermediate mass range, $ 10 MeV \lesssim M_h \lesssim 30 MeV$, illustrated by the distribution function in figure \ref{fig:lightedist} for a neutrino mass $M_h = 20 MeV$ in the electron channel. For this $M_h = 20 MeV$ case, the distribution resembles a superposition of the two distinct distributions seen for the lighter ($M_h \lesssim 10 MeV$) and heavier species ($M_h \gtrsim 30 MeV$).

Unfortunately, only limited analytic progress can be made towards an understanding of the full distribution function but this proves enough to shed some light on what governs the transition between the different distribution shapes. In reference \cite{lellolightsterile}, it is shown that under appropriate approximations, the distribution function for light mass neutrinos takes the form

\be
n_{lh}(\tau_0,y) \Big|_{M_h \lesssim 1 MeV}=\frac{\Lambda_{lh}}{y^2} \sum_{k=1}^{\infty} \left[ \frac{1 + (-1)^{k+1} e^{ky}}{1 + e^{y}} \right] \frac{\exp\left(-k y/\Delta_{lh}\right)}{k} \times J_k(\tau_0,y)
\ee where

\be
J_k(\tau_0,y) = 2 \tau_0 \left(\frac{ y}{k \Delta_{lh}}\right)  \exp\left(-\frac{k \Delta_{lh} \tau_0^2}{4 y}\right) + \left( \frac{ y}{k \Delta_{lh}} \right)^{1/2} \left[\frac{2 y}{k \Delta_{lh}} - \frac{M_{\pi}^2}{6 f_{\pi}^2}\right]\Gamma\left(1/2,\frac{k \Delta_{lh} \tau_0^2}{4 y}\right) \,.
\ee where $\Gamma(\nu,z)$ is the incomplete gamma function and the details of this calculation are reproduced in appendix \ref{app:approximation}. As detailed in appendix \ref{app:approximation}, the main approximation employed towards producing this semi-analytic expression is $M_{h}/M_{\pi} \ll 1$. This approximation is equivalent to the assumption that the neutrino is produced \emph{ultra-relativistically} ($E_{h}/T \sim y$) and this proves to be an excellent approximation for lighter species ($M_h \lesssim 1 MeV$).

\begin{figure}[h!]
\begin{center}
\includegraphics[height=3.5in,width=3.2in,keepaspectratio=true]{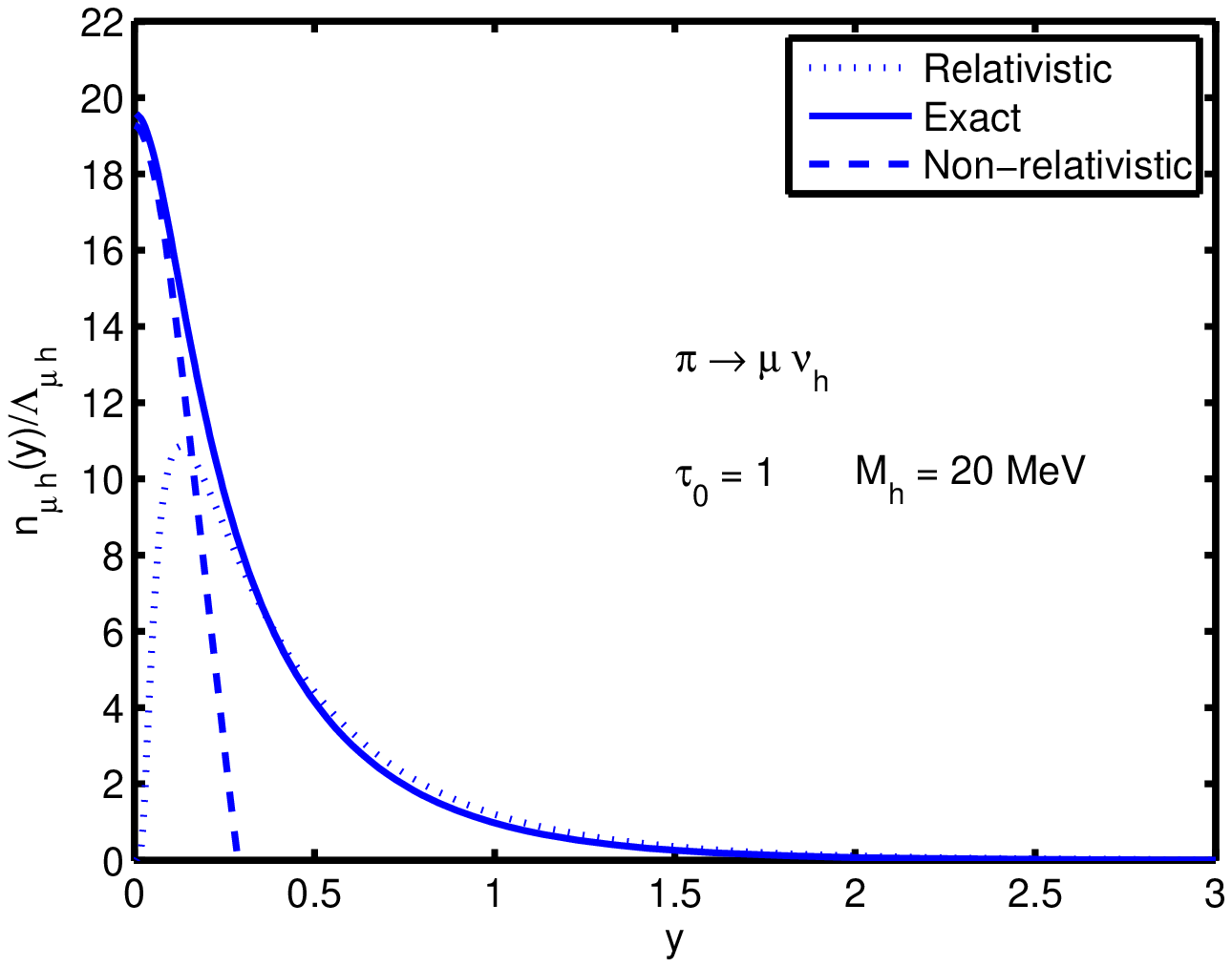}
\includegraphics[height=3.5in,width=3.2in,keepaspectratio=true]{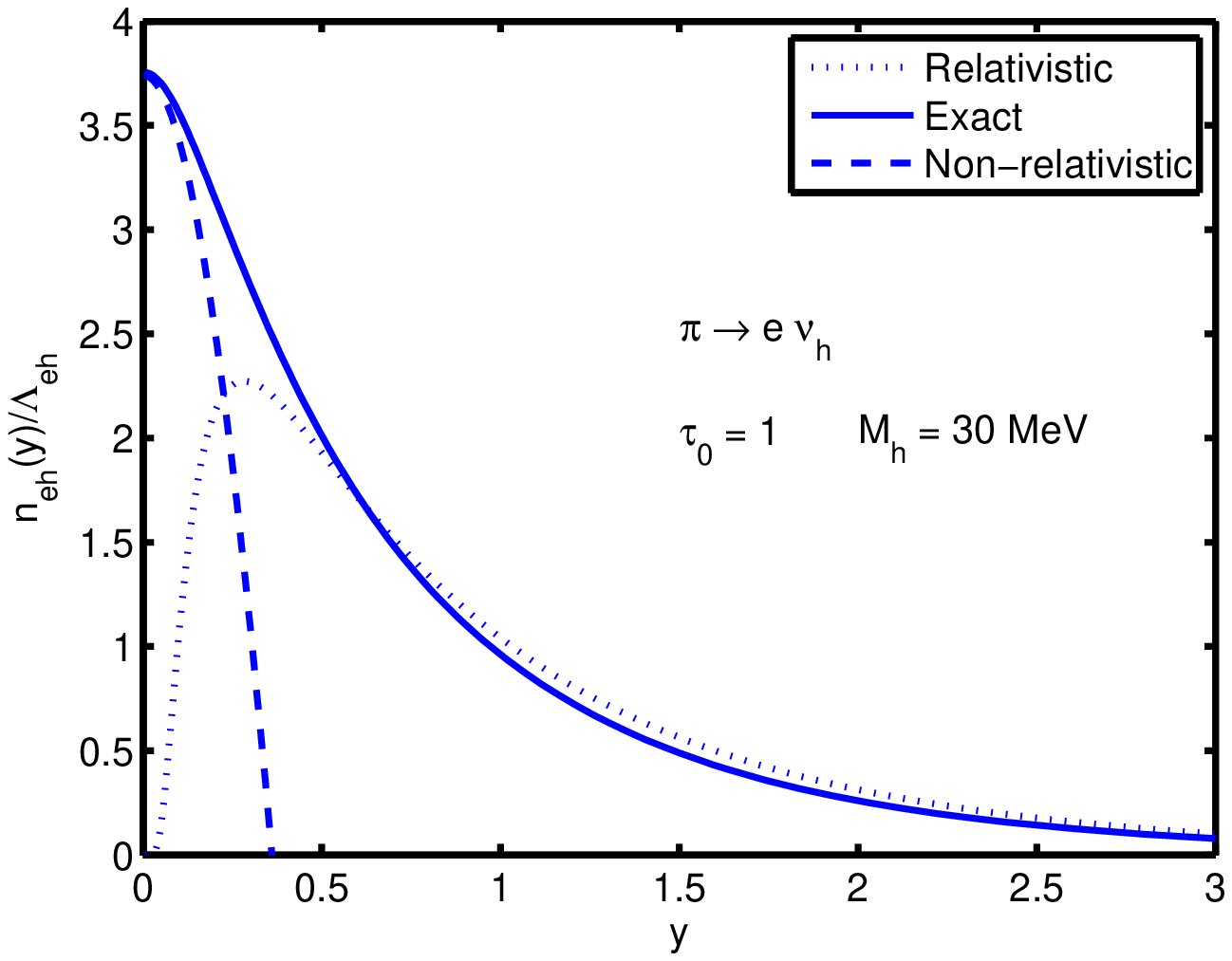}
\caption{Comparison of the exact distribution with a distribution producing ultra-relativistic and non-relativistic heavy neutrinos.}
\label{fig:relnonrelprod}
\end{center}
\end{figure}

The opposite type of approximation, where the neutrino is produced \emph{non-relativistically} ($E_{h}/T \sim M/T + \frac{T}{2M} y^2$), can be made but leads to a very unwieldy expression which is given by Eqs \ref{analytic},\ref{nonrelapp}. The production of non-relativistic heavy neutrinos leads to a natural source of non-trivial behavior at small momentum which explains the low momentum plateau. These arguments lend themselves to the interpretation that neutrinos with $M_h \lesssim 10 MeV$ are solely produced ultra-relativistically while some fraction are produced non-relativistically as the neutrino mass increases. This argument is illustrated in fig. \ref{fig:relnonrelprod} where the distribution functions of \ref{relapp},\ref{nonrelapp}, which are results of ultra-relativistic and non-relativistic production approximations, are plotted against the full distribution. In fig. \ref{fig:relnonrelprod}, the low momentum region is dominated by the non-relativistic result whereas the large momentum region fits quite well to the ultra-relativistic result. Both the ultra-relativistic and non-relativistic results fail outside the appropriate momentum regions: the non-relativistic approximation fails for higher momentum whereas the ultra-relativistic approximations fails at decreasing momentum.

This analysis explains the origin of the features of the distribution functions obtained numerically in figures \ref{fig:lightedist}, \ref{fig:heavyedist}, \ref{fig:mudist}, namely that the ultra-relativistic approximation serves as a good estimate for the whole distribution as the neutrino mass becomes negligible compared to the pion mass while the non-relativistic approximation serves to understand the appearance of the plateau at low momentum in the distribution. This means that for light mass neutrinos ($M_{h} \ll M_{\pi}$) all of the neutrinos that are produced are done so \emph{ultra-relativistically} while the production of a more massive species leads to some fraction of neutrinos produced \emph{non-relativistically} as well. The fraction of neutrinos which are produced relativistically increases as sterile neutrino mass decreases while the fraction which are produced non-relativistically increases as the sterile neutrino mass increases.

\subsection{Non-thermality}

The equation of state parameter, $w(T)$, is given by eq \ref{eos} where $w=1/3,0$ correspond to ultra-relativistic and non-relativistic species respectively. Depending on the temperature of production and decoupling, a heavy neutrino could be ultra-relativistic, non-relativistic ($M \ll T$ , $M \gg T$) or somewhere in between as freeze-out occurs and $w(T)$  determines when a particular species becomes non-relativistic.

The correct equation of state (\ref{eos}) and velocity dispersion (\ref{veldisp}) must be obtained from the total distribution function (\ref{totfrizf}), because these  are \emph{moments} of the distribution function  they are  not simply the addition of the two components.

However, it proves illuminating to \emph{define} an equation of state for each channel

\be \label{eoschan}
w_{l}(T) = \frac{\mathcal{P}}{\rho} = \frac{1}{3} \frac{ \int dy \, \frac{y^4}{ \sqrt{y^2+\frac{M_h^2}{T^2}}} \,f_{lh}(q_c)}{\int dy \, y^2 \sqrt{y^2+\frac{M_h^2}{T^2}}\, f_{lh}(q_c)} \,,
\ee these serve as \emph{proxies} to quantify the ``coldness'' of the species produced by the particular channel:  from the discussion in section (\ref{sec:cosmocons}),   $\sqrt{w(T)}$ is a generalization of the ``adiabatic speed of sound'' for collisionless DM, and in the non-relativistic limit $w(T) \rightarrow \langle \vec{V}^2 \rangle /3$.

The distribution function of a thermalized heavy neutrino  would be given by the standard Fermi-Dirac distribution:

\be
f_{LTE}(y;T ) = \frac{1}{e^{\sqrt{y^2+ M^2_h/T^2}} + 1} \,, \label{thernu}
\ee and we compare the equation of state $w_l(T)$ (\ref{eoschan}) with that obtained from (\ref{thernu}), this comparison  quantifies   the ``non-thermality'' of the distribution functions from each production channel.

\begin{figure}[h!]
\begin{center}
\includegraphics[height=3.5in,width=3.2in,keepaspectratio=true]{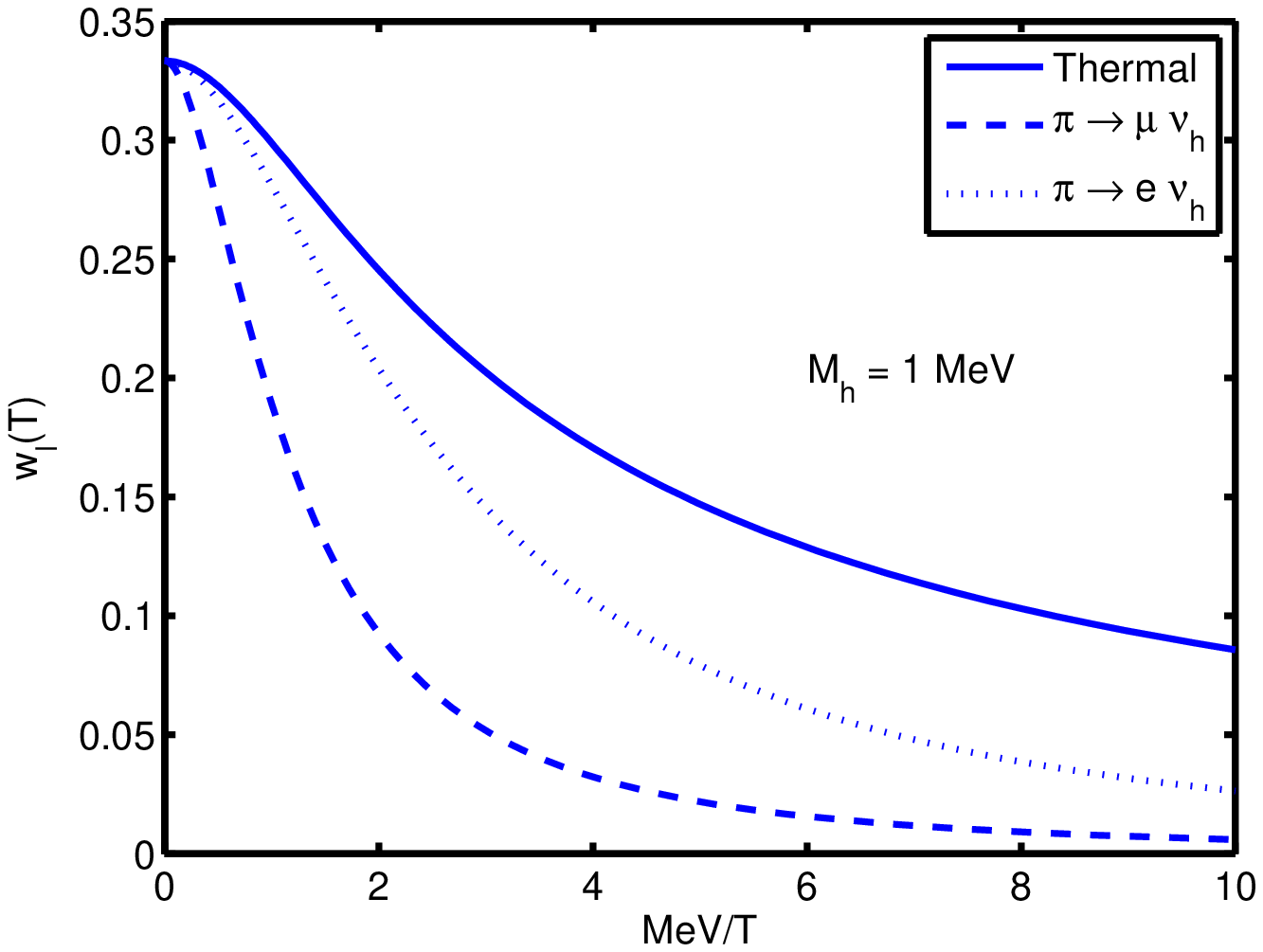}
\includegraphics[height=3.5in,width=3.2in,keepaspectratio=true]{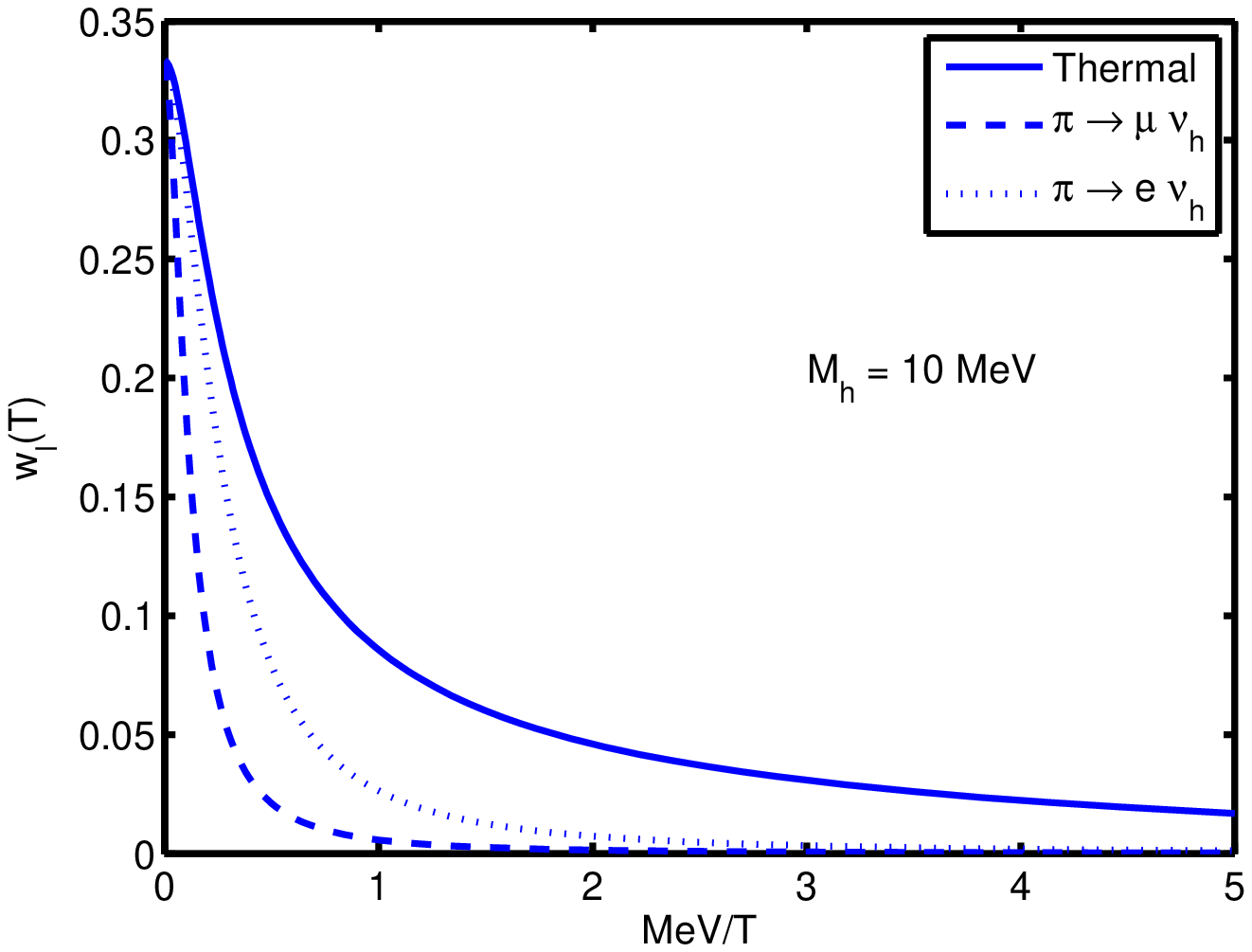}
\includegraphics[height=3.5in,width=3.2in,keepaspectratio=true]{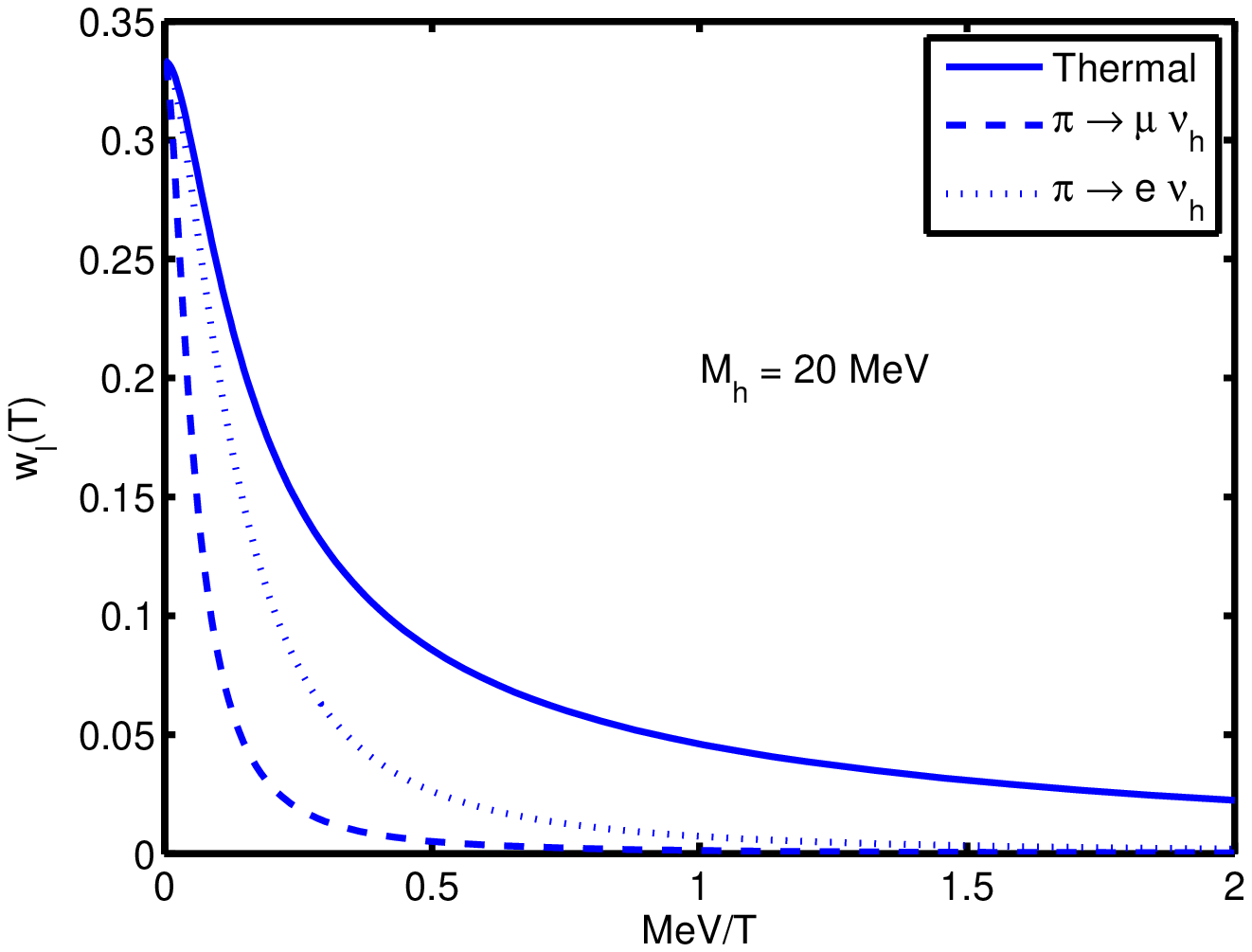}
\includegraphics[height=3.5in,width=3.2in,keepaspectratio=true]{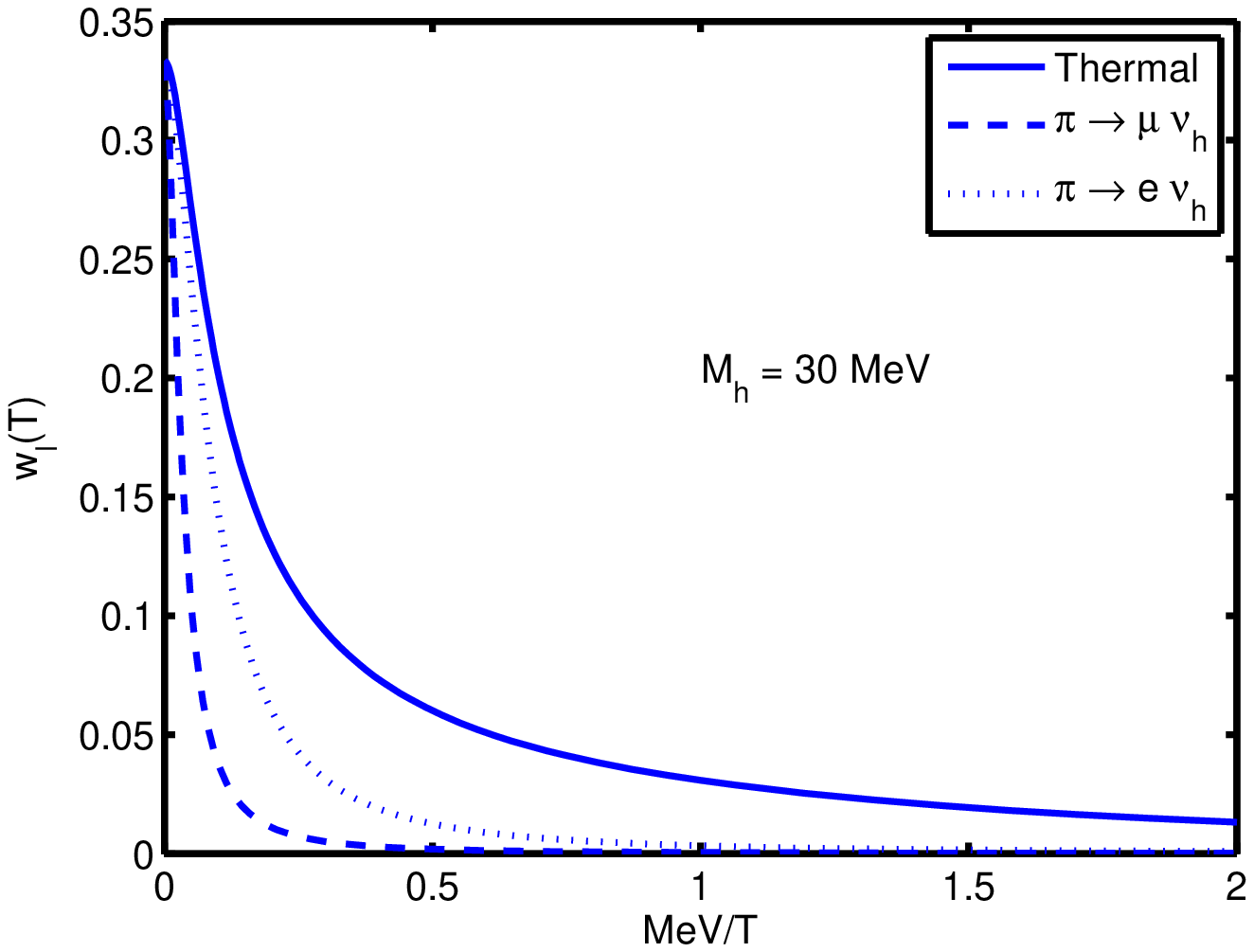}
\caption{$w_l(T)$ for both channels  $\pi \rightarrow \mu \nu_h~;~\pi \rightarrow e \nu_h$ compared to  thermal distribution for  $M_h = 1,10,20,30\, \mathrm{MeV}$ in the kinematic window in which both channels are available.}
\label{fig:lighteleos}
\end{center}
\end{figure}

\begin{figure}[h!]
\begin{center}
\includegraphics[height=3.5in,width=3.2in,keepaspectratio=true]{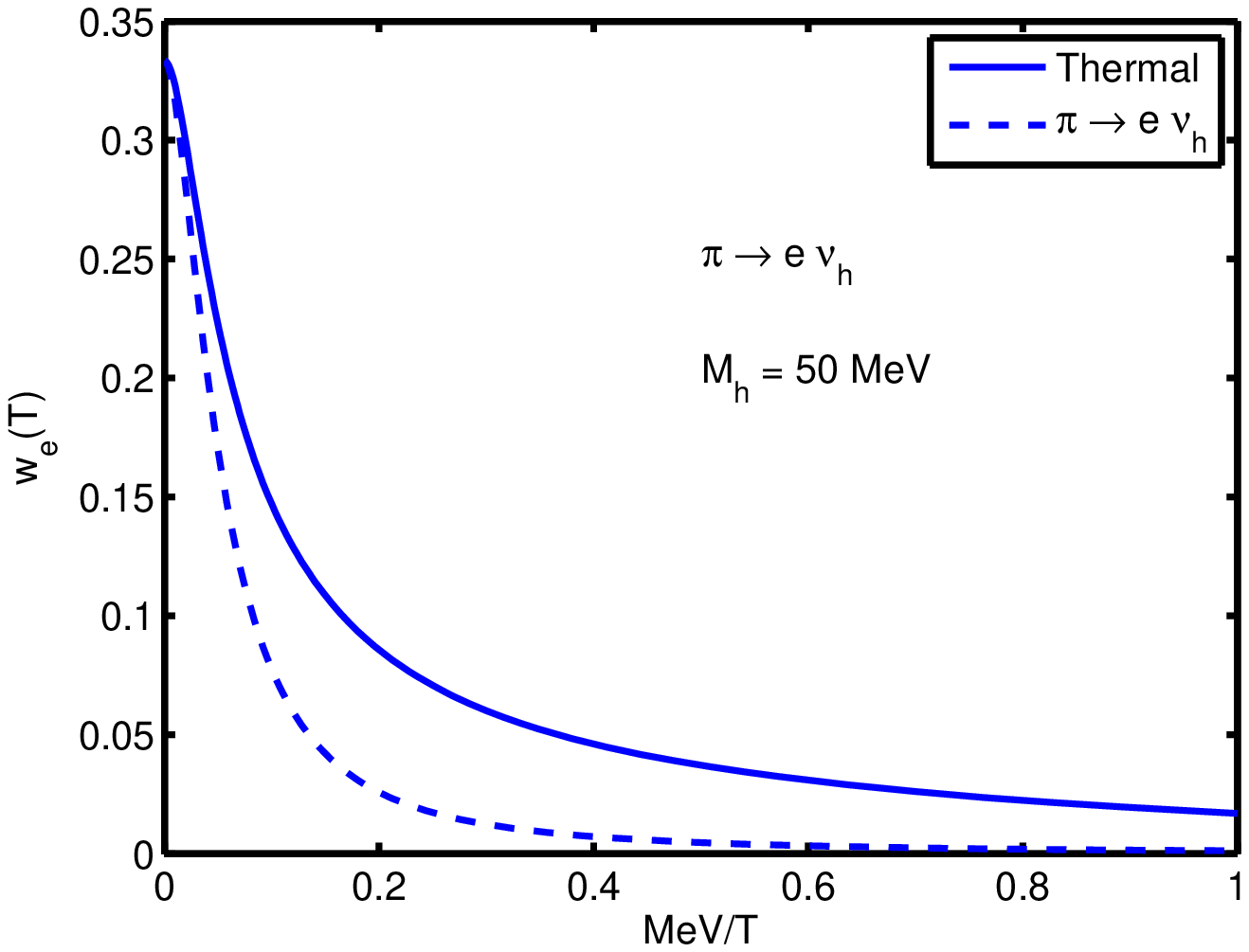}
\includegraphics[height=3.5in,width=3.2in,keepaspectratio=true]{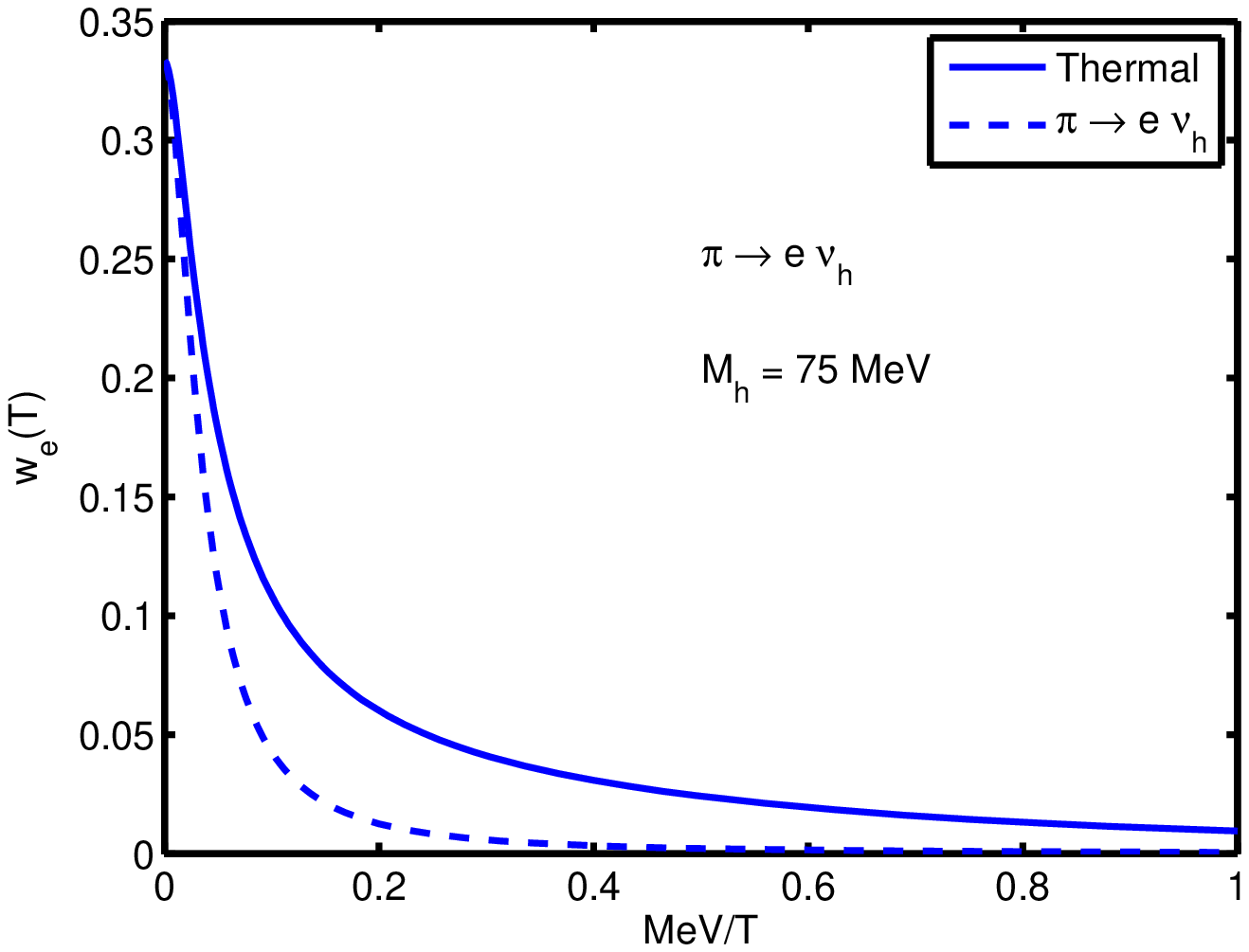}
\includegraphics[height=3.5in,width=3.2in,keepaspectratio=true]{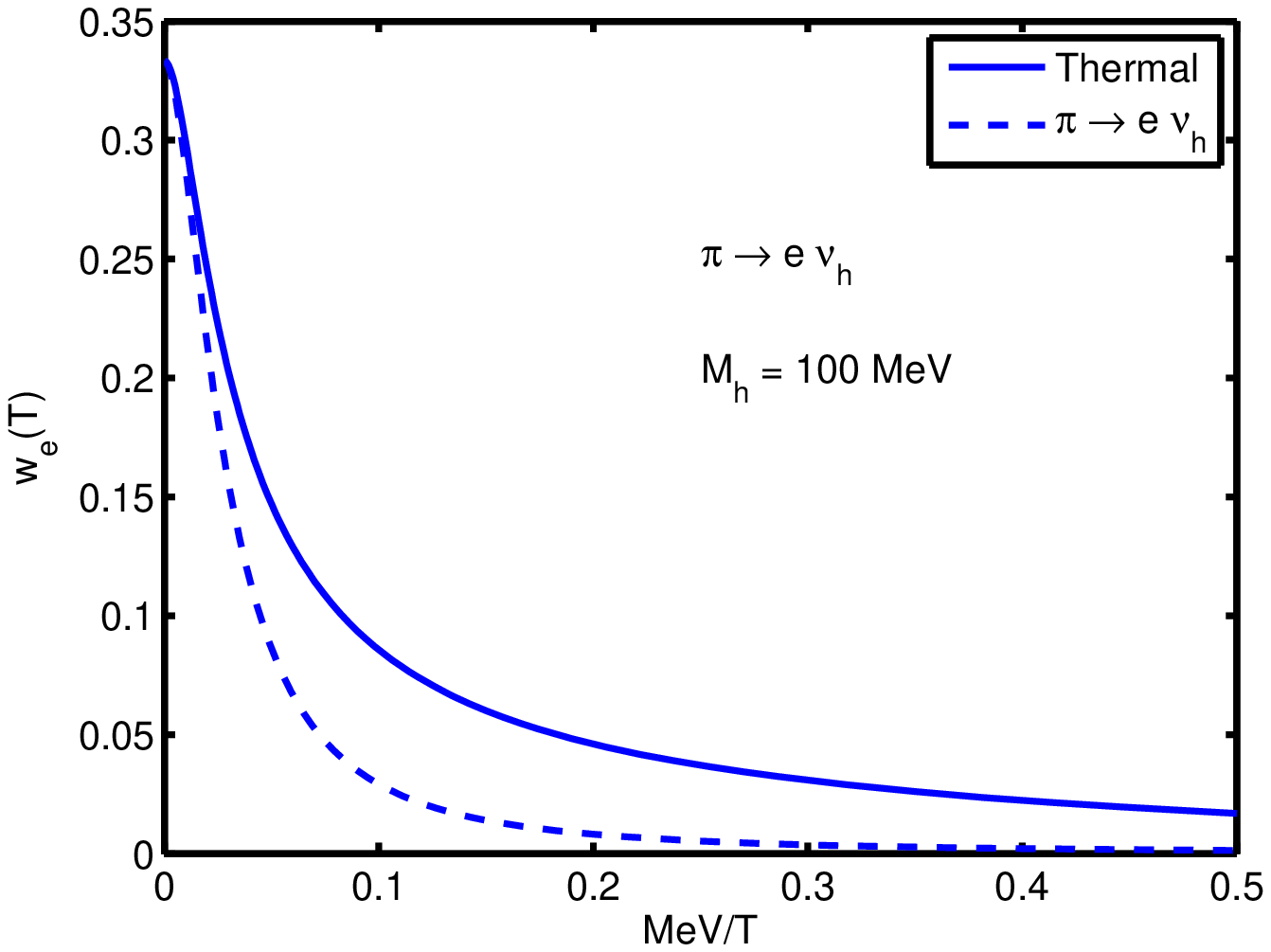}
\includegraphics[height=3.5in,width=3.2in,keepaspectratio=true]{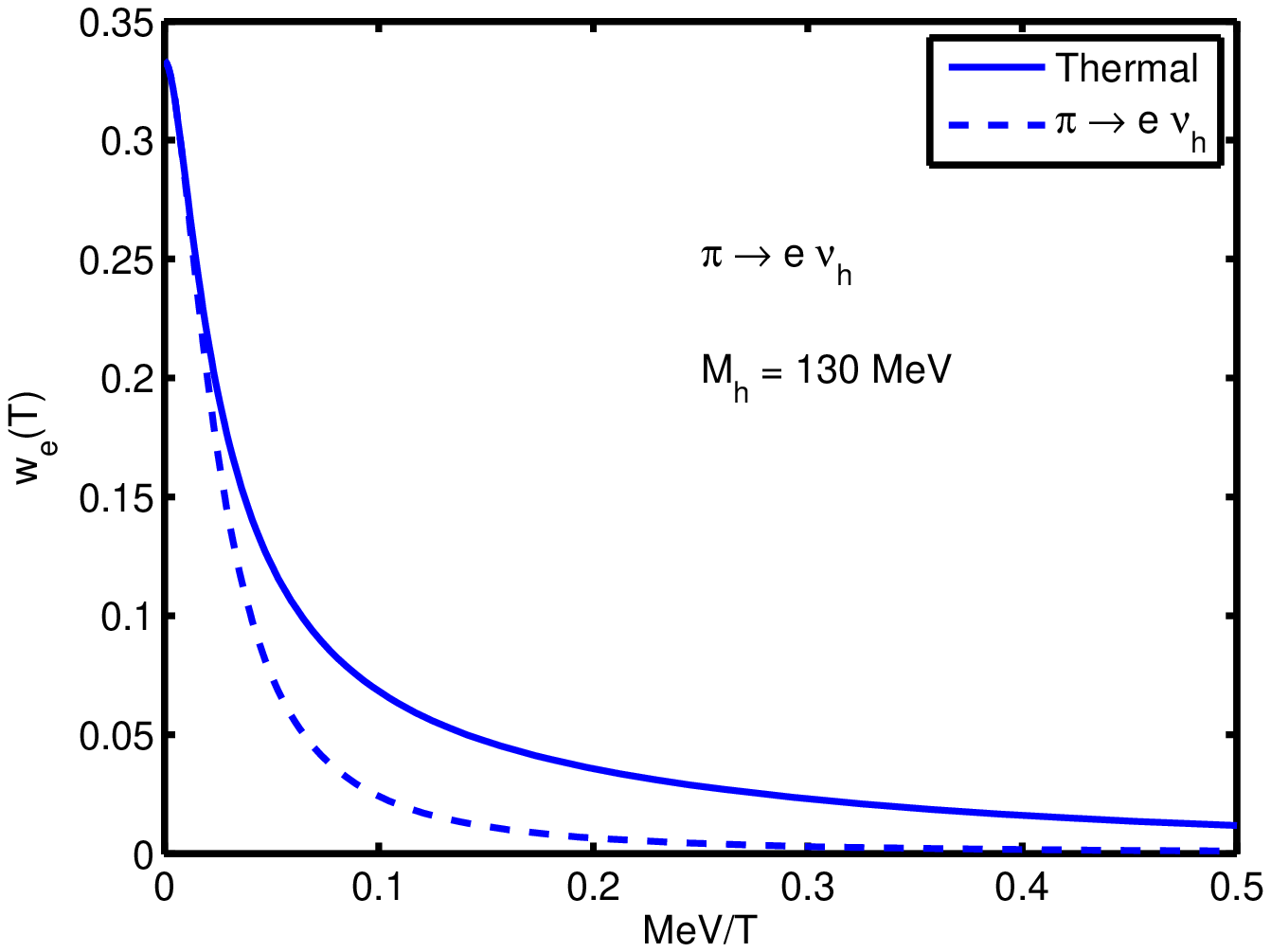}
\caption{$w_l(T)$ for    $\pi \rightarrow e \nu_h$ compared to  thermal distribution for  $M_h = 50,75,100,130\, \mathrm{MeV}$.}
\label{fig:heavyeleos}
\end{center}
\end{figure}

  The equation of state $w_l(T)$ for each channel and for the thermal distribution are compared in figs (\ref{fig:lighteleos},\ref{fig:heavyeleos}) as a function of $1/T$ for a range of different masses.  These figures clearly display the non-thermality of the heavy neutrino species, furthermore, as anticipated by the discussion above, the distribution function from $\pi\rightarrow \mu \nu_h$, namely $f_{\mu h}$ yields a \emph{colder} component than that from the $e$ channel, $f_{eh}$, a direct consequence of the ``kinematic entanglement'' leading to a larger amplitude at small $y$ for $f_{\mu h}$,  and both components are colder and  becoming non-relativistic $w(T) \ll 1/3$ much sooner than the thermal case.

The total equation of state is given by the full distribution function (\ref{tilf},\ref{totfrizf2}), namely
\be \label{eostot}
w (T) = \frac{\mathcal{P}}{\rho} = \frac{1}{3} \frac{ \int dy \, \frac{y^4}{ \sqrt{y^2+\frac{M_h^2}{T^2}}} \,\widetilde{f}_{h}(q_c)}{\int dy \, y^2 \sqrt{y^2+\frac{M_h^2}{T^2}}\, \widetilde{f}_{h}(q_c)} \,,
\ee which depends on the ratio $|H_{eh}|^2/|H_{lh}|^2$ as this ratio varies between $0$ and $\infty$, it follows that $w(T)$  for $0 < M_h < 36\,\mathrm{MeV}$ interpolates between the two dashed lines corresponding to the muon and electron channels in figs. (\ref{fig:lighteleos}), this is shown in fig. (\ref{fig:eosint}). For the mass range $141\,\mathrm{MeV} > M_h > 36\,\mathrm{MeV}$ only the electron channel contributes and $w(T)$ is given by the results displayed in fig.(\ref{fig:heavyeleos}).

\begin{figure}[h!]
\begin{center}
\includegraphics[height=3.5in,width=3.2in,keepaspectratio=true]{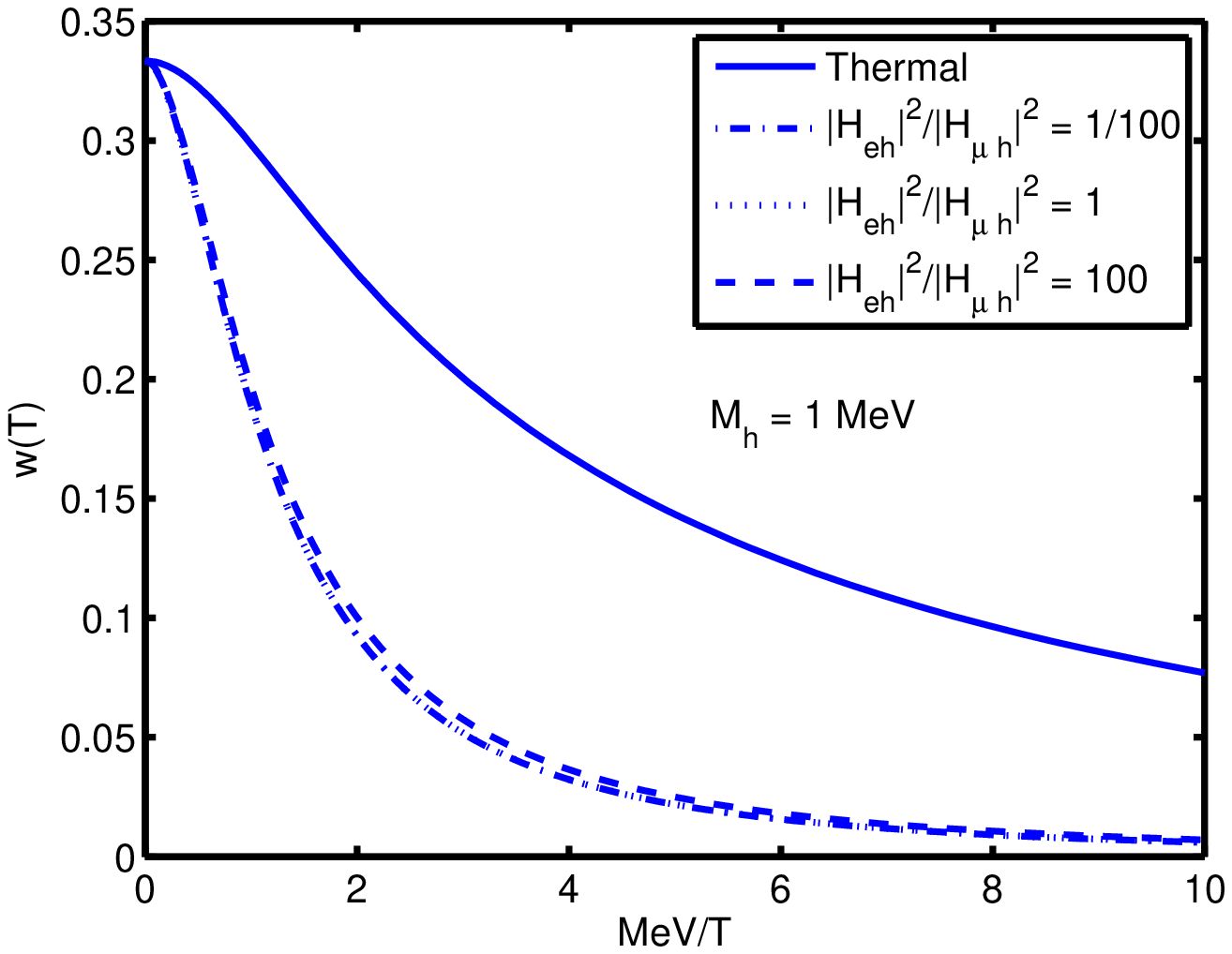}
\includegraphics[height=3.5in,width=3.2in,keepaspectratio=true]{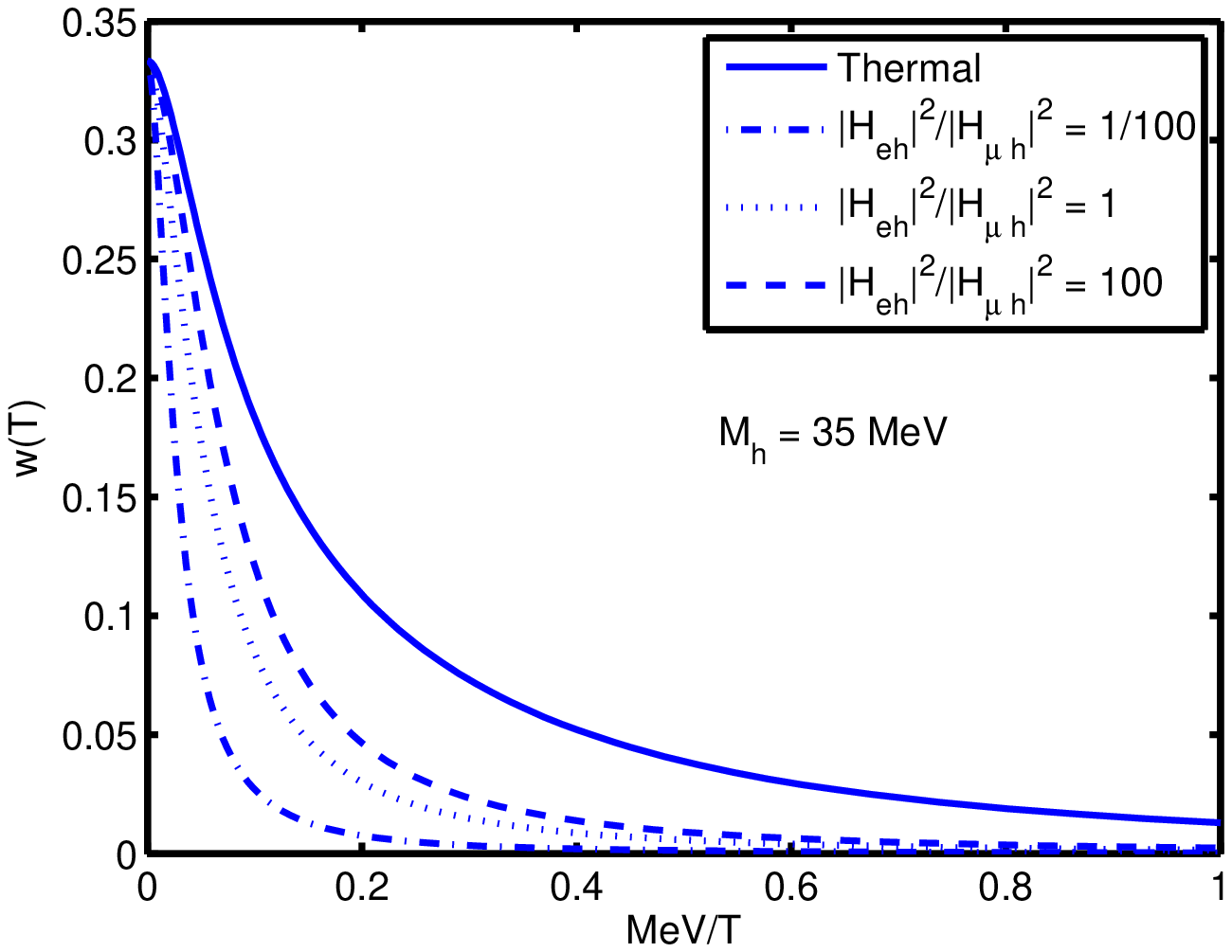}
\caption{Equation of state with full distribution function for the ratios $|H_{eh}|^2/|H_{\mu h}|^2 = 0.01,1,100$ as a function of $(MeV)/T$ for $M_h=1,35 \,\mathrm{MeV}$. $w(T)$ interpolates between the results for $\mu$ and $e$ channels.}
\label{fig:eosint}
\end{center}
\end{figure}

\subsection{Cosmological constraints:  }

Having obtained the distribution function  at freeze-out, we can now implement the general results obtained in section (\ref{sec:cosmocons})  and establish the allowed regions within which the various cosmological constraints discussed in section (\ref{sec:cosmocons}) are satisfied.

\begin{itemize}
\item \textbf{Abundance:} with    $f_h(y)$ and $\widetilde{f}_h(y)$ given by (\ref{totfrizf2}, \ref{tilf}) respectively   and taking $g_{\nu_h}=2$ the abundance constraint (\ref{abucons}) becomes
\be \label{dmbound}
2.81 \Bigg( \frac{M_h}{\mathrm{keV}}\Bigg)\,\Big( \frac{|H_{\mu h}|^2}{10^{-5}}\Big) \int y^2 \widetilde{f_h}(y) \, dy \leq 1  \,.
\ee This is a function of the ratio $  |H_{e h}|^2/|H_{\mu h}|^2$ and also of $M_h$ through the coefficients $C[m_l,m_h]$. This bound    was calculated in \cite{lellolightsterile} for   $M_h \le 1 MeV$ this  mass region is completely dominated by the muon channel, whereas here we allow masses up to $M_h \le M_{\pi} - M_l$ which translates to $M_h \le 142 MeV$ for the electron channel and $M_h \le 36 MeV$ for the muon channel.

\item \textbf{Phase space (Tremaine-Gunn):}

  Using the smallest phase space value of ref \cite{destri} which comes from the Fornax dwarf spheroidal galaxy:  $(\rho/\sigma^3)_{today} = 2.56\times 10^{-4} \,(keV)^4$,  and using the general result (\ref{tg2}) with $g_{\nu_h}=2$ we obtain the constraint

\be \label{dsphbound}
\Big( \frac{M_h}{\mathrm{keV}}\Big)^4\,\Big(\frac{|H_{\alpha h}|^2}{10^{-5}}\Big)    \ge  0.0038 \, \frac{ \Big[ \int dy \, y^4 \widetilde{f}_h(y)  \Big]^{3/2} }{ \Big[\int dy \, y^2 \widetilde{f}_h(y)  \Big]^{5/2} } \,.
\ee

\item \textbf{Stability:}
The ``conservative'' stability constraint (\ref{longlive2}) is independent of the distribution function.

  These bounds and allowed parameter space are shown in fig. (\ref{fig:sterilebounds}) along with the parameters of heavy neutrinos from the recently reported  X-ray signals\cite{bulbul,boyarsky} ($M_h = 7.1\,\mathrm{keV}, |H_{\alpha h}|^2 = 7\times 10^{-11}$). An important observation about this figure is that the inclusion of \emph{both channels} in the total distribution function leads to  (marginal) consistency with the claimed X-ray signal. This differs from ref. \cite{lellolightsterile} which claimed that consistency with the X-ray data was supported in the muon channel but not the electron channel. The main differences between these results and those of ref \cite{lellolightsterile} are that we generalize the results to arbitrary mass (as opposed to $\lesssim 1 MeV$) and we include \emph{both production channels} as opposed to considering each channel independently.

\begin{figure}[h!]
\begin{center}
\includegraphics[height=3.5in,width=3.2in,keepaspectratio=true]{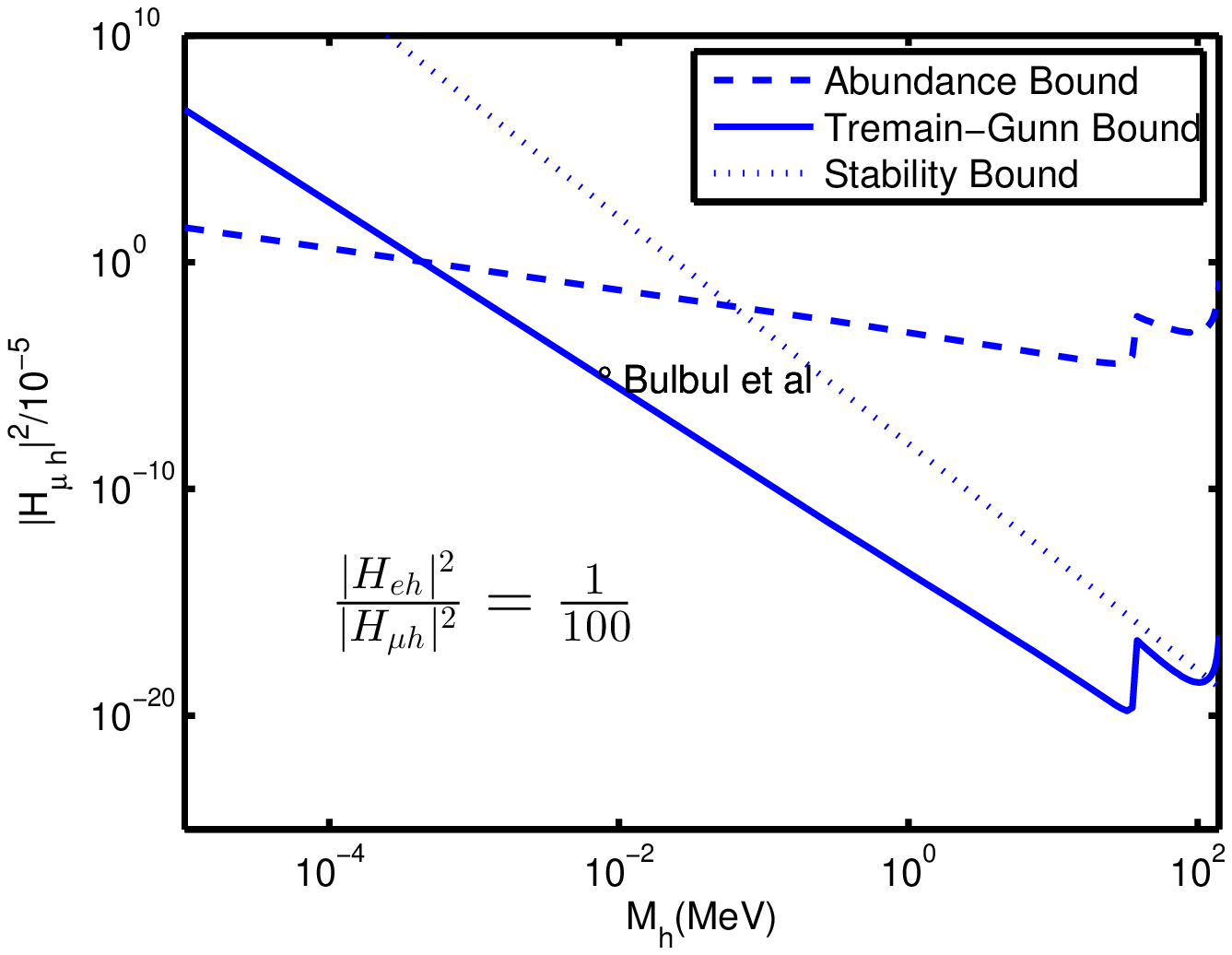}
\includegraphics[height=3.5in,width=3.2in,keepaspectratio=true]{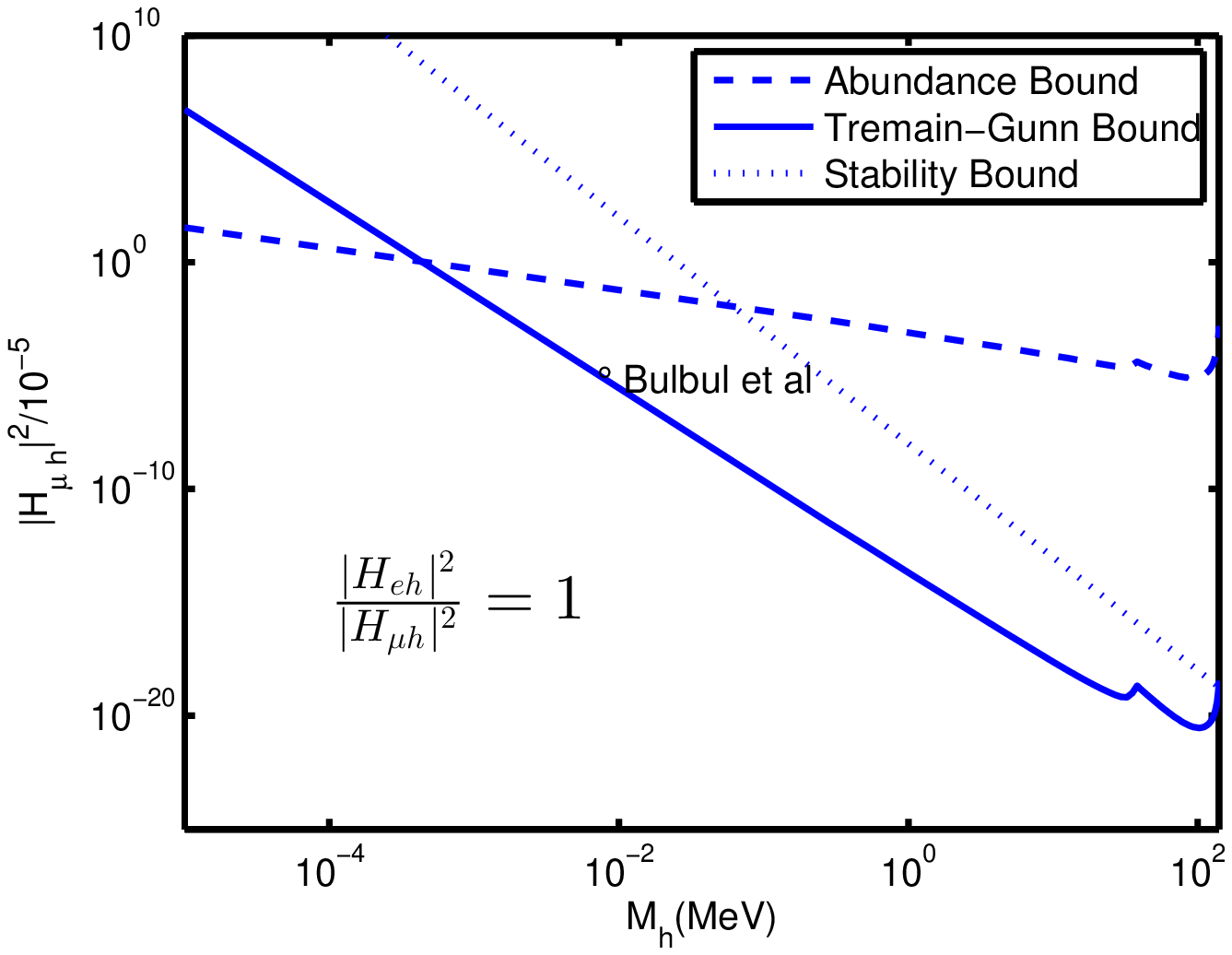}
\includegraphics[height=3.5in,width=3.2in,keepaspectratio=true]{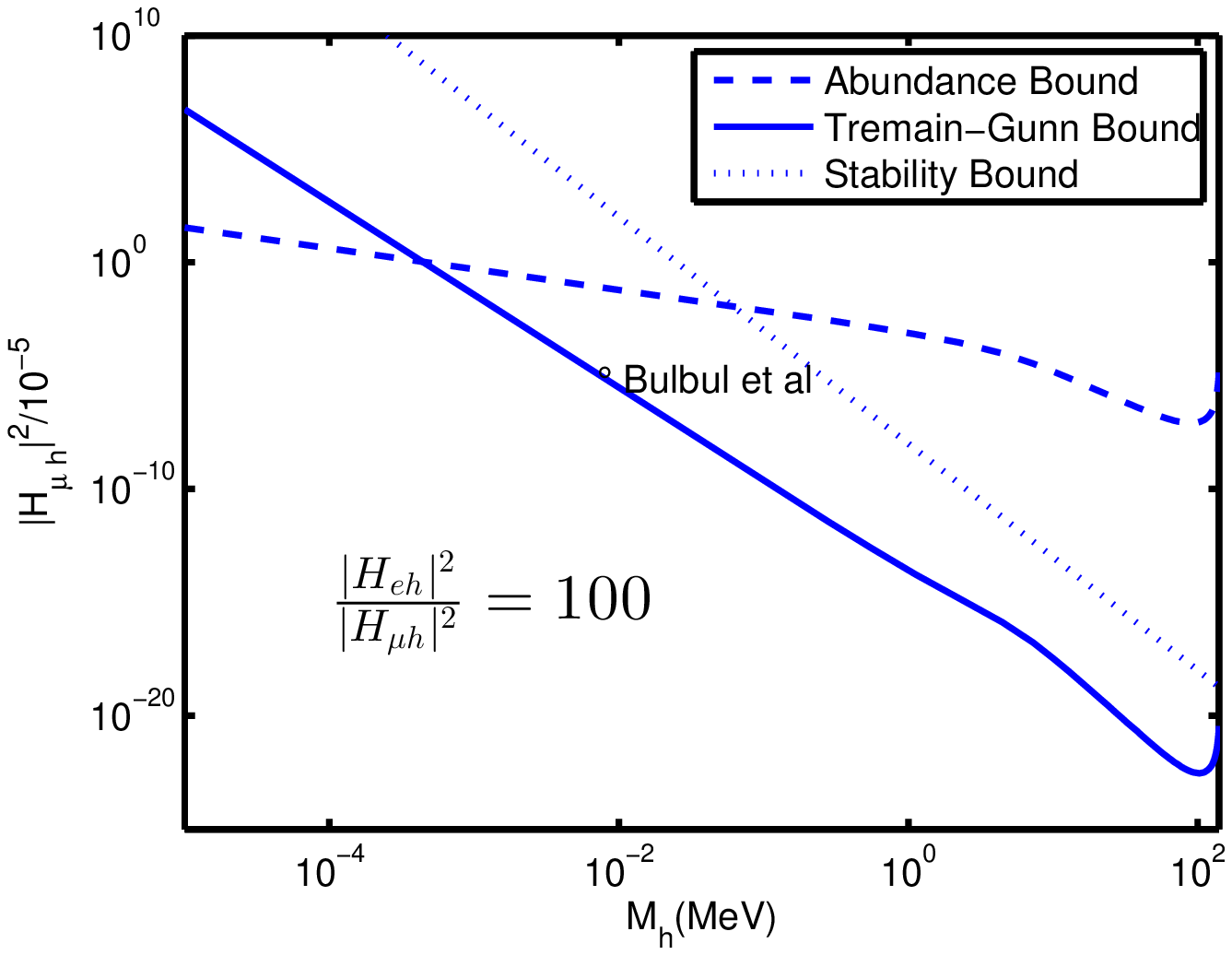}
\caption{The bounds on $M_h-|H_{lh}|^2$  from abundance, stability and phase space constraints. The allowed regions determined from Eqs (\ref{dmbound},\ref{dsphbound} ,  \ref{longlive2}) are shown and the  parameters which potentially explain the $3.5 \,\mathrm{keV}$ signal\cite{bulbul,boyarsky} are also shown. The kink is a consequence of the thresholds. The allowed parameter space is within the region bound by the three lines. }
\label{fig:sterilebounds}
\end{center}
\end{figure}

An interesting aspect in the two top figures are ``kinks'' in the abundance and Tremaine-Gunn (phase space density) line, this is a consequence of the kinematic thresholds   in the full distribution function (\ref{totfrizf}).

\item \textbf{Free streaming scale:} the free streaming wavevector and length scale are given by eqns. (\ref{kaifes}-\ref{lamfes}) respectively,  namely

    \be
\lambda_{fs,\nu_h}(0) \simeq  5.3 \,  \left(\frac{keV}{M_h}\right) \, \sqrt{\frac{ \int dy \, y^4 \widetilde{f}_h(y)}{ \int dy \, y^2 \widetilde{f}_h(y)} } ~ \mathrm{kpc}\,. \label{lamfespi}
\ee again this is a non-linear function of the mass and for a species $\nu_h$ produced by a single channel it would be independent of the mixing angle, however if there are several channels, as is the case in $\pi$ decay, it depends on the ratio of mixing angles. Fig.(\ref{fig:lambfs}) displays $\lambda_{fs,\nu_h}(0)$ as a function of $M_h$ for various ratios.

\begin{figure}[h!]
\begin{center}
\includegraphics[height=3.5in,width=3.2in,keepaspectratio=true]{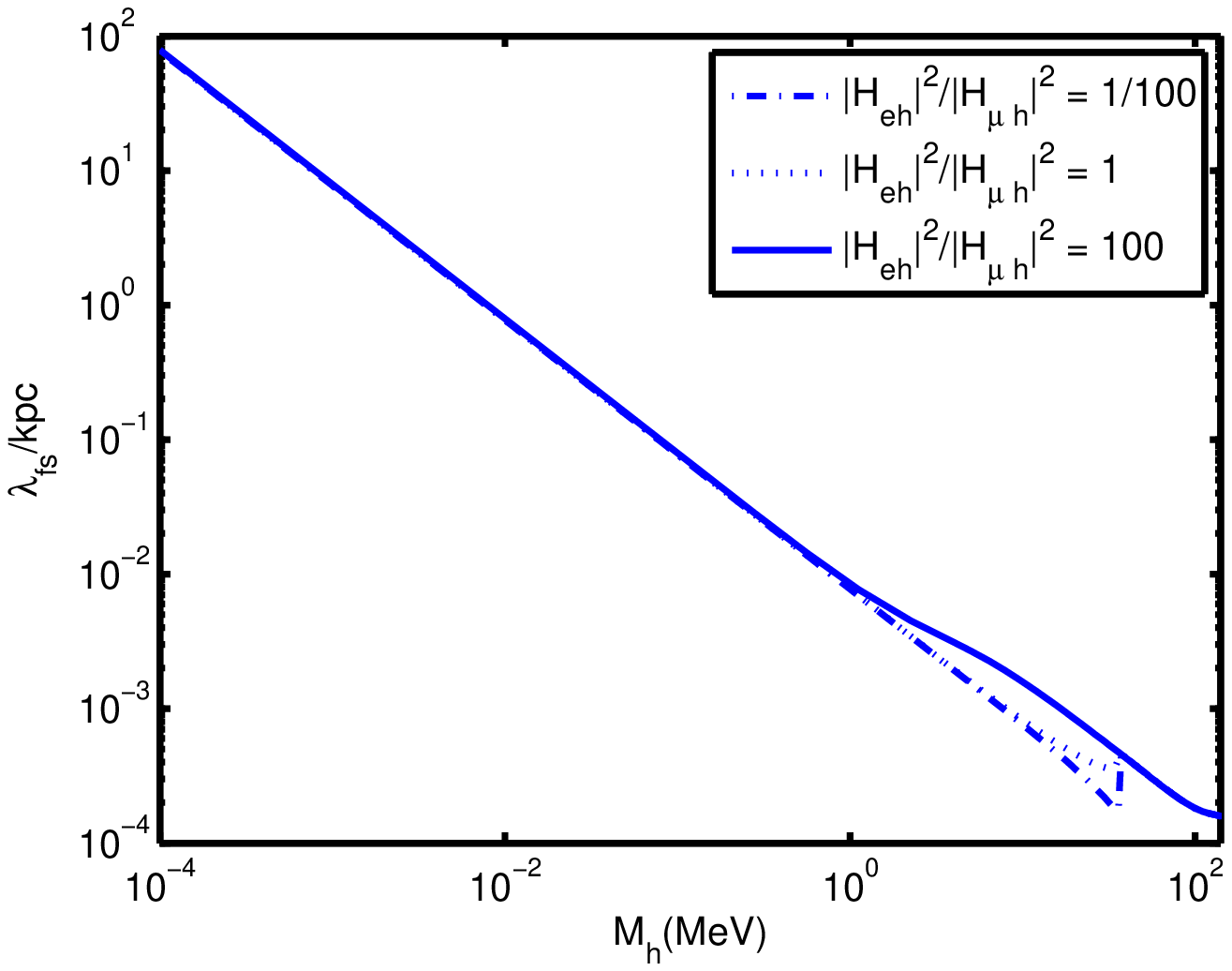}
\includegraphics[height=3.5in,width=3.2in,keepaspectratio=true]{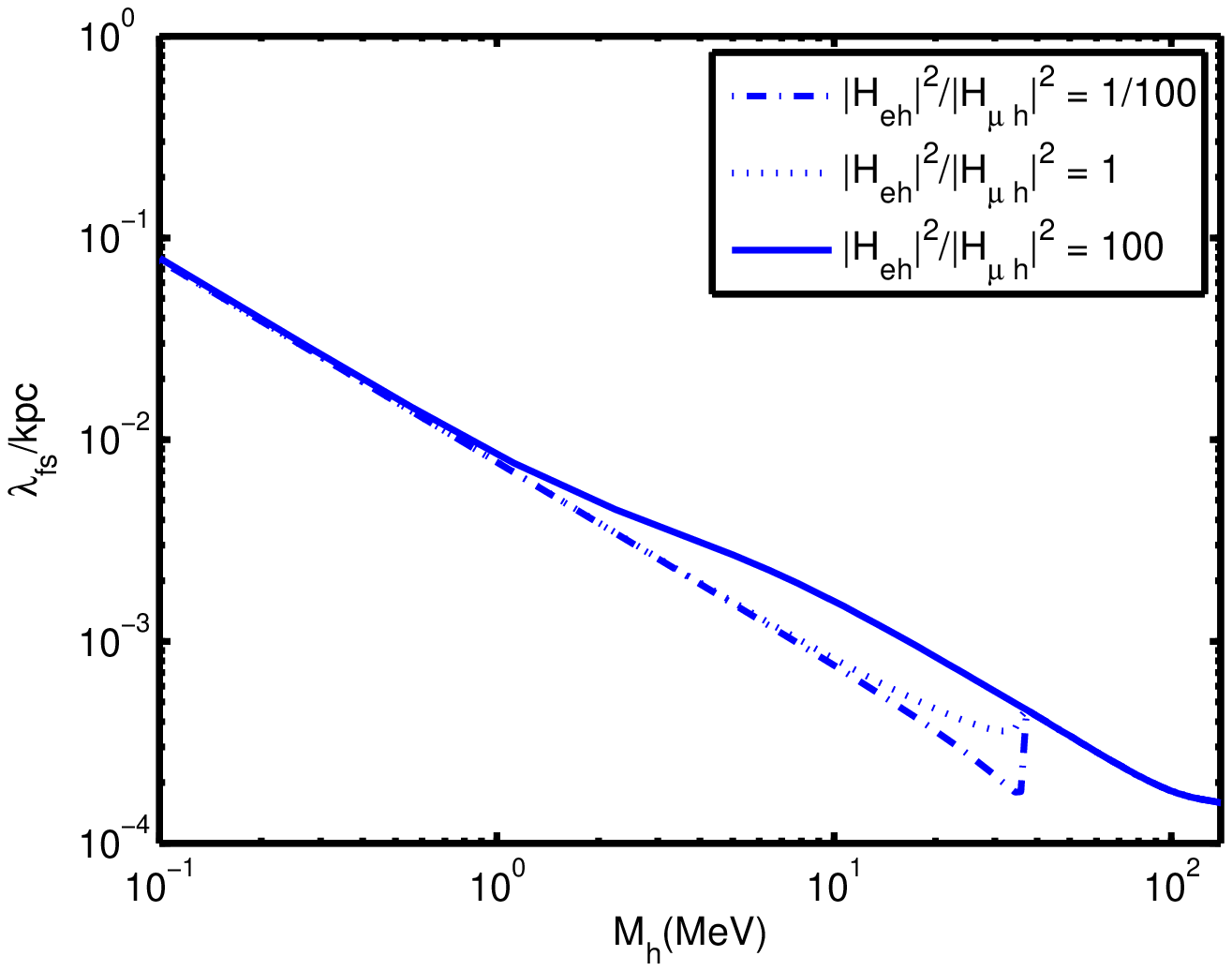}
\caption{$\lambda_{fs}(0)/(kpc)$ as a function of $M_h$ for the ratios $|H_{eh}|^2/|H_{\mu h}|^2 = 0.01,1,100$. Right panel zooms in to highlight the kink because of thresholds and interpolation.}
\label{fig:lambfs}
\end{center}
\end{figure}

For $M_h \lesssim 1 \,\mathrm{MeV}$ the $\mu$ (coldest) channel dominates, and $\lambda_{fs}(0)$ is insensitive to the ratio, however, as $M_h \simeq 30 \,\mathrm{MeV}$ the contribution of the $e$ channel becomes substantial and dominates above the threshold at $M_h \simeq  36\, \mathrm{MeV}$, the kink in the figure is a result of this threshold. Therefore $\lambda_{fs}(0)$ interpolates between that of the $\mu$ and electron channels as a function of the ratio, just as the equation of state interpolates between the colder ($\mu$) and warmer electron dominated components.

\end{itemize}

\subsection{Other processes in the same temperature range}
In this section we focused on a detailed presentation of heavy neutrino production from pion decay after the QCD hadronization transition, primarily as a clear example where the finite temperature corrections of the pion decay constant and mass have been previously studied in the literature. However, as the discussion in section (\ref{sec:qk}) highlights, there are many processes that produce heavy neutrinos via charged and neutral current vertices provided the kinematics is favorable. In the temperature range just explored $T \simeq 150 \,\mathrm{MeV}$ muons are thermally populated and muon decay $\mu \rightarrow e \nu_m \nu_h$ is also a production mechanism that is available provided $M_h$ is in the kinematic window for the three body decay. However, this process, is subleading in the temperature range $T \lesssim M_\pi$, this is because the ratio of decay rates
\be \frac{\Gamma_\pi}{\Gamma_\mu} = \frac{\tau_\mu}{\tau_\pi} \simeq 10^2 \ee \emph{however} while pions are only available as a production channel below $T_{QCD}$, muons, on the other hand, are thermally populated at larger temperatures therefore they contribute substantially to the production of heavy neutrinos with masses within the kinematic window. Their contribution to the total abundance of heavy neutrinos merits a deeper study, along with all the other mechanisms described in section (\ref{sec:qk}). Furthermore, light neutrinos are thermally populated in a much wider temperature range and $\nu_{m1} \nu_{m2} \nu_{m3} \rightarrow \nu_h $ is the three body ``fusion'' process described by the gain term (\ref{gain4}) that yields heavy neutrinos at temperatures $T \geq M_h$. The inverse process, $\nu_h  \rightarrow \nu_{m1} \nu_{m2} \nu_{m3}$ is the decay process described by the loss term (\ref{loss4}), which contributes even at zero temperature and describes the decay of the heavy neutrino (\ref{nuh3nus}). The detailed study of all of these processes clearly defines an extensive program to assess reliably the production of heavy neutrinos in cosmology.

\subsection{Comparison with Dodelson-Widrow\cite{dodwid}:}
Although the Dodelson-Widrow (DW)\cite{dodwid} (non-resonant) mechanism of sterile neutrino production via active-sterile oscillations has been recently shown to be inconsistent with cosmological data at $> 99\%$ confidence level\cite{kaplinghat2},   it is illustrative to compare the results obtained above for the non-equilibrium distribution function to the (DW) case for further understanding of its cosmological consequences.

The (DW) distribution function is
\be f_{DW}(y) = \frac{\beta}{e^y+1}\,, \label{fDWofy}\ee where $\beta$ is determined by saturating the DM abundance, namely\cite{dodwid,kaplinghat2}

\be \beta = \Omega_{DM}h^2 \,\Big(\frac{94\, eV}{M_h} \Big)= \frac{11.3\,eV}{M_h} \,.\label{betaDW}\ee Furthermore, we will focus our comparison on the mass scale $M_h\simeq 7\,keV$ because this scale is of observational relevance\cite{bulbul,boyarsky} and we consider the $\mu$ channel since for $M_h \simeq 7\,keV$ this channel features the largest branching ratio. The distribution function obtained from pion decay, $f_h(y)$ is defined by eqn. (\ref{frizf}). An important ingredient in the comparison are the following results:
\bea \frac{1}{\beta} \int dy\,y^2 f_{DW}(y) & = &  1.803 \label{intDW}\\
\frac{1}{\Lambda_{\mu h}} \int dy\,y^2 f_{h}(y) & = &  1.830 \label{intfh} \,\eea therefore the total integral of the distribution functions \emph{divided by their prefactors} is \emph{approximately the same}, however as gleaned from fig.\ref{fig:mudist} $f_h(y)/\Lambda_{\mu h}$ is strongly peaked at small momenta, and as discussed in the text the distribution function is fairly insensitive to $M_h$ for $M_h \lesssim 1\,MeV$. We provide two different manners to compare the distributions; i) fixing $\beta$ and $\Lambda_{\mu h}$ to saturate the DM density in both cases ii) for fixed values of $M_h$ and mixing angles $|H_{\mu h}|^2$ extracted from the analysis of ref.\cite{kaplinghat2} to obtain $\Lambda_{\mu h}$ but consistent with the (DW) scenario we keep the value of $\beta$ that saturates $\Omega_{DM}$.

\vspace{2mm}

\textbf{Comparison 1:} both $\beta$ and $\Lambda_{\mu h}$ are fixed to yield the total DM density, namely the fraction (\ref{fractionh}) is $\mathcal{F} =1$ in both cases. This yields
\be \beta = 1.611 \times 10^{-3} ~~,~~ \Lambda_{\mu h} = 1.600 \times 10^{-3} \label{betlam}\ee with these values we display both distribution functions in fig.\ref{fig:compare1}.

\begin{figure}[h!]
\begin{center}
\includegraphics[height=3.5in,width=3.2in,keepaspectratio=true]{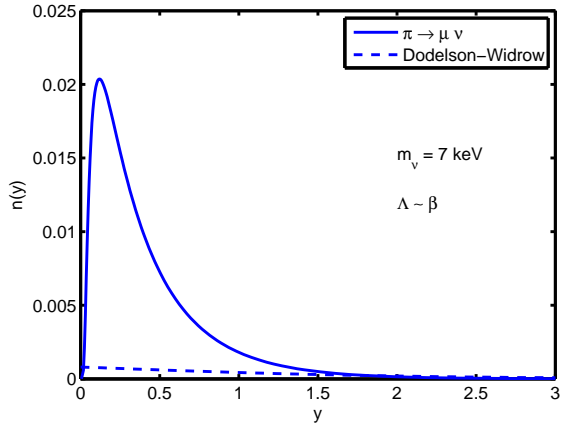}
\caption{The distribution functions $f_{DW}(y)$ (dashed line) and $f_{h}(y)$ solid line, for $\beta \simeq \Lambda$. In this case both distribution functions are fixed to saturate the DM density. $M_h = 7 \, keV$}
\label{fig:compare1}
\end{center}
\end{figure}

Although the integrals of the distribution functions are the same, obviously $f_h(y)$ is sharply localized at smaller momenta, therefore yielding a colder distribution.

\vspace{2mm}

\textbf{Comparison2:} for this case we keep $\beta= 1.61\times 10^{-3}$ so that $f_{DW}$ saturates the abundance bound, but $\Lambda_{\mu h}$ is now extracted from the upper bounds on the confidence band for the (DW) species of fig. (4) in ref.\cite{kaplinghat2}, identifying the mixing matrix element $|H_{\mu h}|^2$ with $\sin^2(2\theta)$ in this reference. We read from this fig. the values
\be M_{h} \simeq 7 \,keV~~;~~ |H_{\mu h}|^2 \simeq 10^{-7} \label{params}\ee with these values we find
\be \Lambda_{\mu h} = 3.18 \times 10^{-4} \label{lamparam}\ee the comparison between the distribution functions is displayed in fig.\ref{fig:compare2}. In this case the distribution function $f_h(y)$ yields a fraction $\mathcal{F}=0.205$ to the DM abundance.

\begin{figure}[h!]
\begin{center}
\includegraphics[height=3.5in,width=3.2in,keepaspectratio=true]{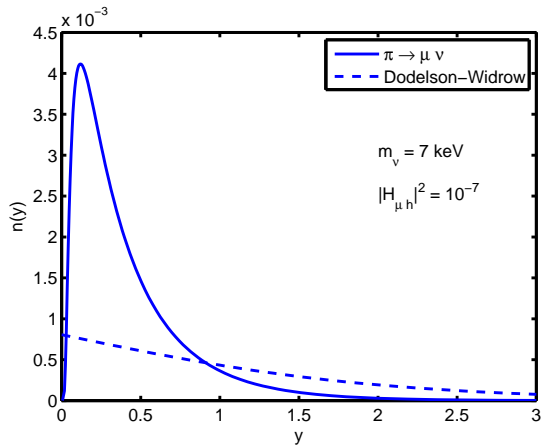}
\caption{The distribution functions $f_{DW}(y)$ (dashed line) and $f_{h}(y)$ solid line, for $\beta = 1.61\times 10^{-3}~;~\Lambda_{\mu h} = 3.18 \times 10^{-4}$. In this case the (DW) distribution function is fixed to saturate the DM abundance, whereas $\Lambda_{\mu h};M_h = 7 \, keV$ are fixed by the upper limits of the confidence band for (DW) in fig. (4) of ref.\cite{kaplinghat2}. For this case $f_h$ yields a fraction $\mathcal{F}=0.205$ of the DM density. }
\label{fig:compare2}
\end{center}
\end{figure}

Again it is clear that even when the distribution function yields a smaller fraction, it features a larger contribution at smaller momenta and falls off faster at larger momenta, this feature makes this species \emph{colder} than (DW) with an abundance that while smaller than (DW) is fairly substantial.

A further illuminating comparison is obtained from the free streaming length (\ref{lamfes}), which is \emph{independent} of the normalization factors $\beta,\Lambda_{\mu h}$ respectively. We find
\bea \lambda^{(DW)}_{fs}(0) & = & 2.7116 \, kpc\label{lamfsDW} \\
\lambda^{h}_{fs}(0) & = & 1.0231 \, kpc \,,\label{lamfshfi} \eea the low momentum enhancement of $f_h(y)$ yields a much colder distribution with a much shorter free streaming length as compared to the (DW) case.

We conclude that while the (DW) mechanism seems to be ruled out as a \emph{sole} production channel of sterile neutrinos, the comparison with the non-equilibrium function obtained from pion decay in the dominant channel offers a useful insight into the properties of sterile neutrinos produced via this mechanism and suggests that the abundance from this alternative scenario could be a substantial contribution to the DM component.

\subsection{Comments:} We have implemented the results on the finite temperature dependence of the pion decay constant and pion mass from a substantial body of work on chiral perturbation theory including resonances and linear and non-linear sigma models including contributions from vector mesons\cite{tytgat1,nicola1,nicola2,jeon,harada}. Taken together this body of results offer a consistent description of the temperature dependence. However, we recognize that there remain uncertainties inherent in an effective description of hadronic degrees of freedom but questioning the validity of ChPT is beyond the scope of this work. Recently a lattice study\cite{newlattice} reported results some of which are at odds with those of refs.\cite{tytgat1,nicola1,nicola2,jeon,harada} such as an increase in the pion decay constant and a decrease of pion temperature near the QCD crossover scale. While a confirmation or rebuttal of these results and/or a resolution of the controversy is awaiting, we can speculate on the impact of the results of this reference, if these hold up. First: if the pion decay constant is \emph{larger} this obviously results in a larger production rate, secondly: if the pion mass is smaller, this also leads to a larger production rate (less thermal suppression of pions in the medium) and \emph{crucially} to a delayed freeze out with a longer stage of production as pions remain populated in the medium for a longer time. Lastly: \emph{if} the pion mass falls below the muon channel threshold at high temperature, the production is solely through the electron channel resulting in a \emph{warmer} distribution, as the pion mass increases towards smaller temperature, the muon channel opens up and the muon production channel with a \emph{colder}
component becomes available thus confirming the argument on the ``mixed'' contributions to the distribution function. All of these aspects \emph{bolster} the case for consideration of pions as an important production mechanism, and the last point in particular,  bolsters the argument on kinematic entanglement. We thus conclude that the features imprinted on the non-equilibrium distribution function from the various decay channels and \emph{mixed} nature of the distribution function  is  a robust qualitative prediction.

\section{Unstable heavy neutrinos: cascade decays into stable DM}\label{sec:cascade}

Unstable heavy neutrinos with lifetimes much smaller than $1/H_0$ do not play a \emph{direct} role as a viable DM candidate, however they can decay via a \emph{cascade} into lighter stable heavy neutrinos that could be viable candidates. To discuss this scenario in more concrete terms, let us consider a hierarchy of two heavy neutrinos $\nu_{h_1},\nu_{h_2}$ with $M_{h_1} \gg M_{h_2}$ and assume that the heavier, $\nu_{h_1}$ is in the kinematically allowed window  that allows its  production on-shell from $\pi$ decay, namely $\pi \rightarrow l \nu_{h_1}$. If so after being produced, $\nu_{h1}$ will cascade decay into $\nu_{h_2}+\mathrm{leptons}$ on a time scale $\simeq \tau_{\nu_{h1}}$, namely the intermediate heavier $\nu_{h_1}$ yields yet another production channel to the lighter $\nu_{h_2}$,
\be \pi \rightarrow l \nu_{h_1} \rightarrow l \nu_{h_2} l_1 l_2 \label{cascade} \,, \ee where $l_1,l_2$ are other charged or neutral leptons. In this process the intermediate $\nu_{h_1}$ goes on shell and the cascade is mediated by resonant decay. In this scenario the production rate of the stable(r) species $\nu_{h_2}$ is given by\cite{cascade,han,heavysterile}
\be \Gamma^<_{\pi \rightarrow \nu_{h2} l l_1 l_2} = \Gamma^<_{\pi \rightarrow \nu_{h1} l}\times \mathrm{Br}(\nu_{h_1} \rightarrow \nu_{h_2} l_1 l_2)\,, \label{casca} \ee  where $\mathrm{Br}(\nu_{h_1} \rightarrow \nu_{h_2} l_1 l_2)$ is the branching ratio. If $\nu_{h_1}$ decays into a lighter heavy $\nu_{h_2}$ it can also decay into the active-like- light neutrinos $\nu_m$, the decay amplitude for $\nu_{h1} \rightarrow \nu_{m_1} \nu_{m_2} \nu_{m_3} \propto H U$, whereas the amplitude to decay into $\nu_{h_1}\rightarrow \nu_{h_2} l_1 l_2 \propto H^2 $ therefore $Br \propto H^2$ and $\Gamma^<_{\pi \rightarrow \nu_{h_2} l l_1 l_2} \propto H^4$ namely is suppressed by an extra factor $|H|^2$. If the lifetime of $\nu_{h_1}$ is $\gtrsim 10^3 \,secs$ it can decay into active-like neutrinos well after Big Bang Nucleosynthesis (BBN) avoiding the constraints on the number of relativistic active neutrinos during BBN providing a late injection of neutrinos into the cosmic neutrino background well after BBN. A lifetime $\gtrsim 10^{11}\,\mathrm{secs}$ would inject a lighter heavy neutrino as a DM candidate after matter radiation equality, just when density perturbations begin to grow under gravitational collapse. Two specific examples illustrate these possibilities: a) consider $M_h \lesssim 1\,\mathrm{MeV}$ from the discussion in section (\ref{sec:stable}) the decay channel with largest branching ratio ($\simeq 99\%$) is the ``invisible'' channel $\nu_h \rightarrow 3\nu_m$ (see eqn. (\ref{nuh3nus})) with a lifetime
\be \tau \simeq \frac{10^{5}}{|H_{m h}|^2 }\, \Bigg(\frac{\mathrm{MeV}}{M_h}\Bigg)^5 \,s \,,\label{lonlif}\ee
 with $M_h \lesssim 1\,\mathrm{MeV}~;~|H_{m h}|^2 \lesssim 10^{-6}$ the heavy neutrino decays into active-like neutrinos after matter-radiation equality ``injecting'' a non-LTE component in the cosmic neutrino background after neutrino decoupling. b) The decay rate into a lighter $\nu_{h_2}$ is suppressed by another power of $|H_{mh}|^2$, therefore a $\nu_{h_1}$ with $M_{h_1} \lesssim 10 \,\mathrm{MeV}~;~|H_{m h}|^2 \lesssim 10^{-6}$ can decay into $\nu_{h_2}$ with  $M_{h_1} \simeq \mathrm{few}\,\mathrm{keV}$ also after matter-radiation equality, now providing a heavy neutrino with a $\mathrm{keV}$ mass range as a DM candidate just at the time when density perturbations begin to grow. This latter case may yield an excess of positrons, however, to assess whether this is observationally significant requires a deeper study.

Obviously these are \emph{conjectures} that merit a far deeper analysis, however this scenario is similar to that posited in ref.\cite{dienes} of a hierarchy of heavy degrees of freedom decaying in a cascade into lighter species that may act as stable(r) DM candidates.

\section{Summary, Discussion and Further Questions }
The main premise of our study is that if sterile neutrinos are a suitable extension beyond the Standard Model in which these mix with active neutrinos via an off-diagonal mass matrix, diagonalization of the mass matrix to the mass basis implies that heavy neutrinos mass eigenstates couple to standard model leptons via charged and neutral current interactions. The same processes that produce active-like light neutrinos also produce heavier neutrinos if kinematically allowed, albeit with a much smaller branching ratio determined by small mixing angles. We study the production of heavy neutrinos \emph{via standard model charged and neutral current interactions} under a minimal set of assumptions:   small mixing angles with flavor neutrinos, standard model particles are in local thermodynamic equilibrium. We obtain the quantum kinetic equations that describe their production to leading order in the small mixing angles and give the general solution in terms of gain and loss rates that obey detailed balance.  A wide range of charged and neutral current  processes available throughout the thermal history of the Universe lead to cosmological production of heavy neutrinos including the possibility of production from \emph{collective excitations} and ``rare'' processes in the medium  such as plasmon decay, and ``inverse'' processes such as $\gamma \nu_m \rightarrow \nu_h$.

  We discuss the general conditions for thermalization and argue that  heavy neutrinos with lifetimes $> 1/H_0$ (the Hubble time scale) freeze-out with non-equilibrium distribution functions. We generalize the concept of \emph{mixed DM} to the case in which a single species of heavy neutrinos is produced by different channels and argue that in each channel the heavy neutrinos produced are \emph{kinematically entangled} with the lepton produced in the reaction. If the distribution function freezes out of local thermal equilibrium it maintains memory of this kinematic entanglement in the form of a colder or warmer distribution function as compared to other channels. We quantify the ``coldness'' by obtaining the equation of state parameter for each channel, which serves as a ``proxy'' for the velocity dispersion of the DM particle when it becomes non-relativistic. If several channels contribute to the production of a particular species, the total distribution function is a \emph{mixture} with components produced from the different channels, these  may be colder or warmer as a consequence of the kinematic entanglement. The concentration of each component depends on the kinematics and the ratio of mixing angles in each channel.

We summarize the abundance, phase space density and \emph{stability} constraints that a suitable DM candidate must fulfill in terms of the total distribution function at freeze-out and discuss clustering properties such as primordial velocity dispersions and free streaming lengths. We compared the quantum kinetic framework with other treatments available in the literature, recognizing the necessity for a consistent non-perturbative treatment in the case of possible MSW resonances in the medium.

An important conclusion of this analysis is that, whereas many efforts focus on some temperature scale and on particular production processes,  in order to reliably assess the feasibility of a particular heavy neutrino DM candidate, \emph{all} possible production channels of this species must be carefully analyzed throughout the thermal history of the Universe. An immediate consequence of this principle is to alter the ``standard'' production mechanisms - Dodelson-Widrow, Shi-Fuller, scalar decays - all of which assume a vanishing initial population. The example of pion decay provides a unambiguous source of sterile neutrino population which will modify the standard results in a prescription described in \cite{merleini}. Further work is in progress on this front.

We argued that the final distribution function of heavy neutrinos after freeze-out is indeed a \emph{mixture} of the various contributions to it from the different production processes that are kinematically available.  Only in the case of a fully thermalized population will the ``memory'' from the different processes be erased, but non-equilibrium distribution functions will have imprinted in them the kinematic entanglement from the different processes.

As an explicit example we studied the production of heavy neutrinos   from charged pion decay after the QCD crossover into the hadronized phase within the effective field theory of weak interactions of charged pions, including finite temperature corrections to the pion decay constant and mass. Pion decay is one of the main   sources of neutrino beams in accelerator experiments and pions being the lightest hadrons formed after the QCD crossover,   their decay surely contributed to the cosmological production of heavy neutrinos. While not claiming that this process is more or less important than others (a claim that requires a detailed assessment of the other production mechanisms)  it provides a wide kinematic window $M_h \lesssim 140 \,\mathrm{MeV}$ from two different channels and offers a clear example of kinematic entanglement: the distribution function from the $\mu$ channel is distinctly \emph{colder} than that of the electron channel and the total distribution is a mixture of both. We obtain the allowed region of parameters that fulfill the abundance, phase space density and stability constraints, these are displayed in figs. (\ref{fig:sterilebounds}) for various values of the ratio $|H_{eh}|^2/|H_{\mu h}|^2$, the boundaries of these regions reveal the thresholds for the different channels, a hallmark of   kinematic entanglement. The equation of state (or alternatively velocity dispersion) and free streaming length interpolate between the colder   component and the warmer component from the $\mu$ and $e$ channels respectively, as a function of $M_h$  and the \emph{ratio} $|H_{eh}|^2/|H_{\mu h}|^2$,  clearly displaying the \emph{mixed} nature of the total distribution function.

We \emph{conjecture} that heavy neutrinos with lifetimes $\ll 1/H_0$ may decay into active-like neutrinos after neutrino decoupling injecting light neutrinos out of equilibrium into the cosmic neutrino background, and for $M_h \lesssim 10\,\mathrm{MeV}$ and $|H_{l h}|^2 \lesssim 10^{-6}$ \emph{may} decay after matter radiation equality into another heavy but lighter and stabler neutrino that may be a suitable DM candidate.

Furthermore, we have provided a detailed comparison between the non-equilibrium (non-thermal) distribution function obtained from pion decay and that of the Dodelson-Widrow scenario. The comparison was carried out within two different scenarios: i) saturation of DM abundance for each case, and ii) the parameters $M_h,|H_{lh}|$ were extracted from the upper bounds of the confidence band in the analysis of ref.\cite{kaplinghat2}. In the second case the distribution function from pion decay yields a substantial fraction of DM, in both cases the distribution function from pion decay  yields much shorter free streaming lengths. This comparison suggests that production via pion decay \emph{could} yield a substantial contribution to the DM abundance with a species that is colder than a thermal species but without invoking resonant production via a leptonic asymmetry.

The wide range of production mechanisms available throughout the thermal history of the Universe explored in this study  suggests the necessity of an exhaustive program to assess their contributions to DM from heavy neutrinos.

\vspace{2mm}

\textbf{Further Questions.}

Our study raises important questions that merit further and deeper investigation: \textbf{ i:)} we recognize that the possibility of MSW resonances in the medium would require going beyond the leading order in the mixing matrix element to obtain the quantum kinetic equations (this is also a caveat of the approach in ref.\cite{asaka}), one possible avenue would be to obtain the non-equilibrium effective action for the neutrino sector by tracing over the remaining degrees of freedom of the standard model (quarks, charged leptons, vector bosons) assumed to be in LTE. \textbf{ ii:)} we argued that \emph{collective excitations in the medium} could lead to the production of heavy neutrinos at high temperature, for example  via plasmon decay, this is a cooling  mechanism of stars in advanced stages of stellar evolution, and its high temperature counterpart in the early Universe \emph{may} be a suitable production mechanism of heavy neutrinos. This is an intriguing possibility that requires a thorough analysis implementing the hard thermal loop program\cite{htl,pisarskihtl,lebellac} to obtain the plasmon dispersion relations and couplings to neutrinos. \textbf{ iii:)} if there is a hierarchy of heavy neutrinos there is the possibility of  mixed DM: the contribution from different heavy neutrinos providing cold and  warm components depending on their masses and distribution functions. In this case (unlike the case where mixed DM arises solely from one component but from various production channels) it is far from clear what is the \emph{effective free streaming length and coarse grained phase space density}. Since the free streaming length determines the cutoff scale in the linear power spectrum of density perturbations it is important to obtain a reliable assessment of its interpretation in the case of mixed DM, is it possible that a very heavy neutrino  (cold DM) and a lighter one (hot DM) can mimic  warm DM?, is it possible that such mixture would lead to the same power spectrum as one single species of mass $\simeq 7 \,\mathrm{keV}$?. In order to shed light on these questions, the collisionless Boltzmann equation for several DM components in an expanding cosmology must be studied. Similar questions apply to the \emph{effective} coarse grained phase space density, as this quantity is observationally accessible (or inferred) from the kinematics of dwarf spheroidal galaxies and provides a powerful constraint on a DM candidate independently of abundance. We expect to report on some answers to these questions in later studies.

\acknowledgments The authors gratefully acknowledge support from NSF through grants PHY-1202227, PHY-1506912. L. L. is partially supported by the U.S. Department of Energy, Office of Science, Office of Workforce Development for Teachers and Scientists, Office of Science Graduate Student Research (SCGSR) program. The SCGSR program is administered by the Oak Ridge Institute for Science and Education for the DOE under contract number DE-AC05-06OR23100. L.L. thanks the U.S. Department of Energy for additional support under contract DE-SC0012704. He would also like to thank R. Pisarski for useful discussions. The authors would like to thank an anonymous referee for useful comments on the manuscript.

\appendix

\section{Quantum Kinetic Equation} \label{kinetics}

In this appendix we set up the quantum kinetic equation describing the sterile neutrino population arising from the reaction $\pi^{\pm} \rightleftharpoons l^{\pm} \nu_s (\bar{\nu}_s)$. The kinetics were originally set up in \cite{lellolightsterile} but is repeated here for completeness. This process will occur when the plasma is at temperatures below the QCD phase transition and it is acceptable to use an effective field theory which treat pions as the fundamental degrees of freedom. The low-energy effective interaction Hamiltonian responsible for this process is given by
eqn. (\ref{piHeff}) including finite temperature corrections to $f_\pi$.

The population buildup of the heavy neutrinos is described with a quantum kinetic equation given by form (\ref{quakin}) where the gain and loss terms are calculated by writing the quantum mechanical transition amplitudes from an initial state $i$ to a final state $f$, $|\mathcal{A}_{fi}|^2$, and integrating over kinematic region  of phase space. With the effective Hamiltonian (\ref{piHeff}), the relevant reactions are displayed in fig \ref{fig:kinetic}.

\begin{figure}[h!]
\includegraphics[height=4.0in,width=4.0in,keepaspectratio=true]{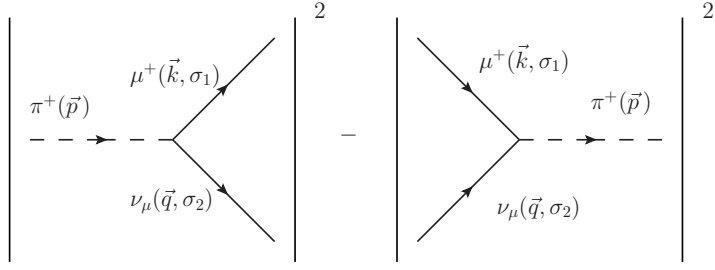}
\caption{The gain/loss terms for the quantum kinetic equation describing $\pi^{+} \rightarrow \bar{\mu} \nu_{\mu}$.}
\label{fig:kinetic}
\end{figure} The gain terms are due to the decay process $\pi^{+} \rightarrow \bar{l} \nu_{l}$ which starts from an initial state with $n_\pi(\vp)$ pion quanta and $n_\alpha(\vk),n_h(\vq)$ charged lepton and heavy neutrino mass eigenstate  respectively while the final state has $n_\pi(\vp)-1,n_\alpha(\vk)+1,n_h(\vq)+1$ quanta for each respective species. The Fock states for the decay process are given by

\be
|i \rangle = | n_{\pi}(p), n_{\bar{l}}(k), n_h(q) \rangle ~~;~~ |f \rangle = | n_{\pi}(p) - 1, n_{\bar{l}}(k)+1, n_{h}(q) +1 \rangle \,.
\ee In a similar fashion, the loss terms are obtained from the recombination $\bar{l} \nu_{l} \rightarrow \pi^+$ which has an initial state with $n_\pi(p),n_l(k),n_h(q)$ pion, charged lepton and neutrino quanta respectively. The final state is populated with $n_\pi(p)+1,n_l(k)-1,n_h(q)-1$ of the appropriate quanta and the Fock states for the loss process are given by

\be
|i \rangle = |n_{\pi}(p), n_{\bar{l}}(k), n_h(q) \rangle ~~;~~ |f \rangle = | n_{\pi}(p) + 1, n_{\bar{l}}(k)-1, n_h(q) - 1 \rangle .
\ee   Where the relation between  the neutrino  mass and flavor eigenstates are given by (\ref{relanus2}):

A standard calculation yields the transition amplitudes at tree-level and the matrix elements for the gain term (decay) is given by ( $\mathcal{H}_I$ is the interaction Hamiltonian density)

\bea
 \mathcal{A}_{fi} |_{gain} & = & -i\int d^4x \langle n_{\pi}(p) - 1, n_{\bar{l}}(k)+1, n_h(q) +1 |  \mathcal{H}_I(x) | n_{\pi}(p), n_{\bar{l}}(k), n_h(q) \rangle \\
\nonumber & = & i \sqrt{2} G_F V_{ud} f_{\pi}   H^*_{lh} \frac{2 \pi}{\sqrt{V}} \frac{ \bar{\mathcal{U}}_{\nu_h}(q,\sigma_1) \slashed{p} \mathcal{P}_L \mathcal{V}^{l}(k,\sigma_2) }{\sqrt{8 E_{\pi}(p) E_{l}(k) E_{\nu}(q) } } \\
\nonumber & \times & \delta_{\vp,\vk+\vq} \, \delta(E_{\pi}(p) - E_{\nu}(q) - E_{l}(k) \sqrt{n_{\pi}(p)} \sqrt{1-n_{l}(k)} \sqrt{1-n_{\nu}(q)} \,
\eea where $\mathcal{U},\mathcal{V}$ are Dirac spinors and $\sigma_{1,2}$ are helicity quantum numbers, while the transition amplitude  for the loss term (recombination) is given by

\bea
 \mathcal{A}_{fi} |_{loss} & = & -i\int d^4x \langle n_{\pi^+}(p) + 1, n_{\bar{l}}(k)-1, n_h(q) - 1 | \mathcal{H}_I(x) |n_{\pi}(p), n_{\bar{l}}(k), n_h(q)\rangle \\
 \nonumber & = & i  \sqrt{2}\, G_F V_{ud} f_{\pi}  H_{lh} \frac{2 \pi}{\sqrt{V}} \frac{ \bar{\mathcal{V}}^{\, l}(k,\sigma_2) \slashed{p} \mathcal{P}_L \mathcal{U}^{\nu_i}(q,\sigma_1) }{\sqrt{8 E_{\pi}(p) E_{l}(k) E_{\nu}(q) } }\\
 \nonumber & \times & \delta_{\vp,\vk+\vq \,} \delta(E_{\pi}(p) - E_{\nu}(q) - E_{l}(k)) \sqrt{n_{\pi}(p)+1} \sqrt{n_{l}(k)} \sqrt{n_{h}(q)} \,.
\eea  Summing over the final states   leads to the transition probability per unit time ($V$ is the quantization volume)

\bea
 \frac{1}{\mathcal{T}} \sum_{\vk,\vp,\sigma_1,\sigma_2} \overline{|\mathcal{A}_{fi} |^2}_{gain} & = &  \frac{(2\pi)}{2E_h(q)}   \int \frac{d^3 p}{(2\pi)^3} \frac{ 2 |H_{\alpha h}|^2 |V_{u d}|^2   G_F^2 f_{\pi}^2}{2E_{\pi}(p) 2 E_{l}(k)}\,Tr[\slashed{p} \mathcal{P}_L (\slashed{q} +M_h ) \slashed{p} \mathcal{P}_L (\slashed{k} -M_{l}) ]
    \nonumber \\
   & \times &    n_{\pi}(p) (1-n_{l}(k))(1-n_{\nu}(q))  \delta(E_{\pi}(p) - E_{h}(q) - E_{l}(k)) \\
 \nonumber \\ & = &   \frac{(2\pi)}{2E_h(q)}   \int \frac{d^3 p}{(2\pi)^3} \frac{ 2 |H_{\alpha h}|^2 |V_{u d}|^2   G_F^2 f_{\pi}^2}{2E_{\pi}(p) 2 E_{l}(k)}\, 2[2 (p \cdot q) (p \cdot k) - p^2 (q \cdot k) ] \nonumber \\ & \times &  n_{\pi}(p) (1- n_{l}(k))(1-n_{\nu}(q))   \delta(E_{\pi}(p) - E_{\nu}(q) - E_{l}(k))
\eea where  $\mathcal{T}$  is the total interaction time, not to be confused with temperature, and $ k= |\vp-\vq| $.    The loss term is calculated in the same way with the replacement $n_\pi \rightarrow 1 + n_\pi$ and $1-n \rightarrow n$ for leptons. With these steps and explicitly evaluating the energy/momentum conservation leads to the quantum kinetic equation governing the sterile neutrino population:

\be \frac{d n_h(q;t)}{dt} = \frac{d n_h(q;t)}{dt}\big|_{\textrm{gain}}-\frac{d n_h(q;t)}{dt}\big|_{\textrm{loss}}\,,\label{quakinpidecay}\ee
where
\bea \frac{d n_h(q;t)}{dt}\big|_{\textrm{gain}} & = &  \frac{2\pi}{2E_h(q)}\,\int \frac{d^3p}{(2\pi)^3} \, \frac{\overline{|\mathcal{M}_{fi} |^2}}{2E_\pi(p)2E_l(k)}\,
   n_{\pi}(p) (1 - n_{\bar{l}}(k) )(1 - n_{h}(q;t))  \nonumber \\ & \times &  \delta(E_\pi(p)-E_l(k)-E_h(q))  ~~;~~k = |\vp-\vq|\,,  \label{gainpidec}\eea

 \bea \frac{d n_h(q;t)}{dt}\big|_{\textrm{loss}}&  =  &  \frac{2\pi}{2E_h(q)}\, \int \frac{d^3p}{(2\pi)^3} \, \frac{\overline{|\mathcal{M}_{fi} |^2}}{2E_\pi(p)2E_l(k)}\,
  (1+n_{\pi}(p))  n_{\bar{l}}(k) n_{h}(q;t )\nonumber \\ & \times & \delta(E_\pi(p)-E_l(k)-E_h(q))  ~~;~~k = |\vp-\vq|\,, \label{losspidec}\eea and the averaged $\overline{|\mathcal{M}_{fi}|^2}$ matrix element is given by
 \be \overline{|\mathcal{M}_{fi} |^2} = 4 |H_{\alpha h}|^2 |V_{u d}|^2   G_F^2 f_{\pi}^2 \,  [2 (p \cdot q) (p \cdot k) - p^2 (q \cdot k) ] \,. \ee
 Therefore
\be  \frac{d n_h(q;t)}{dt} = \Gamma^<(q) (1-n_h(q;t))-\Gamma^>(q)n_h(q;t)\,,  \label{kineqpidec}\ee with the gain and loss rates given by
 \bea \Gamma^<(q)  & = &  \frac{2\pi}{2E_h(q)}\,\int \frac{d^3p}{(2\pi)^3} \, \frac{\overline{|\mathcal{M}_{fi} |^2}}{2E_\pi(p)2E_l(k)}\,
   n_{\pi}(p) (1 - n_{\bar{l}}(k) ) \nonumber \\ & \times &  \delta(E_\pi(p)-E_l(k)-E_h(q)) \label{gainpidec2}\\
   \Gamma^>(q)  & = &  \frac{2\pi}{2E_h(q)}\,\int \frac{d^3p}{(2\pi)^3} \, \frac{\overline{|\mathcal{M}_{fi} |^2}}{2E_\pi(p)2E_l(k)}\,(1+n_{\pi}(p))  n_{\bar{l}}(k)  \nonumber \\ & \times &  \delta(E_\pi(p)-E_l(k)-E_h(q))~~;~~ k= |\vp-\vq| \,.   \label{losspidec2}\eea
  Performing the angular integration using the delta function constraint we find

\bea   \Gamma^<(q)  & = & \frac{ |H_{\alpha h} |^2 |V_{u d}|^2 G_F^2 f_{\pi}^2 }{8 \pi} \frac{M^2_{\pi}(M^2_{l} + M^2_{h} ) - (M^2_{l} -M^2_{\nu})^2}{q E_{h}(q)}
 \nonumber \\ & \times  & \int^{p_+}_{p_-} \frac{dp \, p}{\sqrt{p^2 + M_{\pi}^2 } } \Big[ n_{\pi}(p) (1 - n_{\bar{l}}(|\vp-\vq|)) \Big] \label{gamalesfi2}\eea
 \bea   \Gamma^>(q)  & = & \frac{ |H_{\alpha h} |^2 |V_{u d}|^2 G_F^2 f_{\pi}^2 }{8 \pi} \frac{M^2_{\pi}(M^2_{l} + M^2_{h} ) - (M^2_{l} -M^2_{\nu})^2}{q E_{h}(q)}
 \nonumber \\ & \times  & \int^{p_+}_{p_-} \frac{dp \, p}{\sqrt{p^2 + M_{\pi}^2 } } \Big[ (1+ n_{\pi}(p)) n_{\bar{l}}(|\vp-\vq|) \Big] \label{gamalgreatfi2}\eea
    The integration limits $p_{\pm}$ are obtained from the constraint

\be
[ (|\vp|-|\vq|)^2 + M^2_{l} ]^{1/2} \leq E_{\pi}(p) - E_{\nu}(q) \leq [ (|\vp|+|\vq|)^2 + M^2_{l} ]^{1/2}  .
\ee The solutions are given by

\be
p_{\pm} = \left| \frac{E_{h}(q)}{2M_h^2} [ (M_{\pi}^2 - M^2_{l} + M_h^2)^2 -4M^2_{\pi}M^2_{h} ]^{1/2} \pm \frac{q(M^2_{\pi} - M^2_{l} + M^2_{h} )}{2M^2_{h}} \right|  \,. \label{limits}
\ee Note that these bounds coalesce at threshold when $M_{\pi}^2-M_l^2+M_h^2 = 2M_{\pi}^2 M_h^2$ and the population change vanishes as expected.

  Using the relations
 \be 1+n_\pi(p) = e^{ {E_\pi(p)}/{T}}\,n_\pi(p) ~~;~~ 1-n_l(k) = e^{{E_l(k)}/{T}}\,n_l(k) \label{pisrel}\ee and using the energy delta function constraints, the detailed balance condition follows, namely
 \be \Gamma^<(q) e^{{E_h(q)}/{T}} = \Gamma^>(q)\,.  \label{pidetbal}\ee

 These results are extended to cosmology by replacing the momentum with the physical momentum, $q \rightarrow Q_f = q_c/a(t)$ where physical energy and momentum is measured by an observer at rest with respect to the expanding spacetime.

\section{ Approximate Distributions.} \label{app:approximation}

The exact rate equation \ref{exactrate} is unwieldy and only limited analytical progress can be made. Recasting the rate equation in terms of the definitions of \ref{ratiosm},\ref{delgran},\ref{delsmal} and a dimensionless variable for the neutrino energy

\be
\varepsilon \equiv E_{h}/T  = \sqrt{y^2+ \frac{M_{h}^2}{M_{\pi}^2}\tau^2} = y \sqrt{1 + u} ~~;~~ u = m_h^2 \frac{\tau^2}{y^2} \label{enervar}
\ee leads to the following form of the rate equation which is more suitable to manipulations:

\bea
&& \frac{1}{\Lambda_{lh}} \frac{dn_{lh}}{d\varepsilon}(y,\varepsilon)   =  \left(\frac{1}{m_{h}}\right)^3   \frac{1}{y } \frac{ [(\varepsilon^2-y^2)^{1/2} - \frac{M_{h}^2}{6 f_{\pi}^2}(\varepsilon^2-y^2)^{-1/2}]e^{-\varepsilon}}{1+e^{-\varepsilon}} \times   \\ && \Bigg\{
\ln \left(\frac{ 1 - \exp\left(-\frac{1}{2m_h^2}(\delta_{lh} y + \Delta_{lh} \varepsilon)\right) } {1-\exp\left(-\frac{1}{2m_h^2}(-\delta_{lh} y + \Delta_{lh} \varepsilon)\right) } \right)   - \ln \left(\frac{ 1 + \exp\left(-\frac{1}{2m_h^2}(\delta_{lh} y + (\Delta_{lh}-2m_h^2) \varepsilon)\right) } {1+\exp\left(-\frac{1}{2m_h^2}(-\delta_{lh} y + (\Delta_{lh}-2m_h^2) \varepsilon)\right)} \right) \Bigg\} \nonumber \,.
\eea The population build up is obtained by integrating

\be
n(\varepsilon_0,y) = \int^{\infty}_{\varepsilon_0} d\varepsilon' \frac{dn}{d\varepsilon}(\varepsilon',y) ~~;~~ \varepsilon_0 = \sqrt{y^2 + m_h^2 \tau_0^2} = y\sqrt{1+u_0}\,.
\ee where the initial population of $\nu_h$ has been neglected and the value of $\varepsilon_0$ is set by the temperature at which pions appear in thermal equilibrium, assumed almost immediately after the hadronization transition. It is assumed that pion thermalization happens instantaneously at the QCD phase transition and we set $\tau_0 \sim 1$ which is justified by the lattice results of \cite{hotqcd} which suggest a continuous transition that allows for thermalization on strong interaction time scales.

The arguments of the exponentials inside of the logarithms can be shown to be $<0$ so that the logarithms can be expanded consistently in a power series. Upon expanding the logarithms and integrating the rate, the population becomes

\be
\frac{1}{\Lambda_{lh}} n_{lh}(\tau_0,y) = \left(\frac{1}{m_{h}}\right)^3 \frac{1}{y} \sum_{k=1}^{\infty} \frac{2}{k} \sinh\left(\frac{k \delta_{lh}}{2 m_h^2} y \right) \left( \mathcal{I}_{lh}(k,y) - \frac{M_{h}^2}{6f_{\pi}^2} \mathcal{J}_{lh}(k,y) \right) \label{analytic}
\ee where the expressions $\mathcal{I}_{lh}(k),\mathcal{J}_{lh}(k)$ are given by

\bea \label{uvar1}
\mathcal{I}_{lh}(k,y) =  \frac{y^2}{2}\sum_{j=0}^{\infty} (-1)^j & \Bigg[ & \int^{\infty}_{u_0} du \frac{\sqrt{u}}{\sqrt{1+u}} \exp\left(-\left(j+1+\frac{k \Delta_{lh}}{2 m_h^2}\right)y\sqrt{1+u}\right)  \\
&+& (-1)^{k+1}  \int^{\infty}_{u_0} du \frac{\sqrt{u}}{\sqrt{1+u}} \exp \left(-\left(j+1+\frac{k \Delta_{lh}}{2 m_h^2}-k\right) y\sqrt{1+u}\right) \Bigg] \nonumber
\eea

\bea \label{uvar2}
\mathcal{J}_{lh}(k,y) =  \frac{1}{2}\sum_{j=0}^{\infty} (-1)^j & \Bigg[ & \int^{\infty}_{u_0} du \frac{1}{\sqrt{u}\sqrt{1+u}} \exp \left(-\left(j+1+\frac{k \Delta_{lh}}{2 m_h^2} \right)y\sqrt{1+u} \right)  \\
&+& (-1)^{k+1} \int^{\infty}_{u_0} du \frac{1}{\sqrt{u}\sqrt{1+u}} \exp \left(-\left(j+1+\frac{k \Delta_{lh}}{2 m_h^2}-k\right) y\sqrt{1+u} \right) \Bigg] \,. \nonumber
\eea These expressions can be written in terms of the incomplete modified Bessel functions \cite{incmodbes} but this proves to be an unilluminating exercise. However, it is possible that slightly simpler expressions can be found under appropriate approximations such as the assumption that the neutrinos are produced either ultra-relativistically or non-relativistically over the duration of production. The dimensionless variable $u = m_h^2 \,\tau^2/y^2$, which was introduced in Eq \ref{enervar}, is the obvious quantity which controls how much of an ultra-relativistic or non-relativistic dispersion relation is obeyed by the heavy neutrino; specifically $u \ll 1$ implies a light, ultra-relativistic species while $u \gg 1$ implies a very heavy, non-relativistic species.

From figs \ref{fig:lightedist},\ref{fig:heavyedist},\ref{fig:mudist}, it can be seen that $y,\tau$ take the values $1 < \tau < 10$ and $0 < y < 3$ from the beginning of production until decoupling which implies that $0 < y/ \tau < 3$ for any neutrino produced. The production of purely ultra-relativistic neutrinos implies that $u \ll 1$ or that $m_{\nu}/m_{\pi} \ll y / \tau$ for all values of $\tau,y$ throughout production until freezeout. This assumption will hold for values of $m_{\nu}/m_{\pi} \ll 1$ or for very large values of $y$. For ultra-relativistic production, \ref{uvar1} \ref{uvar2} simplify to

\bea
\mathcal{I}_{lh}(k,y) & - & \frac{M_{h}^2}{6f_{\pi}^2} \mathcal{J}_{lh}(k,y) \Big|_{UR} =  \left(\frac{1}{m_h}\right)^3 \frac{1 + (-1)^{k+1}e^{ky} }{ 1+e^y}   \exp\left(-\frac{k \Delta_{lh}}{2 m_h^2}y \right)  \\
 & * & \left(\frac{2 \tau_0}{k \Delta_{lh}} \exp\left(-\frac{k \Delta_{lh} \tau_0^2}{4 y}\right) + \left( \frac{1}{k \Delta_{lh} y} \right)^{1/2} \left(\frac{2 y}{k \Delta_{lh}} - \frac{M_{\pi}^2}{6 f_{\pi}^2}\right)\Gamma\left(1/2,\frac{k \Delta_{lh} \tau_0^2}{4 y}\right) \right) \,. \nonumber
\eea where $\Gamma(\nu,z)$ is the incomplete gamma function. Further assuming $M_{h}/M_{\pi} \ll 1$ leads to another simplification

\be
2 \sinh\left(\frac{k \delta_{lh}}{2 m_h^2} y \right) \rightarrow \exp\left(\frac{k \Delta_{lh}}{2 m_h^2} y\right) \exp\left(-k y/\Delta_{lh}\right) \,.
\ee With these approximations, the form of the distribution for production of purely ultra-relativistic neutrinos becomes

\be \label{relapp}
n_{lh}(\tau_0,y) \Big|_{UR}=\frac{\Lambda_{lh}}{y^2} \sum_{k=1}^{\infty} \left[ \frac{1 + (-1)^{k+1} e^{ky}}{1 + e^{y}} \right] \frac{\exp\left(-k y/\Delta_{lh}\right)}{k} \times J_k(\tau_0,y)
\ee where

\be
J_k(\tau_0,y) = 2 \tau_0 \left(\frac{ y}{k \Delta_{lh}}\right)  \exp\left(-\frac{k \Delta_{lh} \tau_0^2}{4 y}\right) + \left( \frac{ y}{k \Delta_{lh}} \right)^{1/2} \left[\frac{2 y}{k \Delta_{lh}} - \frac{M_{\pi}^2}{6 f_{\pi}^2}\right]\Gamma\left(1/2,\frac{k \Delta_{lh} \tau_0^2}{4 y}\right) \,.
\ee This expression is equivalent to the distribution which was obtained in \cite{lellolightsterile} and explicitly depends on the time at which production begins, $\tau_0$. In \cite{lellolightsterile} it is shown that this approximate distribution is in excellent agreement with the distribution obtained from the exact numerical solution provided that $M_h \lesssim 1 MeV$.

From figs \ref{fig:lightedist},\ref{fig:heavyedist},\ref{fig:mudist}, it is clear that for massive neutrinos with $M_h \gtrsim 10MeV$ there is a very differently shaped distribution function at small values of $y$ compared with that seen from heavy neutrinos with masses below an MeV (as studied in \cite{lellolightsterile}). This switch in shape between the distributions can be understood in terms of the production of \emph{non-relativistic neutrinos} such that $u \gg 1$. The assumption that $u\gg 1$ implies that $m_{\nu}/m_{\pi} \gg y/\tau$ for all values of $\tau,y$ and, since $m_{\nu}/m_{\pi}<1$, this approximation is clearly only valid for a range of values for $y \ll 1$ which explains the plateau in the distribution function for the heavier species. Under this assumption, the expressions \ref{uvar1},\ref{uvar2} become

\bea \label{nonrelapp}
\mathcal{I}_{lh}(k,y) \Big|_{NR} & = & \sum_{j=0}^{\infty} (-1)^j \frac{ \exp \left[-(j+1+\frac{k \Delta_{lh}}{2m_h^2})y\sqrt{1+u_0} \right] }{(j+1+\frac{k \Delta_{lh}}{2m_h^2})^2} \Bigg[ \left(1+ (j+1+\frac{k \Delta_{lh}}{2m_h^2})y\sqrt{1+u_0}\right) \nonumber \\
 +(-1)^{k+1} & e^{ky\sqrt{1+u_0}} &   \left( \frac{j+1+\frac{k \Delta_{lh}}{2m_h^2}} {j+1+\frac{k \Delta_{lh}}{2m_h^2}-k}\right)^2\left(1+ (j+1+\frac{k \Delta_{lh}}{2m_h^2}-k )y\sqrt{1+u_0}\right) \Bigg] - \frac{y^2}{2} \mathcal{J}_k  \nonumber \\
\mathcal{J}_{lh}(k,y) \Big|_{NR} & = &  \sum_{j=0}^{\infty} \frac{(-1)^j }{2} \Bigg[ \Big( \exp \left[(j+1+\frac{k \Delta_{lh}}{2m_h^2})y \right] E_1((j+1+\frac{k \Delta_{lh}}{2m_h^2})y(\sqrt{1+u_0}+1))  \nonumber \\
 & + & \exp \left[-(j+1+\frac{k \Delta_{lh}}{2m_h^2})y \right] E_1((j+1+\frac{k \Delta_{lh}}{2m_h^2})y(\sqrt{1+u_0}-1)) \Big) \nonumber \\
 & + & (-1)^{k+1} \Big( \exp \left[(j+1+\frac{k \Delta_{lh}}{2m_h^2}-k )y\right] E_1 ((j+1+\frac{k \Delta_{lh}}{2m_h^2} - k )y(\sqrt{1+u_0}+1))  \nonumber \\
 & + & \exp \left[-(j+1+\frac{k \Delta_{lh}}{2m_h^2} -k)y \right] E_1((j+1+\frac{k \Delta_{lh}}{2m_h^2}-k)y(\sqrt{1+u_0}-1)) \Big)
  \Bigg] \nonumber \\
\eea where $E_1(x)$ is the exponential integral. In the ultra-relativistic limit, exploiting $M_{h} \ll M_{\pi}$ was used in order to ignore several terms up to leading order in $M_{h}/M_{\pi}$; this allowed the sum, $\sum_j$, from \ref{uvar1},\ref{uvar2} to be evaluated analytically. With heavier sterile neutrinos, such an approximation is unavailable and we are forced to retain the expression in the form of \ref{nonrelapp} if we desire any nontrivial momentum dependence.

\end{document}